\documentclass[apj,onecolumn]{emulateapj}
   \usepackage{graphicx}
   \usepackage{amsmath}
   \usepackage{color}
   \usepackage{epstopdf}
   \usepackage{pdfsync}

\newcommand{\beq}{\begin{equation}}
\newcommand{\eeq}{\end{equation}}

\def\fsc{\alpha_f} 
\def\lambar{\lambda\llap {--}}

             \font\sevenrm=cmr7

          \font\sixrm=cmr6       

\def\dover#1#2{\hbox{${{\displaystyle#1 \vphantom{(} }\over{
   \displaystyle #2 \vphantom{(} }}$}}
\def\calFErber{{\cal F}_{\hbox{\sevenrm Erber}}}
\def\rns{R_{\hbox{\sixrm NS}}}
\def\rlc{R_{\hbox{\sixrm LC}}}
\def\psilc{\Psi_{\hbox{\sixrm LC}}}
\def\muB{\mu_{\hbox{\sixrm B}}}
\def\erg{\varepsilon}
\def\eesc{\varepsilon_{\rm esc}}
\def\thetakB{\theta_{\hbox{\sevenrm kB}}}
\def\delthetaapp{\Delta\theta_{\hbox{\sevenrm app}}}
\def\betavec{\overrightarrow{\beta}}
\def\omegaS{\omega_{\hbox{\sixrm S}}}

\def\rS{r_{\hbox{\sixrm S}}}
\def\thetaS{\theta_{\hbox{\sixrm S}}}
\def\phiS{\phi_{\hbox{\sixrm S}}}

\def\ergO{\erg_{\hbox{\sixrm O}}}
\def\rO{r_{\hbox{\sixrm O}}}
\def\thetaO{\theta_{\hbox{\sixrm O}}}
\def\phiO{\phi_{\hbox{\sixrm O}}}

\def\omegaSE{\omega_{\hbox{\sixrm S,E}}}
\def\rE{r_{\hbox{\sixrm E}}}
\def\thetaE{\theta_{\hbox{\sixrm E}}}
\def\phiE{\phi_{\hbox{\sixrm E}}}
\def\betaE{\beta_{\hbox{\sixrm E}}}
\def\betaEvec{\overrightarrow{\beta_{\hbox{\sixrm E}}}}
\def\gammaE{\gamma_{\hbox{\sixrm E}}}
\def\deltaE{\delta_{\hbox{\sixrm E}}}
\def\PsiE{\Psi_{\hbox{\sixrm E}}}

\def\teq#1{$\, #1\,$}                         
\def\citesmart#1{\citeauthor{#1}\, (\citeyear{#1})}
\def\citesmartp#1{\citeauthor{#1}\, \citeyear{#1}}
\def\citesmartpp#1{(\citeauthor{#1}\, \citeyear{#1})}

\begin{document}

\newcommand{\vol}[2]{$\,$\bf #1\rm , #2.}         
        
\newcommand{\twofigureoutpdf}[6]{\begin{figure}\centerline{}
   \centerline{\includegraphics[width=#3truein]{#1}
                 \hspace{0.05truein} \includegraphics[width=#3truein]{#2}}
   \vspace{#4truein} \caption{#5} \centerline{} \label{#6} \end{figure}}
\newcommand\citesmartptwo[2]{({\citeauthor{#1}\, \citeyear{#1};}{\ \citeauthor{#2}\, \citeyear{#2}})}
\newcommand\citesmartpthree[3]{({\citeauthor{#1}\, \citeyear{#1};}{\ \citeauthor{#2}\, \citeyear{#2};}{\ \citeauthor{#3}\, \citeyear{#3}})}
\newcommand\citesmartpfour[4]{({\citeauthor{#1}\, \citeyear{#1};}{\ \citeauthor{#2}\, \citeyear{#2};}
{\ \citeauthor{#3}\, \citeyear{#3};}{\ \citeauthor{#4}\, \citeyear{#4}})}

\title{MAGNETIC PAIR CREATION TRANSPARENCY IN GAMMA-RAY PULSARS}

   \author{Sarah A. Story and Matthew G. Baring}
   \affil{Department of Physics and Astronomy, MS 108,\\
      Rice University, Houston, TX 77251, U.S.A.\\
      {\rm ss16@rice.edu, baring@rice.edu}}

\begin{abstract}
Magnetic pair creation, \teq{\gamma\to e^+e^-}, has been at the core of
radio pulsar paradigms and central to polar cap models of gamma-ray
pulsars for over three decades. The {\it Fermi} gamma-ray pulsar population
now exceeds 140 sources and has defined an important part of {\it Fermi}'s
science legacy, providing rich information for the interpretation of
young energetic pulsars and old millisecond pulsars. Among the
population characteristics well established is the common occurrence of
exponential turnovers in their spectra in the 1--10 GeV range.  These turnovers are too
gradual to arise from magnetic pair creation in the strong magnetic
fields of pulsar inner magnetospheres.  By demanding insignificant
photon attenuation precipitated by such single-photon pair
creation, the energies of these turnovers for {\it Fermi} pulsars can be used
to compute lower bounds for the typical altitude of GeV band emission. 
This paper explores such pair transparency constraints below the
turnover energy, and updates earlier altitude bound determinations
of that have been deployed in various {\it Fermi} pulsar papers.  For low
altitude emission locales, general relativistic influences are found to
be important, increasing cumulative opacity, shortening the photon
attenuation lengths, and also reducing the maximum energy that permits
escape of photons from a neutron star magnetosphere. Rotational aberration 
influences are also explored, and are found to be small at low altitudes, 
except near the magnetic pole.  The analysis
presented in this paper clearly demonstrates that including
near-threshold physics in the pair creation rate is essential to
deriving accurate attenuation lengths and escape energies.  The altitude bounds are
typically in the range of 2-7 stellar radii for the young {\it Fermi} pulsar
population, and provide key information on the emission altitude in
radio quiet pulsars that do not possess double-peaked pulse profiles.
The bound for the Crab pulsar is at a much higher altitude, with the putative
detection by MAGIC out to 350--400 GeV implying a lower bound of 310km
to the emission region, i.e., approximately 20\% of the light cylinder
radius. These results are also extended to the super-critical field
domain, where it is found that emission in magnetars originating
below around 10 stellar radii will not appear in the {\it Fermi}-LAT band.
\end{abstract}


\maketitle

\vspace{-15pt}
\section{INTRODUCTION}
 \label{sec:intro}

The {\it Fermi} Gamma-ray Space Telescope has revolutionized our
understanding of high-energy emission from pulsars.  Prior to the launch
of {\it Fermi}, there were only 7 high-confidence detections \citep{THHU97} of
gamma-ray pulsars from the EGRET telescope aboard the Compton Gamma-ray
Observatory (CGRO), of which all but Geminga had a radio counterpart.
Except for Geminga, which is extremely bright, EGRET was not sensitive
enough to perform blind searches, the process of discerning pulsation in
pulsars using their gamma-ray data alone, i.e. without the guide of an
existing radio ephemeris.  Furthermore, the maximum observed photon
energy, typically in the range 1--10 GeV, was just outside the upper end
of EGRET's sensitive energy range.  With the launch of {\it Fermi}, a
wealth of new data became available.  In just five years, the gamma-ray
pulsar sample increased from 7 to over 120 pulsars \citep[][lists 117 
in the second {\it Fermi}-LAT pulsar catalog]{PSRCat2},
including over three dozen millisecond pulsars and over
35 pulsars discovered in {\it Fermi} blind searches \citep{Blind Search 1,
Catalog, Blind Search 2, PSRCat2}.  The overwhelming majority of these
blind search pulsars have been shown to have no discernible radio
counterparts, with upper limits to fluxes at the \teq{30\mu}Jy level
\citep{PSRCat2}.  {\it Fermi}'s increased sensitivity allows the detection of
fainter pulsars, and this combined with better time resolution has given
us more detailed pulse shapes than EGRET could provide.  The energy
window centered on a few GeV is now easily observable for the first
time.  This has yielded clear observations of spectral cutoffs and
determinations of their shapes in the vast majority of pulsars of all
classes: old millisecond ones, young radio-quiet and young radio-loud
rotators.  Such revelations have made it possible to resolve some
long-standing questions about the origins of pulsar high-energy
emission.

Prior to the launch of {\it Fermi}, there were two competing predictions for
the shape of the pulsar spectral cutoff.  Outer gap models, driven by
curvature radiation physics, predicted a simple exponential cutoff 
\citep[see, for example,][]{CR94}, corresponding to the emission by electrons
possessing a maximum Lorentz factor. A similar picture exists for slot
gap models \citep{MH04} that extend polar cap-driven emission to high
altitudes. In contrast, polar cap models \citep{DH96} based on low
altitude photon emission, magnetic pair creation \teq{\gamma\to e^+e^-} 
and pair cascading predict a
super-exponential cutoff due to the very strong dependence of the pair
production rate on photon energy.   EGRET data were equally consistent
with either cutoff scenario \citep{RH07}.  With far greater statistics,
early {\it Fermi}-LAT observations  of the Vela pulsar clearly exhibited
a simple exponential cutoff \citep{Vela 1}, and subsequent observations
of Vela and other pulsars have corroborated this shape, demonstrating
that exponential cutoffs are present in the
phase-resolved spectroscopic data \citep{Vela 2}.  Super-exponential
spectral turnovers in {\it Fermi} GeV band data can be ruled out to high
degrees of significance.  This fact can be used to place a physical
lower bound on the altitude of origin for the high-energy emission.  The
magnetic pair creation process is strongly height-dependent and should
dominate at low altitudes.  Since the signature of strong pair creation
- a super-exponential cutoff in the spectrum - is not observed, the
emission altitude must be high enough that attenuation due to
single-photon pair production is not expected.

Even though magnetic pair creation-driven cutoffs do not occur in the
{\it Fermi} pulsar sample, performing calculations of magnetic pair production
transparency is still a worthwhile exercise.  The associated physical
lower bounds for the emission height should be considered as a
complement to geometric determinations of the emission height from
gamma-ray and gamma-ray/radio peak separation in caustic scenarios
\citep{Watters,PGHG10,VJH12}. In particular, magnetic pair creation
altitude bounds can help constrain magnetospheric geometry in pulsars
that do not possess two distinct gamma-ray peaks \citep[about 30\% of the
blind search pulsars:][]{Blind Search 2} and are radio quiet; such pulsars are 
not as easily amenable to altitude diagnostics using caustic geometry analysis.  
Furthermore, pair production rates stemming from opacity computations are 
important for the understanding of pulsar wind nebula energetics.  The
Goldreich-Julian currents alone cannot carry enough energy to account
for PWN luminosities \citep{RG74,deJager07,BAA11}, and to achieve the
required energy deposition, there must be prolific pair creation
occurring in the pulsar magnetosphere.  Single-photon magnetic pair
creation is very efficient at low altitudes and can produce large pair
multiplicities \citep{DH82,MH03} approaching, but still somewhat lower than, 
those needed to achieve the required nebular energy deposition.

Pair opacity calculations date from early pulsar theory, such as in
the work of \cite{AS79}.
\citet{HEF90}, working on early gamma-ray burst theory, recognized that
$\gamma-B$ attenuation posed a major problem for the escape of
gamma-rays from the neutron star surface.  Their calculations, which
ignored general relativistic (GR) and aberration effects, showed that
for the escape probability to be significant at soft gamma-ray energies,
emission must be strongly collimated around the local magnetic field.  For
the higher-energy gamma-rays seen by {\it Fermi}, relativistic beaming
guarantees that photons will be emitted essentially parallel to the
local magnetic field.  In \citet{HBG97}, although the focus was on
photon splitting, the authors carried out single-photon pair production
attenuation calculations for comparison purposes.  These calculations
included detailed consideration of threshold effects in the computation
of photon attenuation lengths and escape energies, the latter defining
the critical energies above which the magnetosphere is opaque to 
photon passage for a given emission locale.  In an extension of
this analysis, \citet{BH01} illustrated the character of magnetic pair
creation and photon splitting opacities by exploring the dependence of photon
escape energy on the colatitude of emission for each process, for photons originating at
the neutron star surface.  They also discussed cascading and the
conditions under which pair creation (and therefore, arguably, radio
emission) should be effectively quenched.  Most recently, \citet{Lee10}
tackled the problem of $\gamma-B$ attenuation in detail.  Their work,
which produced lower bounds for emission altitudes as a function of
photon energy, incorporated potentially critical aberration and GR
corrections, but largely ignored the threshold behavior of the
\teq{\gamma\to e^+e^-} rate.

The physics that determines the form of the $\gamma-B$ attenuation
coefficient is discussed in some detail in
Section~\ref{sec:pair_physics}.  An early offering that described this
first-order QED process in a manageable form was in the seminal work by
\citet{Erber}, which provided a simple asymptotic form of the
attenuation coefficient.  \citet{TE74} subsequently dealt in detail with
the differences in photon polarization modes.  Near the pair creation
threshold, the simple asymptotic approximations obtained in these works
become less accurate, differing on average by over two orders of
magnitude from exact pair production rates in fields below around 4
TeraGauss. \citet{DH83} provided an empirical approximation to threshold
behavior, while formally precise forms were offered in
the works of \citet{B88} and \cite{BK07}; none of these is quite as simple as the form
highlighted in \citet{Erber}.  These threshold corrections are important
to address in pair opacity computations involving regions near the stellar
surface, when the local field is near-critical or higher, i.e. especially for
magnetars.

In this work, we have taken an analytical approach to the problem of
pair creation opacity whenever possible.  We present magnetic pair
creation transparency conditions as a function of colatitude and height
of emission for photons emitted parallel to the local magnetic field, as
is approximately the case for curvature emission.   Our integrals
for the magnetic pair creation optical depth are computed
for a variety of photon energies and surface polar magnetic fields
\teq{B_p}. We have included, analytically where possible, corrections
for threshold conditions on magnetic pair creation, gravitational
redshift, general relativistic magnetic field distortion, and aberration
due to neutron star rotation. In Section~\ref{sec:flatspace} it is found
that in flat spacetime, the maximum energy \teq{\eesc} of a photon
that can escape the magnetosphere is a declining function of
the emission colatitude \teq{\thetaE} and the field \teq{B_p}.  In
particular, for \teq{B_p < 4 \times 10^{12}}Gauss, the relationship
\teq{\eesc B_p\sin\thetaE\sim} {\it constant} \;\; is borne out, in agreement 
with \citet{AS79} and 
\citet{CCH96}, a direct consequence of the asymptotic form \citep{Erber}
of the pair production rate.  When the surface polar field exceeds
around \teq{B_p\sim 10^{13}}Gauss, the threshold influences become
profound, and the dependence of \teq{\eesc} on \teq{B_p} weakens
substantially.  If one fixes the escape energy, the altitude at which a photon
can be emitted and emerge from the magnetosphere unscathed by magnetic
pair attenuation is a monotonically increasing function of colatitude
\teq{\thetaE}.

Including general relativity effects (see Section~\ref{sec:GR}) reduces
the attenuation length for pair creation, lowers the escape energies for
surface emission locales by 20--30\% for \teq{B_p < 4 \times
10^{12}}Gauss (and around a factor of two for \teq{B_p \sim 4 \times
10^{13}}Gauss) and raises the minimum altitudes of emission by at most
10--20\%.  For emission points above two stellar radii, GR influences
are generally insignificant. Including aberration effects (see
Section~\ref{sec:aberration}) dramatically
raises the minimum altitudes \teq{r_{\rm min}} for pair transparency at
small colatitudes above the magnetic pole.  For most emission azimuthal 
angles, the minimum altitude of emission increases monotonically 
with colatitude.  In addition, \teq{r_{\rm min}} quickly maps over
to the flat spacetime, non-rotating magnetosphere results when
\teq{\thetaE\gtrsim 10^{\circ}} --- then aberration influences are
largely minimal in the inner magnetosphere because the co-rotation
speeds are far inferior to \teq{c}. This monotonic trend for \teq{r_{\rm
min}} continues right up to above the magnetic equator, because of
the relative ease with which photons cross field lines when propagating
at high magnetic colatitudes.  In particular, we do not reproduce 
the putative decline of \teq{r_{\rm min}} as \teq{\thetaE} approaches
\teq{90^{\circ}} that is claimed in \citet{Lee10}, and attributed
therein to the influences of aberration.

Our pair transparency computations determine that the emission
altitude lower bounds calculated for {\it Fermi}-LAT pulsars are far
below the altitudes of emission calculated with geometric
(pulse-profile) methods.  Moreover, the detection of pulsed emission
\citep{Aliu_Crab} from the Crab pulsar at 120 GeV by VERITAS puts its minimum
altitude of emission at about 20 neutron star radii, and this
increases to around 31 stellar radii (20\% of the light cylinder radius) if the 
pulsed detection up to \teq{350-400}GeV by MAGIC \citep{Aleksic Crab} is adopted.  In addition,
applying our results to supercritical field domains, we find that escape
energies in magnetars are generally below around 30 MeV, thereby
precluding emission in the {\it Fermi}-LAT band unless the altitude is
above around 10 stellar radii.


\section{REACTION RATES FOR MAGNETIC PAIR CREATION}
 \label{sec:pair_physics}

The form of the magnetic pair creation rate is a 
central piece of the pair attenuation calculation.  The physics of this 
purely quantum process has been understood since the early work of 
\citet{Toll52} and \citet{Klep54}.  This one-photon conversion 
process, \teq{\gamma\to e^+e^-} is forbidden in field-free regions due
to four-momentum conservation.  In the presence of an electromagnetic 
field, there is a lack of translational invariance orthogonal to the 
field, so that momentum perpendicular to {\bf B} does not have to be 
conserved; it can be absorbed by the global field structure.  In 
quantum electrodynamics (QED), this process is first order in the 
fine structure constant \teq{\fsc = e^2/\hbar c}, possessing a Feynman diagram with just a 
single vertex.  Accordingly, within the confines of QED perturbation theory, 
it is the strongest photon conversion process in strong-field environments, 
and its rate only becomes significant when the field strength 
approaches the quantum critical field \teq{B_{\rm cr}=m_e^2c^3/(e\hbar)
=4.413\times 10^{13}}Gauss, at which the cyclotron energy equals 
\teq{m_ec^2}.  Since energy is conserved, the absolute threshold 
for \teq{\gamma\to e^+e^-} is \teq{2m_ec^2}, and 
because of Lorentz transformation properties along {\bf B}, 
when photons propagate at an angle \teq{\thetakB} to the 
field, the threshold becomes \teq{2m_ec^2/\sin\thetakB} 
for photons with parallel polarization.

In general, the produced pairs occupy excited Landau 
levels in a magnetic field, and since the process generates pairs 
with identical momenta parallel to {\bf B} at the threshold 
(for \teq{B\ll B_{\rm cr}}) for each Landau level configuration of the pairs, 
the reaction rate \teq{{\cal R}} exhibits a divergent resonance at
each pair state threshold, producing a characteristic sawtooth structure  
(\citet{DH83}, hereafter DH83; see also \citet{BK07}).  
Near threshold, there are relatively few kinematically-available pair states; for 
photon energies \teq{\omega m_ec^2} well above
threshold, the number of pair states becomes large. 
Since the divergences are integrable in photon energy space, 
mathematical approximations of the complicated exact rate can 
be developed using proper-time techniques originally due to \citet{Schwin51}. 
These essentially form averages over \teq{\omega} of the resonant contributions,
and provide the user with convenient asymptotic expressions for the 
polarization-dependent attenuation coefficient.  The most 
widely-used expressions of this genre are those derived in \citet{Klep54},
\citet{Erber}, \citet{ST68} and  \citet{TE74}.  
Expressed as attenuation coefficients, they take the general form 
\begin{equation}
   {\cal R}^{\rm pp}_{\parallel,\perp} \; =\; \dover{\fsc}{\lambar_c} 
   \, B\sin\thetakB\,  {\cal F}_{\parallel,\perp} \left(\omega_\perp,\, B\right)
   \quad ,\quad
   \omega_{\perp}\; =\; \omega \sin\thetakB \quad ,
  \label{eq:pp_general}
\end{equation}
where \teq{\lambar_c = \hbar/m_ec} is the Compton wavelength over \teq{2\pi}.
Hereafter, all representations of \teq{{\cal R}} have units of \teq{\mathrm{cm}^{-1}},
and all forms for \teq{{\cal F}} are dimensionless.
Throughout, we shall employ the scaling convention that \teq{B} will be dimensionless, 
being expressed in units of \teq{B_{\rm cr}}, and \teq{\omega} shall represent the 
dimensionless photon energy, scaled by \teq{m_ec^2}, in the local inertial frame of reference.
The factor of \teq{\sin\thetakB} comes from the Lorentz transformation along {\bf B}
from the frame where \teq{\mathbf{k} \cdot \mathbf{B}=0}, to the interaction frame. 
Thus, the rates in Eq.~(\ref{eq:pp_general}) are cast in Lorentz invariant form:
\teq{\omega_{\perp}} and \teq{B} are invariants under such transformations, 
while \teq{\sin\thetakB} is an aberration or time-dilation factor.
The traditional polarization labelling convention adopted here is as follows:
the label \teq{\parallel} refers to the state with the
photon's \it electric \rm field vector parallel to the plane containing
the magnetic field and the photon's momentum vector, while \teq{\perp}
denotes the photon's electric field vector being normal to this plane.

The functional forms for \teq{{\cal F}_{\parallel,\perp}} derived in 
\citet{Erber} and \citet{TE74} are integrals over the individual energies 
of the created pairs, and are applicable only 
to cases where the produced pairs are ultra-relativistic.  In the limit of 
\teq{\omega_{\perp} B\ll 1}, a domain commonly encountered in pulsar applications, 
these integrals can be evaluated using the method of steepest descents, 
and the asymptotic rate functions become (for \teq{\omega_{\perp}\geq 2})
\begin{equation}
   {\cal F}_{\perp}\; =\; \dover{1}{2}\, {\cal F}_{\parallel} 
   \; =\; \dover{2}{3}\, \calFErber\quad ,\quad
   \calFErber \left(\omega_\perp,\, B\right)\; =\; 
              \dover{3\sqrt{3}}{16\sqrt{2}} \exp \left(-\dover{8}{3 \omega_{\perp} B}\right)\quad .
 \label{eq:Erber_asymp}
\end{equation}
This result was established in \citet{Erber}, and demonstrates
that the rate is an extraordinarily rapidly increasing function 
of photon energy, \teq{\sin\thetakB} and the field strength.
Accordingly, one quickly infers that pair conversions by this process,
instigated by photons emitted parallel to the local field,
will cease above around 10 stellar radii from the surface.
As an average over photon polarizations,
\teq{\calFErber} is the simplest form employed in this paper, and is widely 
cited in the pulsar literature, for example in standard polar cap models 
of radio pulsars \citep{Sturr71,RS75}.  
It is also the form that is employed in the pair attenuation calculations 
of \citet{Lee10}.  In the opposite, ultra-quantum limit where 
\teq{\omega_{\perp} B\gg 1}, alternative asymptotic forms with 
\teq{{\cal F}_{\perp, \parallel}\propto (\omega_{\perp} B)^{-1/3}} can be derived 
\citep{Erber,ST68,TE74}.  These are of less practical use since 
for such high photon energies or magnetic fields, the sawtooth structure
of the rates must be treated exactly during photon propagation in the 
magnetosphere.  
 
High energy radiation in pulsar models is usually emitted at very small angles
to the magnetic field, well below pair threshold.  This is true both in
polar cap models \citep{Sturr71,RS75,DH82,DH96} and outer gap scenarios
\citep{CHR86,Romani96}, since the radiating electrons/pairs
are accelerated along the {\bf B}-field to very high Lorentz factors.
Consequently, $\gamma$-ray photons emitted near the neutron star surface
will convert into pairs only after they have
propagated a distance \teq{s} comparable to the field line radius of curvature
\teq{\rho_c}, so that \teq{\sin\thetakB \sim s/\rho_c} at the altitude
of conversion.  Erber's expression for the pair production rate will be
vanishingly small unless \teq{\omega B\sin\thetakB \gtrsim 0.2}, i.e.,
the argument of the exponential approaches unity.  Hence,
for fields \teq{B\ll 0.1} the asymptotic expression in
Eq.~(\ref{eq:Erber_asymp}) can be used in pair attenuation calculations.
However at higher field strengths, namely
\teq{B \gtrsim  0.1}, pair production will occur fairly close to
or at threshold, where Erber's asymptotic expression overestimates
the exact rate by orders of magnitude (e.g., see DH83).  Accordingly, it is
imperative to include near-threshold modifications to the rates, a
serious need that was recognized and addressed in the pair attenuation
calculations of \citet{CCH96}, \citet{HBG97}
and \citet{BH01}, but omitted by \citet{Lee10}.

\def\calFTH{{\cal F}_{\hbox{\sevenrm TH}}}

\citet{DH83} provided a useful empirical fit to the rate to
approximate the near-threshold reductions below Erber's form.
Baring (1988) developed an analytic result
from detailed asymptotic analysis of the exact pair creation formalism.
The origin of this analytic result was a modification of the WKB approximation
\citet{ST68} applied to the Laguerre functions appearing in
the exact \teq{\gamma\to e^+e^-} rate, to specifically treat created pairs
that are mildly relativistic. A slightly different analysis of threshold corrections
was provided more recently by \cite{BK07}, specifically their Eq.~(3.4), yielding the form
\begin{equation}
   \calFTH \left(\omega_\perp,\, B\right) \;=\;
                \dover{3\omega_\perp^2-4}{2\omega_\perp^2 \,
               \sqrt{ (\omega_\perp^2-4)\,  {\cal L}(\omega_{\perp})\, \phi (\omega_{\perp}) }}
               \, \exp \left\{ -\dover{\phi(\omega_\perp)}{4B}  \right\}
                  \quad , \quad
                   \omega_{\perp} \geq 2\quad ,
 \label{eq:B88_asymp}
\end{equation}
for
\begin{equation}
   \phi (\omega_{\perp}) \; =\; 4\omega_{\perp} - \left( \omega_{\perp}^2 - 4 \right) {\cal L}(\omega_{\perp})
   \quad ,\quad
   {\cal L}(\omega_{\perp}) \; =\; \log_e\left( \dover{\omega_{\perp}+2}{\omega_{\perp}-2}\right)\quad .
 \label{eq:phi_calL_def}
\end{equation}
This analytic result will be used in this paper; it improves the Erber form by several orders
of magnitude near threshold \teq{\omega_{\perp}\sim 2}, and in the limit
\teq{\omega_{\perp}\gg 1}, \teq{\phi (\omega_{\perp})\approx 32/(3\omega_{\perp})}
and Eq.~(\ref{eq:B88_asymp}) reduces to Erber's polarization-averaged
form in Eq.~(\ref{eq:Erber_asymp}).  Also, Eq.~(\ref{eq:B88_asymp})
agrees numerically with the empirical approximation of DH83.  The comparable analytic
result in \cite{B88} differs only by a factor of \teq{(\omega_{\perp}-2)/(\omega_{\perp} +2)}
from Eq.~(\ref{eq:B88_asymp}), and therefore is slightly less accurate as an approximation
to the sawtooth structure of the exact pair creation rate near threshold.
Observe that Eq.~(B.5) of
\citet{BK07} presents polarization-dependent forms to partially account
for near-threshold modifications to the polarized rate.  This suggests that
\teq{{\cal F}_{\perp}\approx (\omega_{\perp}^2-4)/(2\omega_{\perp}^2)\, {\cal F}_{\parallel}},
but the accurate treatment of the polarization dependence of pair thresholds,
embodied in Eqs.~(\ref{eq:tpppar}) and~(\ref{eq:tppperp}) below, was
omitted from their approximation.

Technically, Eq.~(\ref{eq:B88_asymp}) can be applied reliably up to fields 
\teq{B\sim 0.5}, and provided \teq{\omega_{\perp}B\lesssim 1}.
When the field is larger, even the near-threshold correction to the 
asymptotic rate becomes inadequate.  Then, the discreteness of the sawtooth 
structure comes into play, as does the polarization-dependence of the 
process, and pair creation proceeds mostly via
accessing the lowest Landau levels.  We model this in a manner identical 
to HBG97, by adding a ``patch'' for the reaction rate when photons with 
parallel and perpendicular
polarization produce pairs only in the ground (0,0) and first excited
(0,1) and (1,0) states respectively.  Here \teq{(j,k)} denotes the Landau
level quantum numbers of the produced pairs.  We implement this 
patch when \teq{\omega_{\perp} < 1 + \sqrt{1+4 B}}.  The exact, 
polarization-dependent, pair production attenuation
coefficient of Daugherty \& Harding (1983) leads to the following forms.
We include only the (0,0) pair state for \teq{\parallel} polarization:
\begin{equation}
   {\cal F}^{\rm pp}_{\parallel} \; =\; \dover{2B}{\omega_{\perp}^2  \vert p_{\hbox{\sevenrm 00}}\vert}
   \,\exp\left(-\dover{\omega_{\perp}^2}{2B}\right)
   \quad , \quad 
   \omega_{\perp} \geq 2\quad ,
  \label{eq:tpppar}
\end{equation}
and only the sum of the (0,1) and (1,0) states for $\perp$ polarization:
\begin{equation}
   {\cal F}^{\rm pp}_{\perp} \; =\; \dover{2B\, E_0(E_0+E_1)}{\omega_{\perp}^2  \vert p_{\hbox{\sevenrm 01}}\vert}
   \,\exp\left(-\dover{\omega_{\perp}^2}{2B}\right)
   \quad , \quad 
   \omega_{\perp} \geq 1 + \sqrt{1+ 2B} \quad ,
  \label{eq:tppperp}
\end{equation}
where 
\begin{displaymath}
   E_0  = (1 + p_{01}^2)^{1/2}\quad , \quad  
   E_1  = (1 + p_{01}^2 + 2B)^{1/2}
\end{displaymath}
for
\begin{displaymath}
   \vert p_{jk}\vert  = \left[ \dover{\omega_{\perp}^2}{4} - 1 - (j+k)B + 
   \left( \dover{(j-k)B}{\omega_{\perp}} \right)^2\right]^{1/2}  \quad ,
\end{displaymath}
which describes the magnitude of the momentum parallel to {\bf B} of each member 
of the produced pair in the specific frame where \teq{\thetakB=\pi/2}, i.e. 
{\bf k} $\cdot$ {\bf B}\teq{=0}.  Observe that because the pair threshold is dependent on the photon 
polarization state, for near-critical and supercritical fields, incorporating 
polarization influences is potentially important for determining conversion mean free paths,
which are usually very small.  In fact, these mean free paths are small enough that 
the pair production rate in this regime thus behaves like a wall at 
threshold, and photons will pair produce as soon as they satisfy the kinematic
restrictions on \teq{\omega} given in equations~(\ref{eq:tpppar})
and~(\ref{eq:tppperp}).  Thus either asymptotic or exact conversion rates
can be employed with little difference in resultant attenuation lengths 
provided the polarization-dependent kinematic thresholds are treated 
precisely.  It will emerge that escape energies are virtually insensitive 
to the photon polarization state in sub-critical fields because these
generally correspond to conversions at higher altitudes.  Then 
the asymptotic rates are appropriate, and their strong sensitivity 
to \teq{\omega} inverts to yield virtual independence of the escape energy
to polarization.  This convenient circumstance does not apply to magnetars, 
for which polarization dependence is more significant due to the 
disparity in pair thresholds for the two photon polarization states.

\section{PAIR CREATION IN STATIC, FLAT SPACETIME MAGNETOSPHERES}
 \label{sec:flatspace}

Although general relativistic effects are expected to be important near
the neutron star surface, we can glean some important insights from
considering the case of photon attenuation in a dipole magnetic field in
flat spacetime.  This was the case dealt with by \citet{HEF90}, \citet{CCH96} and \citet{HA01}, 
among others, and we compare our results to theirs.  Furthermore, the
analytic behavior of the optical depth function is clearest in flat
spacetime with no aberration.  General relativistic and aberration
influences will perturb these results, but the flat spacetime case in
the absence of rotation will provide a useful limit against which to
check the more complex calculations.  We will also confirm a result of
Zhang \& Harding (2010; see also Lee et al. 2010), which indicates that
in flat spacetime the photon escape energy scales with emission altitude
\teq{r} as \teq{r^{5/2}}, in the absence of rotational aberration effects.

To assess the importance of single-photon pair creation in pulsars, we
compute pair attenuation lengths and escape energies as functions of 
the photon emission location, i.e. altitude and colatitude, and also as 
functions of the energy observed at infinity.  Following \citet{GH94}
and \citet{HBG97}, the optical depth for 
pair creation out to some path length \teq{l}, integrated over the photon
trajectory, is 
\begin{equation}
   \tau (l)\; =\; \int_0^l {\cal R}\, ds\quad ,
 \label{eq:atten_tau_def}
\end{equation}
where \teq{\cal{R}} is the attenuation coefficient, in units of cm$^{-1}$, 
as expressed in general form in Eq.~(\ref{eq:pp_general}).  Also, \teq{s} 
is the path length along the photon trajectory in the local inertial frame;
in flat spacetime, all such inertial frames along the photon path are coincident.
With this construct, the probability of survival along the trajectory is
\teq{\exp \{ - \tau (l)\} }, and the criterion \teq{\tau (l) = 1} establishes
a value of \teq{l=L} that is termed the {\it attenuation length}.
A photon will be able to escape the magnetosphere entirely 
if \teq{\tau (\infty )<1}.  In general, this will only be possible for photon 
energies below some critical value \teq{\eesc}, at which
\teq{\tau (\infty )=1}; this defines the {\it photon escape energy} 
\teq{\eesc} as in \citet{HBG97} and \citet{BH01}.  It is the strongly 
increasing character of the pair conversion functions in
Eqs.~(\ref{eq:Erber_asymp}) and~(\ref{eq:B88_asymp}), as functions 
of energy \teq{\omega}, that guarantees magnetospheric transparency 
at \teq{\erg <  \eesc}.  Observe that these formal definitions 
apply both to flat spacetimes here and general relativistic ones in 
Section~\ref{sec:GR}.

\begin{figure}
 \centerline{\includegraphics[width=.55\textwidth]{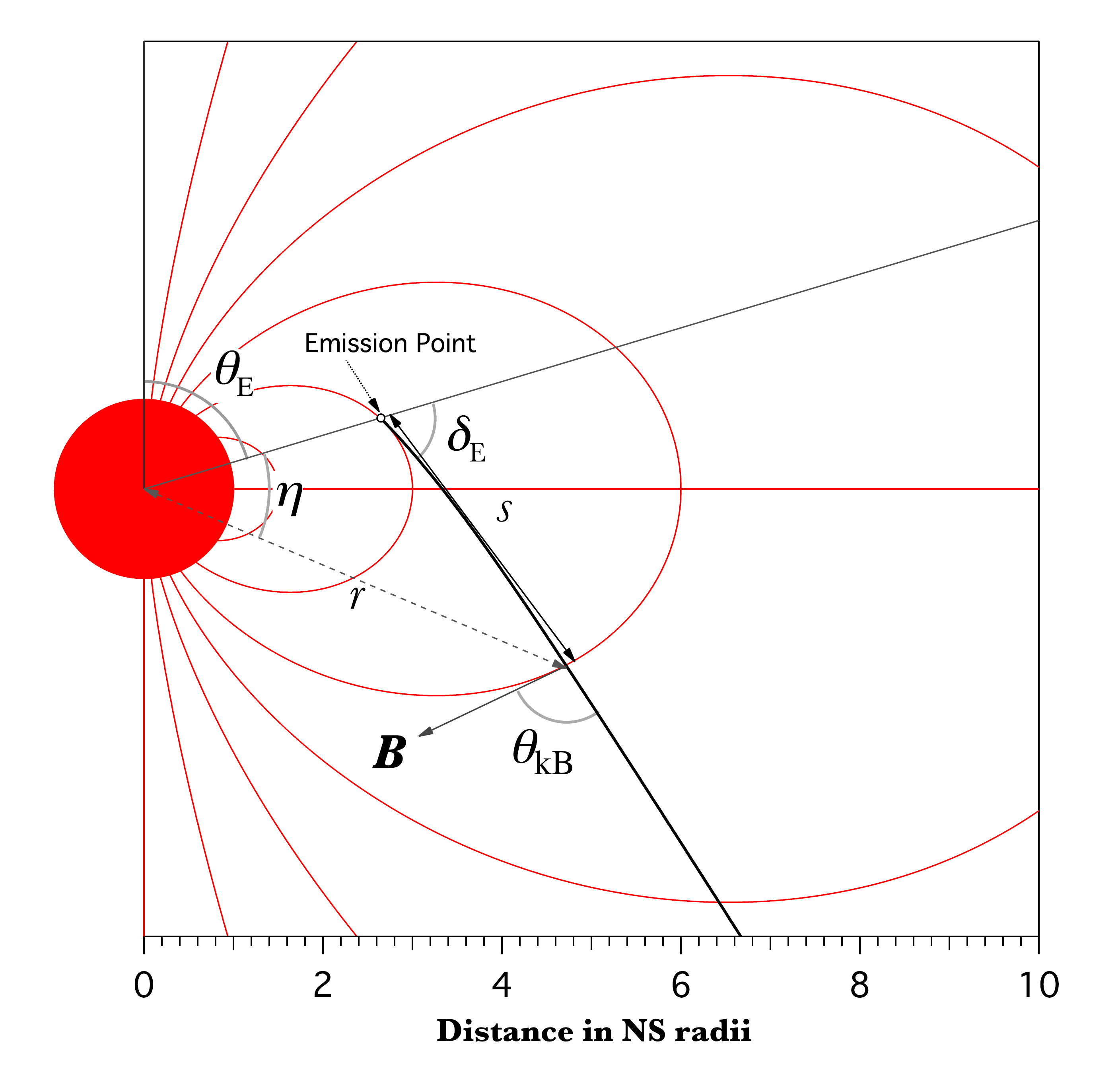}}
\caption{Photon propagation geometry in a dipole magnetic field, 
with red curves representing field lines.  The photon emission point 
is at an altitude \teq{\rE = h \rns} and colatitude \teq{\thetaE}.  
The photon trajectory, represented by the black line, is a straight line 
for flat spacetime and a curved path (shown here) for general relativistic 
considerations.  At any location along the photon path, \teq{\mathbf{k}}
is the photon momentum vector, and \teq{\mathbf{B}} is the local
magnetic field vector; the angle between these two vectors is
\teq{\thetakB}, given in Eq.~(\ref{eq:sin_thetakB}).  All such locations
are defined by the propagation angle \teq{\eta}, with the radial
position \teq{r} relative to the center of the neutron star, and the
distance \teq{s} from the point of emission being described by
Equations~(\ref{eq:chi_def}) and~(\ref{eq:prop_dist}), respectively.}
 \label{fig:prop_geometry}
\end{figure}

The geometry for general spacetime trajectories used in the computation of 
\teq{\tau} is illustrated in Fig. \ref{fig:prop_geometry}.  While slight curvature in the photon 
path is depicted so as to encapsulate the general relativistic study in 
Sec.~\ref{sec:GR}, this curvature can be presumed to be zero for the 
present considerations of flat spacetime.  Each of the angles 
in this diagram can be defined once the emission colatitude \teq{\thetaE} 
and emission altitude \teq{\rE=h \rns} are specified.  The instantaneous 
colatitude \teq{\theta} with respect to the magnetic axis is
\begin{equation}
   \theta\; =\; \eta+\thetaE\quad .
 \label{eq:theta_fs}
\end{equation}
This defines the propagation angle \teq{\eta}, which is the angle between 
the radial vector at the time of emission and the radial vector at the present 
photon position.  The photon trajectory initially starts parallel to the magnetic field, 
since gamma-rays in pulsars are necessarily emitted by ultra-relativistic 
electrons that move basically along field lines.  Standard models of electron 
acceleration invoke electrostatic potentials parallel to the local {\bf B} (e.g. 
Sturrock 1971; Ruderman \& Sutherland 1975; Daugherty \& Harding 1982), 
and velocity drifts across {\bf B} due to pulsar rotation are generally much 
smaller than \teq{c} for young gamma-ray pulsars.  Accordingly, gamma-rays 
produced by primary electrons of Lorentz factor \teq{\gamma_e}
are beamed to within a small Lorentz cone of half angle \teq{\sim 1/\gamma_e}
centered along {\bf B}.  This restriction conveniently simplifies the trajectory 
parameter space, so that the angle between the radial direction and the 
photon trajectory at the point of emission, \teq{\deltaE} (Gonthier \& Harding 1994 
name this \teq{\delta_0}), is determined only by the colatitude \teq{\thetaE} 
at the point of emission.  The magnetic field vector at any point in a flat 
spacetime dipole magnetosphere is given by
\begin{equation}
   \mathbf{B} \; =\; \dover{B_p \rns^3}{2r^3}
       \left\{2\cos\theta\,\hat{r}+\sin\theta\,\hat{\theta}\right\}\quad .
 \label{eq:dipole_fs}
\end{equation}
where \teq{B_p} is the surface polar magnetic field, i.e., that at \teq{r=\rns}
and \teq{\theta =0}.  The geometry of Fig. \ref{fig:prop_geometry} then simply sets
\begin{equation}
   \tan\deltaE \; =\; \dover{1}{2} \tan\thetaE\quad .
 \label{eq:deltae_def}
\end{equation}
This result is, of course, independent of the altitude of emission.  
One remaining piece of the geometry is the relationship between the altitude along 
the photon path, and the angle \teq{\eta}.  This is simply derived using 
the trigonometric law of sines.  Given \teq{\deltaE}, the dimensionless distance 
from the center of the neutron star \teq{\chi = r/\rE}, scaled by the altitude of emission,
satisfies
\begin{equation}
   \chi\; \equiv\; \dover{r}{\rE} \; =\;  \dover{\sin\deltaE}{\sin(\deltaE-\eta)}\quad .
 \label{eq:chi_def}
\end{equation}
This is the locus of a straight line in polar coordinates, and it is trivially 
determined that \teq{\eta\to \deltaE} as \teq{r\to\infty}.  The
photon momentum vector \teq{\mathbf{k}} along this path satisfies \teq{\hat{\mathbf{k}}
\equiv \mathbf{k}/\omega = \cos (\deltaE - \eta) \hat{r} + \sin (\deltaE - \eta ) \hat{\theta}}.

For flat spacetime geometry with no aberration influences, it is convenient to 
restate the optical depth integral in Eq.~(\ref{eq:atten_tau_def}) using 
the propagation angle \teq{\eta} as the integration variable:
\begin{equation}
   \tau (l) \; =\; \dover{\fsc}{\lambar_c} \int_0^{\eta (l)}  
           B \sin\thetakB\, {\cal{F}}(\omega_\perp,B) \dover{ds}{d\eta}\, d\eta\quad .
 \label{eq:tau_beta}
\end{equation}
The propagation distance \teq{s} is easily found using 
the trigonometric law of sines, and thereby yields the change of variables 
Jacobian \teq{ds/d\eta} in Eq.~(\ref{eq:tau_beta}):
\begin{equation}
   s \; =\; \dover{\rE\, \sin\eta}{\sin(\deltaE-\eta)} 
   \quad \Rightarrow\quad
   \dover{ds}{d\eta} \; =\; \dover{\rE\,\sin\deltaE}{\sin^2(\deltaE-\eta)}\quad .
 \label{eq:prop_dist}
\end{equation}
Therefore, the relationship for \teq{\eta (l)} is the inversion of Eq.~(\ref{eq:prop_dist})
for \teq{s=l}, i.e., \teq{\tan\eta = \sin\deltaE /(\cos\deltaE + \rE/l)}.  The integrand 
in Eq.~(\ref{eq:tau_beta}) includes a dependence on the angle \teq{\thetakB}
between the photon trajectory and the local magnetic field, particularly through 
the attenuation coefficient function \teq{{\cal F}}.  The photons start with 
\teq{\thetakB = 0}, and this angle increases at first linearly as the photon 
propagates outward.  The angle \teq{\thetakB} is given geometrically by
\begin{equation}
   \sin\thetakB \; =\; \dover{\vert \mathbf{k} \times \mathbf{B}\vert }{|k| . |B|}
      \; =\;  \dover{{\hat k}_rB_{\theta} - {\hat k}_{\theta} B_r}{|B|} 
      \; =\; \dover{\sin\theta\cos(\deltaE-\eta)-2\cos\theta\sin(\deltaE-\eta)}{\sqrt{1+ 3\cos^2\theta }}
 \label{eq:sin_thetakB}
\end{equation}
at every point along the photon's path.  Using Eq.~(\ref{eq:deltae_def}) 
simply demonstrates that the right hand side of this 
expression approaches zero as \teq{\eta = \theta-\thetaE\to 0}.  Note also
that by forming \teq{\cos\thetakB} and using Eq.~(\ref{eq:chi_def}), one can show 
routinely that this result is equivalent to Eq.~(5) of Baring \& Harding (2007).
In the limit of small colatitudes near the magnetic axis, one simply derives 
\teq{\sin\thetakB\approx 3\eta/2}, which can be combined with 
\teq{r/\rE\approx 1+2\eta/\thetaE} to yield \teq{\thetakB\approx 
3\thetaE (r/\rE -1)/4}.  This dependence closely approximates the low altitude values 
for \teq{\thetakB} in flat spacetime exhibited in Fig.~5a of Gonthier \& Harding (1994). 
This completes the general formalism for pair creation optical depth determination 
in Minkowski metrics.

\subsection{Optical Depth for Emission Near the Magnetic Axis}
\label{sec:tauintegration}

In order to better understand the character of the optical depth integral, 
it is instructive to consider the case of a photon emitted at very small 
colatitudes.  This situation is representative of much of the relevant 
parameter space for young gamma-ray pulsars; for example, the Crab pulsar 
has a polar cap half-angle of about \teq{4.5^\circ} and the Vela pulsar 
has a polar cap half-angle of about \teq{2.8^\circ}.
For these photons emitted very close to the magnetic axis, 
\teq{\eta} and \teq{\thetaE} are small.  In this limit,
we have the approximations
\begin{equation}
   B\; \approx\; \dover{B_p (\deltaE-\eta)^3}{h^3 \deltaE^3}
   \quad ,\quad
   \sin\thetakB \;\approx\; \dover{3}{2}\,\eta
 \label{eq:BsinthetakB_approx}
\end{equation}
for \teq{\rE=h\rns}.  We also have \teq{ds/d\eta\approx \rE\deltaE/(\deltaE -\eta)^2}
using Eq.~(\ref{eq:prop_dist}), with \teq{\deltaE\approx \thetaE/2}.  
These results can be inserted into Eq.~(\ref{eq:tau_beta}), and 
the integration variable changed to \teq{x=\eta/\deltaE}, 
yielding an approximation for the optical depth in axial locales:
\begin{equation}
   \tau (l) \;\approx\; \dover{3\thetaE}{4} \,\dover{B_p}{h^2} \dover{\fsc \rns}{\lambar}  
      \int_{8/(3\thetaE\erg)}^{x_+}  x( 1 - x ) \,
   {\cal F}  \left(\dover{3}{4}\, \erg\thetaE x,
               \, \dover{B_p}{h^3} \left[ 1 - x \right]^3 \right) dx
   \;\; ,\quad
   x_+ \; =\; \dover{l}{l+h\rns} \;\; .
 \label{eq:tau_small_thetae}
\end{equation}
This form is applicable to any choice of the pair conversion function \teq{{\cal F}}.
Observe that here the the local energy \teq{\omega} has been replaced 
with the energy \teq{\erg} seen by an observer at infinity; the two are equivalent in 
flat spacetime with no rotation, but when we consider general relativity and aberration, 
the distinction will become important.  The upper limit \teq{x_+} is the value of 
\teq{x=\eta /\deltaE} that realizes a  path length \teq{l}, 
and is well approximated by \teq{l/\left(l+h\rns\right)} near the magnetic axis. 
The lower limit defines the threshold condition, so that if \teq{\erg\thetaE\leq 8/3}, 
propagation in flat spacetime out of the magnetosphere never moves the photon 
above the pair threshold at \teq{\omega_{\perp}=2}, 
and \teq{\tau=0} over the entire photon trajectory.
For the particular choice of Erber's (1966) attenuation coefficient in 
Eq.~(\ref{eq:Erber_asymp}), the integral for the optical depth assumes 
a fairly simple form:
\begin{equation}
   \tau_{\hbox{\sevenrm Erb}} (l) \;\approx\; \dover{9\sqrt{3}\, \thetaE}{64\sqrt{2}} \,\dover{B_p}{h^2} \dover{\fsc \rns}{\lambar}  
      \int_{8/(3\thetaE\erg)}^{x_+}  x( 1 - x ) \, \exp \left\{ - \dover{32h^3}{9 \erg B_p\thetaE}
               \dover{1}{x ( 1 - x )^3} \right\} dx\quad .
 \label{eq:tau_small_thetae_Erber}
\end{equation}
If one considers emission points near the magnetic axis at different altitudes 
along a particular field line with a footpoint colatitude \teq{\theta_f}, then
\teq{\thetaE\approx \theta_f\sqrt{h}} gives the altitude dependence of the 
emission colatitude.  Exploring attenuation opacity along a fixed field line is germane
to treating gamma-ray emission that takes place along or near the last 
open field line, where \teq{\theta_f} is fixed by the pulsar's rotational period.
The escape energy \teq{\eesc}
can be computed by setting \teq{\tau (\infty )=1}, for which \teq{x_+\to 1}.
Imposing this \teq{\tau (\infty) =1} criterion, and presuming \teq{\erg\thetaE\gg 1} in
Eq.~(\ref{eq:tau_small_thetae_Erber}), yields the approximate altitude dependence
\begin{equation}
   \eesc\;\propto\; h^{5/2}
 \label{eq:e_esc_alt_dep}
\end{equation}
for the escape energy.  This is a flat spacetime result for near polar axis locales
that was identified by Zhang \& Harding (2000; see also Lee, et al. 2010).
Deviations from this simple altitude dependence arise
(i) when the footpoint colatitude \teq{\theta_f} is not sufficiently small,
(ii) if the pair conversion occurs not very far from the 
\teq{\omega_{\perp}=2} threshold, and (iii) down near the stellar 
surface where general relativistic effects modify the values of \teq{\omega}, 
\teq{B} and \teq{\thetakB}.

For significantly sub-critical \teq{B_p}, a complete asymptotic expression 
for the optical depth after propagation to high altitudes 
can be determined using the method of steepest 
descents to compute the integral for \teq{\tau (l)}, since 
the integrand in Eq.~(\ref{eq:tau_small_thetae_Erber}) 
is exponentially sensitive to values of \teq{x}.  
This is precisely the method employed by \citet{AS79} and later 
adopted by \citet{HA01} in developing similar opacity integrations.
The exponential realizes a very narrow peak at \teq{x=1/4}, so that 
for \teq{l\to\infty} and \teq{x_+=1} 
\begin{equation}
   \tau_{\hbox{\sevenrm Erb}} (l) \;\approx\; \dover{3^6}{2^{19}} 
   \left( \dover{3\pi \erg B_p^3\theta_f^3}{2 h^{11/2}} \right)^{1/2}
    \dover{\fsc \rns}{\lambar}  
      \exp \left\{ - \dover{2^{13} h^{5/2}}{3^5 \erg B_p\theta_f}  \right\} \quad .
 \label{eq:tau_Erber_approx}
\end{equation}
This result actually applies for any \teq{x_+ > 1/4}, i.e. 
when \teq{l \gtrsim h\rns/3}.  It is independent of \teq{l} since the 
integrand has sampled beyond the peak and has shrunk to very small values 
when \teq{x} exceeds \teq{1/4} by a significant amount.  
Eq.~(\ref{eq:tau_Erber_approx}) is in agreement with the approximate optical depth computed by 
\citet{HA01} in their Eq.~(8), which was similarly formulated to treat gamma-ray propagation 
above the magnetic pole.  Note that \citet{AS79} provided a more general opacity 
integral by treating magnetic multipole configurations.
Again setting  \teq{\tau_{\hbox{\sevenrm Erb}} (\infty) =1}, and taking logarithms
of Eq.~(\ref{eq:tau_Erber_approx}),
the escape energy \teq{\eesc} for the Erber attenuation coefficient satisfies 
\begin{equation}
   \eesc\; = \; \dover{2^{13} h^{5/2}}{3^5 B_p\theta_f}
      \left\{ \log_e \left( \dover{3^6}{2^{19}}\, \dover{\fsc \rns}{\lambar} \right)
      + \dover{1}{2} \log_e \dover{3\pi \eesc B_p^3\theta_f^3}{2 h^{11/2}} \right\}^{-1}\quad .
 \label{eq:e_esc_Erber}
\end{equation}
While an exact solution for \teq{\eesc} must be determined numerically 
from this transcendental equation, the second logarithmic term on the right 
is only weakly dependent on its arguments.  Therefore, to a good approximation, 
one can infer that \teq{\eesc\propto 1/B_p} and \teq{\eesc\propto 1/\theta_f},
both of which emerge due to the presence of the factor \teq{\omega_{\perp}B} 
in the argument of the exponential in Erber's asymptotic form.

The same protocol can be adopted for pair conversion rates that include 
threshold modifications, specifically Eq.~(\ref{eq:B88_asymp}).
In this case, as the counterpart of Eq.~(\ref{eq:tau_small_thetae}) we have
\begin{equation}
   \tau (l) \;\approx\; \dover{3\thetaE}{4} \,\dover{B_p}{h^2} \dover{\fsc \rns}{\lambar}  
      \int_{8/(3\thetaE\erg)}^{x_+}  x( 1 - x ) \,
   {\cal F}_{\rm BK07}  \left(\dover{3}{4}\, \erg\thetaE x,
               \, \dover{B_p}{h^3} \left[ 1 - x \right]^3 \right) dx \quad ,
 \label{eq:tau_small_thetae_B88}
\end{equation}
again for \teq{x_+ =l/(l+h\rns )}.  We can simplify the ensuing 
analysis by making the substitution 
\teq{\lambda = 3 \thetaE\erg/8 \geq 1} so that locally 
\teq{\omega_{\perp} =2\lambda x} along the trajectory.  The argument of the 
exponential is of the form \teq{h^3 q(x,\,\lambda)/B_p} where
\begin{equation}
   q(x,\,\lambda) \;\equiv\; \dover{\phi (2 \lambda x)}{4(1-x)^3}
   \; =\; -\dover{2\lambda x}{(1-x)^3}
        - \dover{\left(\lambda x\right)^2-1}{(1-x)^3} 
              \log_e \left(\dover{\lambda x-1}{\lambda x +1}\right)
 \label{eq:qfunc_def}
\end{equation}
Using the method of steepest descents once again, we take 
the first derivative of the function in the exponential, and set it 
equal to zero to find the peak of the function.  The solution of 
\teq{\partial q/\partial x = 0} is a transcendental function in 
\teq{\lambda}, but it can be numerically approximated to better 
than 3\% by
\begin{equation}
   \overline{x} \approx \frac{1}{4} + 0.82\lambda^{-5/3}.
\end{equation}
Given $\partial q/\partial x = 0$, the logarithmic term 
\teq{\log_e [(\lambda \overline{x}-1)/(\lambda \overline{x}+1)]} can 
be expressed algebraically, and $q''(x,\lambda)$ can be written in the following form:
\begin{equation}
   q''(\overline{x},\lambda) \; =\; \frac{8\lambda \left[ \lambda^2 
      \left(2\overline{x}^3-3\overline{x}+1\right)-3\overline{x}+3\right]}{
      (\lambda^2\overline{x}(\overline{x}+2)-3)(1-\overline{x})^5 (\lambda^2\overline{x}^2-1)}\quad .
\end{equation}
The integral is then given approximately, as before, by the 
method of steepest descents.  With some cancellation, we then obtain
\begin{equation}
   \tau_{\hbox{\sevenrm BK07}} \;\approx\;  \frac{\fsc \rns}{\lambar}
        \frac{[3(\lambda \overline{x})^2-1] \, (\lambda\overline{x}-1)}{
        2\erg \,\lambda^2 \overline{x} \,  (\lambda \overline{x}+1)}   
          \left[\frac{\pi \chi^3B_p^3}{2 (\overline{x}+2)\, ( 1 - \overline{x} )\, \vert\Upsilon\vert\, h^7}\right]^{1/2}
     \exp \left\{-\frac{h^3}{B_p\chi} \right\}
 \label{eq:B88steep}
\end{equation}
where
\begin{equation}
   \Upsilon\; =\; \left(1-2\overline{x}-2\overline{x}^2\right)+\dover{3}{\lambda^2}
   \quad ,\quad
   \chi\; =\; \dover{(1-\overline{x})^2}{4\lambda}\,
    \Bigl\{ \lambda^2 \overline{x} (\overline{x}+2)-3\Bigr\} \quad .
\end{equation}
Here \teq{\Upsilon} is employed to render the \teq{q''(\overline{x},\, \lambda )} term more compact,
and \teq{\chi = 1/q(\overline{x},\, \lambda )}.
Noting that \teq{\theta_f\approx 8\lambda/(3\erg\sqrt{h})} in this small colatitude limit,
Eq.~(\ref{eq:B88steep}) agrees with the Erber approximation 
in Eq.~(\ref{eq:tau_Erber_approx}) to high precision in the regime 
where \teq{\lambda \rightarrow \infty}, and exhibits the appropriate threshold 
behavior.  Setting \teq{\tau (\infty) = 1} gives a transcendental equation
that can be solved numerically for \teq{\eesc}.  The impressive precision 
of this analytic steepest descents result for the escape energy is apparent 
in Fig.~\ref{fig:Eesc_flatspace} below.

\begin{figure}[h]
 \centerline{
 \includegraphics[width=.7\textwidth]{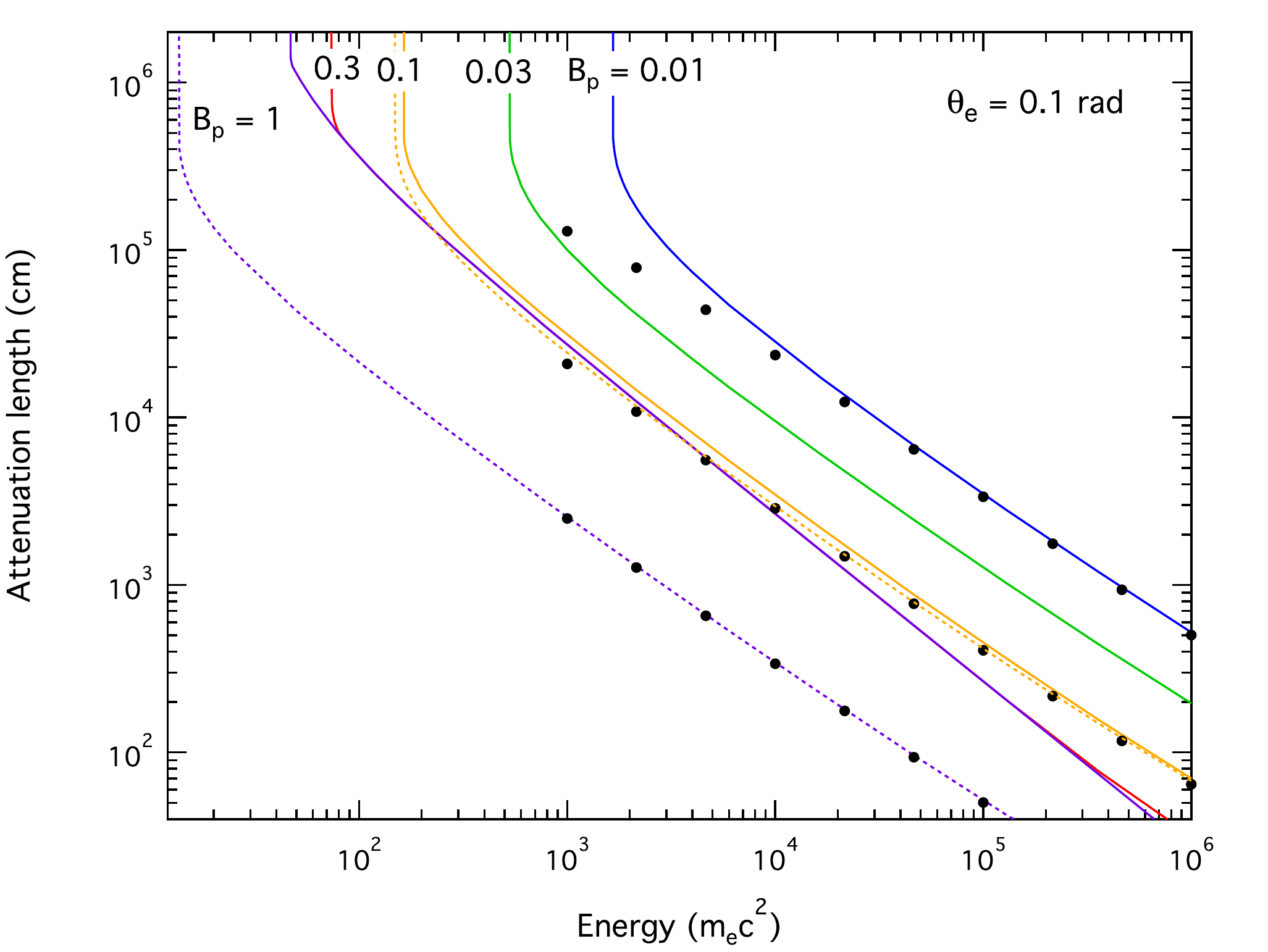}}
 \caption{The pair attenuation length \teq{L} (satisfying the criterion 
\teq{\tau (L)=1}) of photons emitted from the neutron star surface 
(\teq{h=1}) in flat spacetime, plotted as a function of photon energy.  
The emission colatitude is \teq{0.1}radian.  The curves are labeled with 
the surface polar magnetic field, \teq{B_p}, scaled by the quantum critical 
field.  Solid curves represent lengths computed using the integral 
calculation of the optical depth in Eq.~(\ref{eq:tau_small_thetae_B88}), 
i.e. they employ the approximation derived by Baier \& Katkov (2007); 
see Sec.~\ref{sec:pair_physics} for details.  Dashed lines 
depict the attenuation lengths computed from Eq.~(\ref{eq:tau_small_thetae_Erber}) 
using Erber's (1966) reaction rate.  The large black dots display 
the approximate form of the attenuation length encapsulated 
in Eq.~(\ref{eq:tau_small_thetae_Erber_fin}) for the cases 
of \teq{B_p=0.01}, \teq{B_p=0.1} and \teq{B_p=1}, thereby highlighting 
its high level of precision when \teq{L\ll \rns = 10^6}cm.}
 \label{fig:attenlength_flatspace}
\end{figure}

Well above the escape energy, the pair attenuation length \teq{l} is far inferior to 
the neutron star radius.  For surface emission (\teq{h=1}), in this \teq{l\ll \rns} limit,
we can assert \teq{x_+\ll 1} in Eq.~(\ref{eq:tau_small_thetae}), so that series 
expansion in the \teq{x} integration variable yields
\begin{equation}
   \tau (l) \;\approx\; \dover{3\thetaE}{4} \, \dover{\fsc \rns}{\lambar}  \, B_p
      \int_{8/(3\thetaE\erg)}^{x_+}  x \,
   {\cal F}  \left(\dover{3}{4}\, \erg\thetaE x, \, B_p \left[ 1 - 3x \right] \right) dx
   \quad ,\quad
   x_+ \;\ll\; 1 \quad .
 \label{eq:tau_small_thetae_small_x+}
\end{equation}
Analytic reduction of this integral is fairly complicated  for the case of the 
\teq{{\cal F}_{\rm B88}}
rate, but is relatively amenable for the Erber form, which we employ at this juncture.
Inserting Eq.~(\ref{eq:Erber_asymp}) for \teq{{\cal F}}, because of the strong 
exponential dependence of the integrand, the dominant contribution to 
the integral comes from \teq{x\approx x_+}.   Replacing the factor of \teq{x} 
in the integrand that lies outside the exponential by \teq{x_+^3/x^2}, for 
\teq{x_+\approx l/\rns}, yields
\begin{equation}
   \tau (l) \;\approx\; \dover{9\sqrt{3}\, B_p \thetaE}{64\sqrt{2}} \, \dover{\fsc \rns}{\lambar}  
   \, \left( \dover{l}{\rns} \right)^3\, \, \exp \left\{ \dover{32}{3 \erg B_p\thetaE} \right\} 
      \int_{8/(3\thetaE\erg)}^{x_+}  \, \exp \left\{ - \dover{32}{9 \erg B_p\thetaE}
               \dover{1}{x}  \right\} \dover{dx}{x^2} \quad .
 \label{eq:tau_small_thetae_Erber_v3}
\end{equation}
This manipulation affords analytic evaluation of the integral.  
The attenuation length \teq{L} is obtained by setting \teq{\tau (L)=1} in the 
resulting equation for \teq{\tau (l)}, which after rearrangement leads to
\begin{equation}
   \exp \left\{ - \dover{32\rns }{9 \erg B_p\thetaE L} \right\}  
   \;\approx\; \dover{2048\sqrt{2}}{81\sqrt{3}\, \erg B_p^2 \thetaE^2} \, \dover{\lambar}{\fsc \rns} 
   \, \left( \dover{\rns}{L} \right) ^3\, \exp \left\{ -\dover{32}{3 \erg B_p\thetaE} \right\} 
   + \exp \left\{ - \dover{4}{3 B_p} \right\} \quad .
 \label{eq:tau_small_thetae_Erber_v5}
\end{equation}
In general, solutions for \teq{L} are realized when the second exponential 
term on the right hand side of Eq.~(\ref{eq:tau_small_thetae_Erber_v5}) 
can be neglected, at the \teq{<10^{-3}} level.  This simplifies the algebra, 
and taking logarithms, one arrives at
\begin{equation}
    \dover{\rns }{L} \;\approx\; 
   3 + \dover{9 \erg B_p\thetaE}{32} \log_e \left[
   \dover{81\sqrt{3}\, \erg B_p^2 \thetaE^2}{2048\sqrt{2}} \, \dover{\fsc L^3}{\lambar \rns^2}
   \right] \quad .
 \label{eq:tau_small_thetae_Erber_fin}
\end{equation}
This transcendental equation must be solved numerically, though the
general trend is given approximately by \teq{L\propto
(\erg\thetaE)^{-1}} when \teq{L\ll \rns}, since the dependence on
parameters inside the logarithmic term is weak. This compact analytic
derivation nicely describes the attenuation length values for the Erber
rate, as is evident in Fig.~\ref{fig:attenlength_flatspace}.

The general character of the attenuation length solutions for the full
\teq{{\cal F}_{\rm BK07}} pair attenuation rate near threshold is
depicted in Fig.~\ref{fig:attenlength_flatspace}, for a neutron star
radius of  \teq{\rns = 10^6}cm.  Note that computations (not shown) that
also include the polarization-averaged forms for the first two
``sawtooth'' peaks [see Sec.~\ref{sec:pair_physics}] in the exact
attenuation coefficient formula of Daugherty and Harding (1983) generate
attenuation lengths that are almost indistinguishable from those shown,
even for \teq{B_p=1}. The attenuation length curves are
declining functions of the photon energy, generally with the expected
\teq{L\propto 1/\erg} dependence when \teq{L\ll \rns}.  While they
illustrate the particular case of surface emission from a colatitude of
\teq{\thetaE \approx 5.72^{\circ}}, i.e. close to the polar cap
colatitude for the Crab pulsar, at high energies, our calculations also
reveal roughly \teq{L\propto 1/\thetaE} behavior for a range of
non-equatorial emission colatitudes.

Since the pair attenuation rate increases rapidly with magnetic field
strength, the attenuation length declines with increasing \teq{B_p},
roughly as \teq{1/B_p} in accordance with
Eq.~(\ref{eq:tau_small_thetae_Erber_fin}) for \teq{B_p\lesssim 0.1}.
However, as the magnetic field of the pulsar approaches \teq{B_{\rm
cr}}, photon attenuation occurs for angles \teq{\thetakB} closer to the
absolute pair creation threshold of \teq{\erg\sin\thetakB = 2},
independent of the value of \teq{B_p}.  The attenuation length curves
for \teq{B_p = 0.3} and \teq{B_p = 1} then become indistinguishable at
intermediate energies \teq{10^2\lesssim \erg \lesssim 10^4} because the
attenuation coefficients just above the pair creation threshold are so
large that the distances traveled by the photons after they cross the
threshold are minuscule in comparison with the propagation distance
required to reach the threshold.  At the very highest energies, the
\teq{B_p = 0.3} and \teq{B_p = 1} curves begin to diverge again because
the attenuation coefficients drop by several orders of magnitude and the
distance a \teq{B_p=0.3} photon travels after crossing the threshold
before converting becomes comparable to the distance it transits before
reaching the point where \teq{\erg\sin\thetakB = 2}.

The dashed curves display the attenuation lengths for Erber rate
formalism [see Eq.~(\ref{eq:tau_small_thetae_Erber})]. Since the Erber
form significantly overestimates the attenuation coefficient near pair
threshold, it generates shorter attenuation lengths than does the more
precise determination using Eq.~(\ref{eq:tau_small_thetae_B88}).  The
analytic approximation in Eq.~(\ref{eq:tau_small_thetae_Erber_fin}) to
the Erber \teq{\tau (L)=1} formalism is also shown as discrete dots,
demonstrating a good precision in matching the fully numerical curves at
high energies. This approximation provides a useful guide to the generic
character of attenuation in \teq{L\ll \rns} domains.  The vertical
upturns in the curves at low energies define the photon escape energies
\teq{\eesc} for each \teq{B_p} case; such features are the focus of
Section~\ref{sec:eesc_flatspace}, and demarcate the energy domains for
pair creation transparency of the magnetosphere.

\subsection{Pair Creation Escape Energies in Flat Spacetime} 
 \label{sec:eesc_flatspace}

\begin{figure}[h]
 \centerline{
 \includegraphics[width=.7\textwidth]{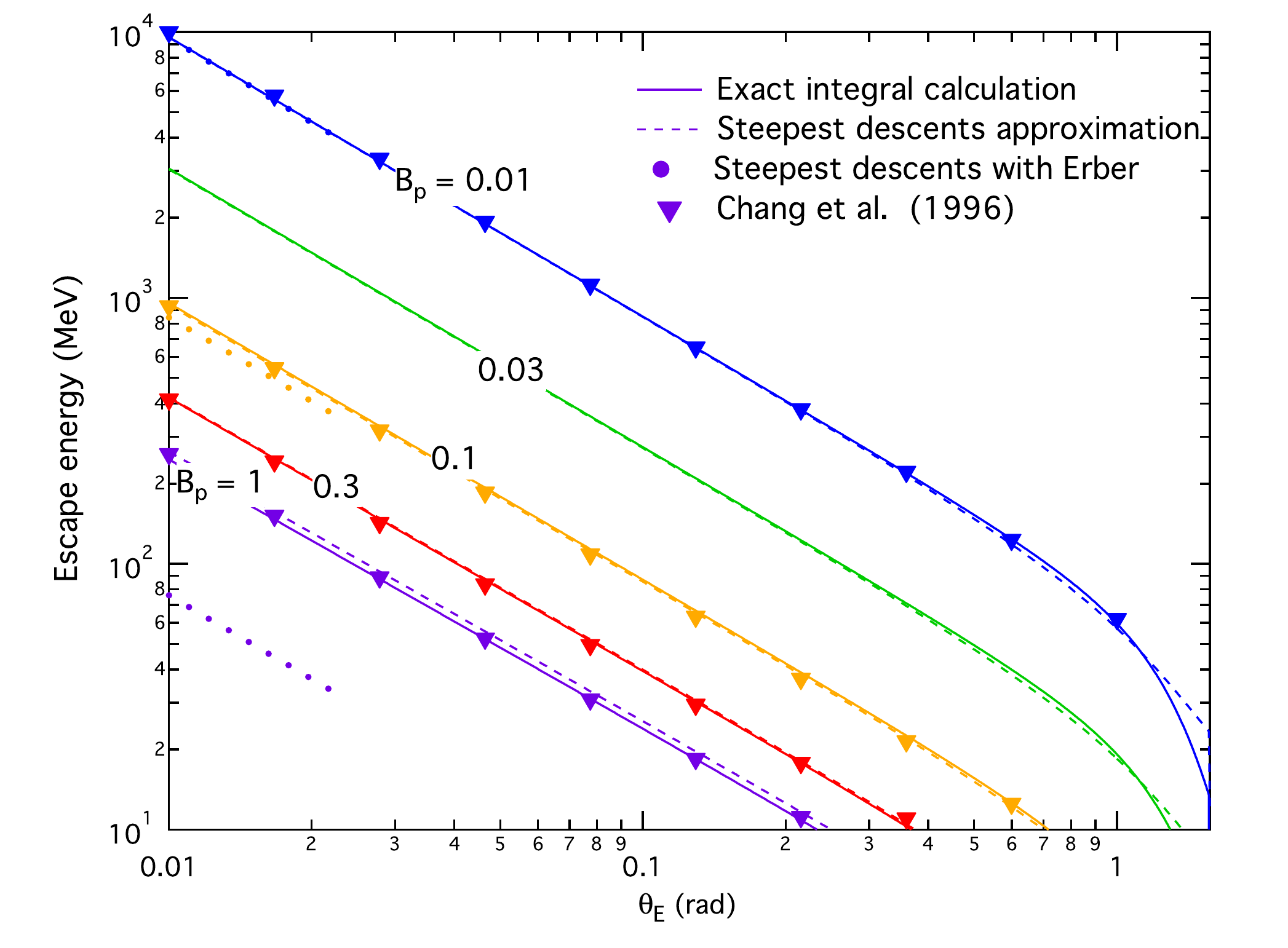}}
 \caption{The maximum energy \teq{\eesc} of photons emitted from the neutron star surface 
(\teq{h=1}) that can escape to infinity in flat spacetime, plotted as a function 
of emission colatitude.  The curves are labeled with the surface polar magnetic field, 
\teq{B_p}, scaled by the quantum critical field.  Solid curves represent 
results for \teq{\tau (\infty )=1} using the integral calculation of the optical depth 
in Eq.~(\ref{eq:tau_small_thetae}).  This determination employs polarization-averaged 
forms for \teq{{\cal F}} that include the first two ``sawtooth'' peaks in the
exact attenuation coefficient formula of Daugherty and Harding (1983), and 
at higher \teq{\omega_{\perp}} uses the approximation derived by \cite{BK07}; 
see Sec.~\ref{sec:pair_physics} for details.  Dashed lines represent the 
\teq{\tau(\infty )=1} determination using the steepest descents approximation in 
Eq.~(\ref{eq:B88steep}), and the short dotted lines depict the 
asymptotic form in Eq.~(\ref{eq:e_esc_Erber}) obtained using Erber's (1966) reaction rate,
but only for \teq{B_p=0.1} and \teq{B_p=1}.
Triangles are taken from the computations illustrated in Fig.~2 of Chang et al. (1996), for
comparison.  Although the steepest descents approximation was calculated 
in the small colatitude limit, it remains extremely good out to moderate colatitudes before 
diverging from the exact determination when \teq{\thetaE \sim 1}.}
 \label{fig:Eesc_flatspace}
\end{figure}

The focus now turns to the escape energies, since they provide upper
bounds to the spectral window of pair transparency for neutron star
magnetospheres. Numerical solutions of Eq. (\ref{eq:tau_beta}) in the
limit \teq{l\to\infty} can help gain a better understanding of where the
effects of magnetic pair creation will be the strongest.  By specifying
emission at the surface (\teq{h=1}), fixing a surface polar magnetic field
\teq{B_p} and then solving for \teq{\eesc} as a function of the colatitude of
emission \teq{\thetaE}, we obtain the plot shown in Fig.
\ref{fig:Eesc_flatspace}. The core results are contained in the solid
curves, which express the \teq{\tau (\infty )=1} criterion using the
\teq{{\cal F}_{\rm BK07}} attenuation coefficient derived by \cite{BK07},
in concert with the polarization-averaged forms for \teq{{\cal F}} 
that include the first two ``sawtooth'' peaks in the exact
attenuation coefficient formula expressed in Eqs.~(\ref{eq:tpppar})
and~(\ref{eq:tppperp}). We can see that for small colatitudes, 
\teq{\eesc\propto 1/\thetaE} and \teq{\eesc \propto 1/B_p}, as expected from 
Eq.~(\ref{eq:e_esc_Erber}).  The dashed lines representing the steepest
descents approximation in Eq. (\ref{eq:B88steep}) clearly illustrate the
remarkable accuracy of this expression over a wide range of colatitudes
and subcritical fields.  The effects of threshold corrections are also
apparent from this Figure.  For the purple \teq{B_p=1} curves, the Erber
approximation (short dotted purple line) to the attenuation coefficient
produces escape energies nearly a factor of 3 below the
threshold-corrected result. For lower magnetic fields, pair creation is
taking place well above threshold, and the Erber curves are much closer
to the threshold-corrected curves. Note that the analysis of
\cite{Lee10} omits consideration of threshold influences in the pair
production rates, and thereby would underestimate \teq{\eesc} when the
critical field is approached.

In comparing with extant flat spacetime computations of escape energies,
our results realize good agreement with Fig.~2 of Ho, Epstein \&
Fenimore (1990), using the Erber asymptotic form of
Eq.~(\ref{eq:Erber_asymp}); this matches their chosen attenuation
coefficient closely.  In this analysis, we assume that photons are
always emitted parallel to the magnetic field, so comparison of our
results is made to the topmost (\teq{\psi_i=0}) curve of their Fig.~2,
the x-axis of which is equivalent to \teq{\log(\deltaE+\thetaE)} in
our variables.  For \teq{B = 2\times 10^{12}}Gauss, the apparent
difference between our numerics and theirs is less than about 15\% ,
though visual precision in reading this plot limits such an estimate.
For the Erber attenuation coefficient in flat spacetime, multiplying the
photon escape energy from a fixed emission altitude and colatitude by
the surface polar magnetic field yields an approximately constant result.

\begin{figure}[h]
 \centerline{\includegraphics[width=.68\textwidth]{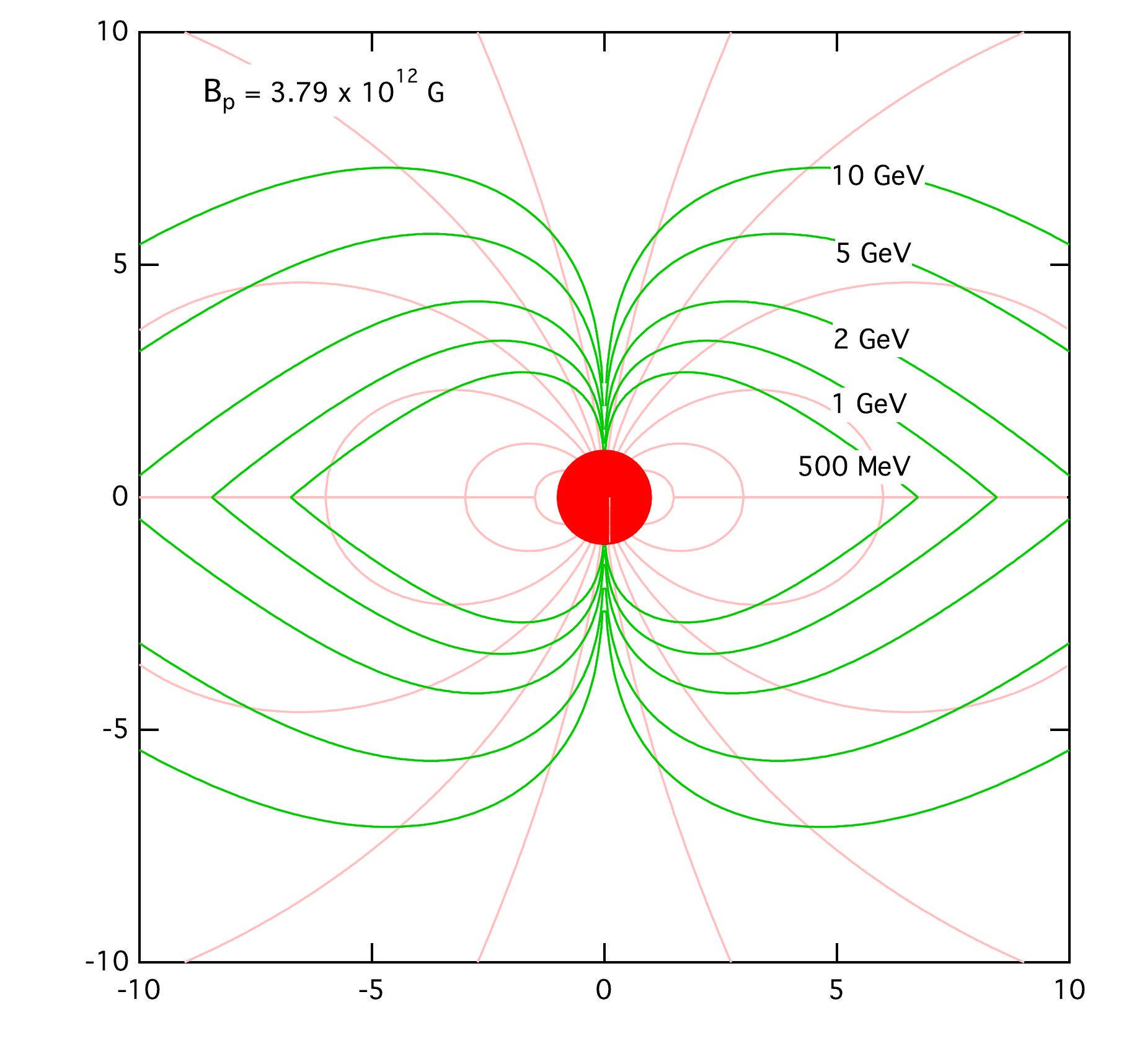}}
\caption{``Pair convertosphere'' diagram for a pulsar with the same 
magnetic field as the Crab, for flat spacetime and the specific case of no 
rotational/aberration influences.  Dipolar field structure (depicted in light red)
underlays each leaf-shaped green curve, which
represents the lowest possible emission point at a given colatitude for 
a photon of a fixed energy (as labelled), below which magnetic pair creation would 
attenuate the photon before it could escape from the neutron star 
magnetosphere.  The photons are always presumed to be emitted parallel to the local field.
The scale is neutron star radii, with the unit radius circle 
in the center representing the neutron star.  General relativistic effects alter these curves 
only very near the neutron star surface, and then only modestly, moving 
them to slightly higher altitudes.}
 \label{fig:pair_convertosphere}
\end{figure}

If, on the other hand, one fixes the surface polar field and the photon
energy, and calculates the lowest altitude \teq{r_{\rm min}} from which
photons of that energy can escape to infinity, one can formulate a
``pair convertosphere" plot like that in Fig.~\ref{fig:pair_convertosphere}, 
which is computed for flat spacetime. The
leaf-shaped curves represent a cross-section through a \teq{\tau = 1}
surface that is symmetric about the magnetic axis.  Inside the surface,
to a first approximation, all photons of the labeled energy will convert
to pairs. Outside the surface, photons can escape and be detected.  At a
fixed colatitude, a higher altitude of emission results in a higher
escape energy (corresponding to shifting the curves in Fig.
\ref{fig:Eesc_flatspace} up in energy).  In a Minkowski metric, all of
these minimum altitude curves drop to below the stellar surface at the
magnetic pole, since there the field line radius of curvature is very
large, and photons do not quickly encounter significant \teq{\thetakB}
during propagation when initially emitted parallel to the local field.
Rotational aberration influences, which will be considered in
Section.~\ref{sec:aberration} below, introduce an azimuthal asymmetry about
the magnetic axis for an inclined pulsar, and significantly distort the
shape of the surfaces near the magnetic poles, but not by much in
equatorial regions. General relativistic influences (discussed in
Section~\ref{sec:GR}) are significant below \teq{2\rns}, and while they
do not appreciably alter the overall morphology of the leaf-shaped
contours, they do force them to high slightly altitudes above the poles.
Rotational aberration distorts the morphology of these 
\teq{\tau = 1} curves somewhat, introducing asymmetry between 
the leading and trailing edges, along the lines of 
the magnetospheric cross section plot in Fig.~3 of \citet{HTE78}.

As the pair convertosphere minimum altitude contours move to equatorial
regions, it is clear that \teq{r_{\rm min}} is a monotonically
increasing function of colatitude \teq{\thetaE}. Photons emitted above
the equator more readily transit across the field lines than in polar
locales, and so have shorter attenuation lengths.  This is because the
magnetic field lines possess shorter radii of curvature in equatorial
zones than in polar regions, at a given emission altitude. Hence, for a
fixed photon energy, in order to compensate and increase \teq{L} to
infinity, the local field strengths sampled must be lowered, forcing the
required \teq{\thetakB} at the instant of conversion to larger values. 
Thus, the minimum altitudes must rise as \teq{\thetaE} does, and the
result is the leaf-shaped morphology in Fig. \ref{fig:Eesc_flatspace}.
For moderately small colatitudes, this trend can be discerned from
Eq.~(\ref{eq:e_esc_Erber}), namely the Erber asymptotic form for the
escape energy.  It is noteworthy that this behavior contradicts that
displayed in Figure~3 of \cite{Lee10}, where their \teq{r_{\rm min}}
values drop for colatitudes \teq{\theta\gtrsim 60^{\circ}}, a decline
that does not appear to depend on aberrational influences in their work.
It is not clear why this behavior is elicited in their computations.
In Section~\ref{sec:aberration}, we demonstrate that rotational/aberrational
influences on escape energy and minimum altitude determinations at these
equatorial colatitudes are comparatively small.

\section{GENERAL RELATIVISTIC EFFECTS}
 \label{sec:GR}
 
Our overall approach to calculating curved spacetime effects on photon 
attenuation will be to integrate the optical depth over path length intervals in the 
local inertial frame (hereafter LIF), with all magnetic fields, angles, energies, 
and distances computed in that frame.  In general, we will use the definitions 
for curved spacetime quantities from \citet[hereafter GH94]{GH94}, with the notation altered 
slightly for clarity.  Our starting point is again Eq.~(\ref{eq:atten_tau_def}), 
therefore requiring specification of the quantities \teq{B}, \teq{\omega} 
and \teq{\thetakB} in the LIF.   The blueshift of the photon energy in the LIF
from its value \teq{\erg\equiv\omega_{\infty}} at infinity (i.e. as observed)
can be accounted for with the simple correction
\begin{equation}
   \omega\; =\; \dover{\erg}{\sqrt{1-\Psi}}\quad ,\quad
   \Psi\; =\; \dover{r_s}{r}\; \equiv\; \dover{2GM}{c^2r}
 \label{eq:blueshift}
\end{equation}
at radius \teq{r}, where \teq{r_s=2GM/c^2} is the Schwarzschild radius
of a neutron star of mass \teq{M}.  The introduction of the dimensionless 
parameter \teq{\Psi} to describe the radial position will expedite the 
path length integration in curved spacetime constructs; we will use 
it as our integration variable instead of \teq{\eta} in Eq.~(\ref{eq:tau_beta}), 
approximately equivalently to the approach of GH94.
The emission altitude \teq{\rE} will be prescribed by \teq{\PsiE=r_s/\rE < 1}.
Note that throughout, we will adopt the convention that \teq{\erg} shall denote 
the dimensionless photon energy as seen by an observer, and 
\teq{\omega} shall signify that in the LIF.

The general relativistic form of a dipole magnetic field in a Schwarzschild metric was 
developed in \citet{WS83}.  It is also expressed in Eq.~(21)
of GH94 in the LIF in terms of the coordinates \teq{r} and \teq{\theta}
for an observer at infinity:
\begin{eqnarray}
   \hbox{\bf B}_{GR} & = & -\, 3\, \dover{B_p\cos\theta}{r_s^2 r} 
      \left[ \dover{r}{r_s} \log_e \left(1-\dover{r_s}{r}\right)+1+\dover{r_s}{2r}\right] \hat{r}\nonumber\\[-5.5pt]
 \label{eq:B_GR}\\[-5.5pt]
   & & +\, 3\, \dover{B_p\sin\theta}{r_s^2 r} 
       \left[ \left( \dover{r}{r_s} -1\right) \log_e \left(1-\dover{r_s}{r}\right)+1-\dover{r_s}{2r}\right] 
       \dover{\hat{\theta}}{\sqrt{1-r_s/r}}\quad .\nonumber
\end{eqnarray}
In flat spacetime, where \teq{r_s\ll r}, \teq{B_p} represents the surface polar field 
at \teq{\theta =0}.  It is more convenient to write this in terms of the scaled 
inverse radius \teq{\Psi= r_s/r }.  To this end we define the functions
\begin{eqnarray}
   \xi_r (x) & = & - \frac{1}{x^3} \left[ \log_e(1-x) + x + \frac{x^2}{2} \right] \nonumber\\[-5.5pt]
 \label{eq:xi_r_theta_def}\\[-5.5pt]
   \xi_{\theta} (x) & = & \frac{1}{x^3\sqrt{1-x}} \left[ (1-x)\, \log_e(1-x) + x - \frac{x^2}{2} \right] \;\; .\nonumber
\end{eqnarray}
Then, the curved spacetime dipole field is expressed via
\begin{equation}
   \hbox{\bf B}_{GR}\; =\; 3\dover{B_p\Psi^3}{r_s^3} \left\{ \xi_r (\Psi)\,\cos\theta\, \hat{r}
      + \xi_{\theta} (\Psi)\,\sin\theta\, \hat{\theta}\right\}\quad .
 \label{eq:dipole_GR}
\end{equation}
In flat spacetime, where \teq{\Psi\ll 1}, the leading terms of the Taylor series expansion
yield \teq{\xi_r(\Psi )\approx 1/3} and \teq{\xi_{\theta} (\Psi )\approx 1/6}, 
so that then Eq.~(\ref{eq:dipole_GR}) reproduces the familar result in 
Eq.~(\ref{eq:dipole_fs}) in the absence of general relativity.  The magnitude 
of the general relativistic field is then
\begin{equation}
   B_{GR} \; =\; 3\dover{B_p\Psi^3}{r_s^3}\, 
   \sqrt{ [\xi_r (\Psi)]^2\,\cos^2\theta + [\xi_{\theta} (\Psi)]^2\,\sin^2\theta}\quad ;
 \label{eq:B_GR_tot}
\end{equation}
this will be employed in the quantum pair creation rates in the local inertial frame.
The ratio of Eq.~(\ref{eq:B_GR_tot}) for altitudes near the surface to its flat 
spacetime value (i.e., \teq{\Psi\to 0}) inferred from Eq.~(\ref{eq:dipole_fs})
reproduces the ratio plotted in Fig.~5c of GH94.

The trajectory of a photon emitted from a point in a neutron star magnetosphere 
will be curved in the frame of an observer at infinity, though for cases of emission 
near the polar cap, this is generally small \citep[see][]{BH01}.  Here 
we incorporate the influence of the slight curvature in the path, so that
calculating \teq{\sin\thetakB} becomes a slightly more complicated exercise 
than it was in the flat spacetime approximation.  First, the photon is emitted parallel 
to the magnetic field in the LIF.  This fixes \teq{\deltaE}, 
the initial angle between the photon trajectory and the radial direction
(depicted in Fig.~\ref{fig:prop_geometry}):
\begin{equation}
   \sin\deltaE\; \equiv\;  \dover{B^{\hat{\theta}}}{B}\biggl\vert_{r=R_e}
   \; =\; \dover{\sin\thetaE\, \xi_{\theta}(\PsiE)}{\sqrt{ \cos^2\thetaE \left[\xi_r(\PsiE)\right]^2
      + \sin^2\thetaE \left[\xi_{\theta}(\PsiE)\right]^2 }}\quad .
 \label{eq:sindelta0}
\end{equation}
When \teq{\PsiE\ll 1}, this reduces to Eq.~(\ref{eq:deltae_def}), though in general,
since \teq{\xi_{\theta} (\PsiE )/\xi_r(\PsiE ) \approx 1/2 + \PsiE/8 + O(\PsiE^2)} 
in this limit, it is easily seen that spacetime curvature increases \teq{\deltaE} 
for proximity to the magnetic pole.  This effect is illustrated in Figure~3b of 
GH94.  The photon's trajectory at infinity emerges parallel 
to a line drawn from the center of the star, displaced from it by a distance \teq{b}.  
This impact parameter \teq{b} is proportional to the ratio of two conserved 
quantities of the unbound photon orbit, the orbital angular momentum
and the energy; consult \citet{PCF83} or 
Chapter 8 of \citet{Weinberg72} for illustrations of such orbits.
Scaling \teq{b} by the Schwarzschild radius, as we have with \teq{r}, 
introduces a new trajectory parameter \teq{\Psi_b= r_s/b} that can 
be related to \teq{\PsiE} and \teq{\deltaE} via
\begin{equation}
   \Psi_b \; =\; \dover{\PsiE}{\sin\deltaE}\,\sqrt{1-\PsiE}
   \; \equiv\; \PsiE\, \sqrt{ (1-\PsiE)\, \left\{ 1+\left[\xi (\PsiE)\right]^2 \cot^2\thetaE \right\}  }\quad ,
 \label{eq:Psi_b_def}
\end{equation}
where
\begin{equation}
    \xi (\Psi )\; =\; \dover{\xi_r(\Psi )}{\xi_{\theta}(\Psi )}\quad .
 \label{eq:xi_def}
\end{equation}
The first identity in Eq.~(\ref{eq:Psi_b_def}) is derived from Eq.~(17) of 
GH94 (correcting a typographical error therein: see Eq.~(A2) of HBG97), who use the notation 
\teq{\delta_0} for \teq{\deltaE}.  Observe that the impact parameter 
can be smaller than the Schwarzschild radius for almost radial 
trajectories initiated near the magnetic polar axis (setting 
\teq{\sin\deltaE\ll 1}), so \teq{\Psi_b} can assume values well in excess of 
unity where the orbit is a capture one, if reversed.  Inserting 
Eq.~(\ref{eq:sindelta0}) to substitute for \teq{\sin\deltaE} then yields \teq{\Psi_b} purely as a function 
of the emission altitude (i.e. \teq{\PsiE}) and colatitude \teq{\thetaE}, and
derives the second identity in Eq.~(\ref{eq:Psi_b_def}), with
\teq{0\leq\xi (\Psi ) \leq 2} on the interval \teq{0\leq\Psi\leq 1}.  

This second form for \teq{\Psi_b}
is needed for the photon trajectory computation, an integral expression 
for which is given in Eq.~(11) of GH94:
\begin{equation}
   \theta(\Psi) \; \equiv\; \thetaE + \Delta\theta \; =\; 
   \thetaE + \int_{\Psi}^{\PsiE} \frac{d\Psi_r}{\sqrt{\Psi_b^2-\Psi_r^2(1-\Psi_r)}}\quad ,
 \label{eq:curved_traj}   
\end{equation}
expressing the functional dependence \teq{\theta (r)}, as viewed by an observer 
at infinity.  An alternative version of this can be obtained from 
Eq.~(8.5.6) of \citet{Weinberg72}; see also \citet{MTW73}.
Since \teq{\Psi \leq\PsiE} in this construction, as the photon propagates 
out from the star, then the change in colatitude \teq{\Delta\theta} is necessarily 
positive as the altitude \teq{r} increases.  Observe that \teq{\Psi_b^2 > \PsiE^2 
(1-\PsiE)} from the second identity in Eq.~(\ref{eq:Psi_b_def}) so that the 
argument of the square root in the integrand of Eq.~(\ref{eq:curved_traj})
is positive-definite.  In the case of a neutron star, generally \teq{\PsiE\lesssim 0.4}, and 
the integral in Eq.~(\ref{eq:curved_traj}) can be approximated extremely accurately
by an analytic form, for non-equatorial emission colatitudes \teq{\thetaE\lesssim \pi/4}; 
see the Appendix for details.  This expedient step removes the trajectory integral from 
consideration, and speeds up optical depth computations immensely.  
In the flat spacetime limit, \teq{\PsiE\ll 1}, the integral for the trajectory in Eq.~(\ref{eq:curved_traj})
can be expressed analytically by replacing the argument of the square root in the denominator 
by \teq{\Psi_b^2-\Psi_r^2}.  Then, forming \teq{\sin\Delta\theta}, the result can be inverted 
to solve for \teq{\Psi} and thereby find the locus for the trajectory:
\begin{equation}
   \Psi\; =\; \Psi_b\, \sin \left( \thetaE - \theta - \arcsin \dover{\PsiE}{\Psi_b}\right)\quad . 
 \label{eq:flatspace_locus}
\end{equation}
This is a polar coordinate form for a straight line, and is easily shown to be 
equivalent to Eq.~(\ref{eq:chi_def}) using the limiting form 
\teq{\Psi_b \approx \PsiE\sqrt{1+4\cot^2\thetaE}\approx \PsiE/\sin\deltaE} 
when \teq{\PsiE\ll 1}.

Given emission locale coordinates (\teq{\PsiE, \thetaE}), for any subsequent 
position (\teq{\Psi, \theta}) along the curved trajectory, we can determine the angle
\teq{\thetakB} of the photon momentum to the local field direction, in the LIF.  This is simply
done by forming a cross product between the photon momentum \teq{\hbox{\bf k}_{GR}} and 
\teq{\hbox{\bf B}_{GR}} using Eq.~(\ref{eq:dipole_GR}) for the field.  The 
photon momentum in the LIF can be derived from the formalism in Section 3 of GH94, 
or by manipulation of the differential form of the trajectory 
equation in Eq.~(\ref{eq:curved_traj}), i.e. setting \teq{k_{\theta}/k_r = d\theta/dr
= - (\Psi /r)\, d\theta/d\Psi} and then normalizing to Eq.~(\ref{eq:blueshift}).  The result is
\begin{equation}
   \hbox{\bf k}_{GR}\; =\; \dover{\erg}{\Psi_b\sqrt{1-\Psi }}
    \left\{ \sqrt{\Psi_b^2 - \Psi^2 (1-\Psi )}\, {\hat r} + \Psi \sqrt{1-\Psi} \, {\hat \theta} \right\}\quad ,
 \label{eq:momentum_LIF}
\end{equation}
which can be simply inferred from Eq.~(A1) of \citet{HBG97}.
>From this, one can form the angle \teq{\deltaE} for the initial angle of the photon 
momentum relative to the radial direction, via \teq{\sin\deltaE = \vert\hbox{\bf k}_{GR}
\times {\hat r}\vert / \vert \hbox{\bf k}_{GR} \vert = \PsiE\sqrt{1-\PsiE}\, /\Psi_b},
a result that is the first identity in Eq.~(\ref{eq:Psi_b_def}).  Forming a cross product 
between the photon momentum and the field vectors, it follows that 
\begin{equation}
   \sin\thetakB \; \equiv\; \dover{\vert \hbox{\bf k}_{GR} \times \hbox{\bf B}_{GR}\vert}{
                          \vert \hbox{\bf k}_{GR}\vert . \vert \hbox{\bf B}_{GR} \vert}
                \; =\; \dover{B^{\hat{\theta}}}{B} \left[1-(1-\Psi)\dover{\Psi^2}{\Psi_b^2}\right]^{1/2}
                                          -\dover{B^{\hat{r}}}{B}\dover{\Psi (1-\Psi)^{1/2}}{\Psi_b}\quad ,
 \label{eq:sintheta_kB}
\end{equation}
an expression that is also routinely obtained by rearranging Eq.~(37) of 
GH94.  Inserting the forms for the field components, 
elementary manipulations yield 
\begin{equation}
   \sin\thetakB \; =\; \dover{\sqrt{\Psi_b^2 - \Psi^2 (1-\Psi )}
        - \Psi \sqrt{1-\Psi}\, \xi (\Psi ) \, \cot\theta }{\Psi_b\, \sqrt{  1+ \left[\xi (\Psi )\right]^2 \cot^2\theta } }
 \label{eq:sinthetakB_GR}
\end{equation}
Employing the second form for \teq{\Psi_b} in Eq.~(\ref{eq:Psi_b_def})
quickly reveals that when \teq{\Psi=\PsiE}, this expression yields
\teq{\sin\thetakB=0}. Using the fact that \teq{\Psi^2 (1-\Psi)} is an
increasing function for \teq{0 < \Psi < 2/3}, and that \teq{\xi (\Psi )}
is a more modestly declining function of \teq{\Psi} on the same
interval, it is routinely established that \teq{\sin\thetakB} increases
as \teq{r} increases from the emission radius, i.e. \teq{\Psi} drops
below \teq{\PsiE}. Numerical comparisons of our computations of
\teq{\sin\thetakB} and the effective pair threshold \teq{2/\sin\thetakB}
with panels (a) and (b) of Fig.~5 of GH94 were performed, yielding
excellent agreement. In the flat spacetime limit \teq{\PsiE\ll 1},
\teq{\Psi_b \approx \PsiE/\sin\deltaE \approx \Psi/\sin (\deltaE-\eta )}
can be deduced using Eq.~(\ref{eq:chi_def}), and then it is
straightforward to demonstrate that Eq.~(\ref{eq:sinthetakB_GR}) reduces
to Eq.~(\ref{eq:sin_thetakB}).

Finally, we choose to change our integration variable from \teq{s} to
\teq{\Psi}. In the LIF, the path length is related to the coordinate
transit time: \teq{ds^2 = (1-\Psi) c^2dt^2} in the Schwarzschild case.
Equivalently, the path length can be connected to the radial and angular
(equatorial) contributions to the Schwarzschild metric via \teq{ds^2 =
dr^2/(1-\Psi ) + r^2\, d\theta^2}. The two forms are equivalent,
yielding the proper time interval \teq{d\tau^2 =0} for light-like
propagation.  Employing Eq.~(18) of GH94, or equivalently taking the
derivative of Eq.~(8.7.2) of \citet{Weinberg72}, yields an expression
for \teq{dt/d\Psi} for the photon's transit along its trajectory,
essentially formulae for Shapiro delay.  Assembling these pieces one
quickly arrives at the change of variables
\begin{equation}
   \dover{ds}{\rE} = -\, \dover{\PsiE\, \Psi_b\, d\Psi}{\Psi^2 \sqrt{(1-\Psi )\, \left\{\Psi_b^2-\Psi^2(1-\Psi)\right\} }}\quad .
 \label{eq:ds_dPsi}
\end{equation}
The optical depth integration for the case of including general relativity then takes the form 
\begin{equation}
   \tau (\Psi )\; =\; \rE\,\PsiE \int_{\Psi}^{\PsiE}
   \dover{{\cal R}(\omega,\, \sin\thetakB, B_{GR})
         \, \Psi_b\, d\Psi_r}{\Psi_r^2 \sqrt{(1-\Psi_r ) \left\{\Psi_b^2-\Psi_r^2(1-\Psi_r )\right\} }}\quad ,
 \label{eq:tau_Psi}
\end{equation}
where the arguments of the scaled quantum pair creation rate
\teq{{\cal R}} are given by Eqs.~(\ref{eq:blueshift}),~(\ref{eq:B_GR_tot})
and~(\ref{eq:sinthetakB_GR}).  With this construct, we can formally define the attenuation 
length \teq{L} as in \citet{HBG97} and \citet{BH01} via
\begin{equation}
   \tau \left(\Psi_L\right) = 1\quad ; \quad s\left(\Psi_L\right) = L \quad .
 \label{eq:atten_length}
\end{equation}
\teq{L} is approximately the cumulative LIF distance that a photon of a given energy 
will travel from its emission point before converting to an electron-positron pair.
When \teq{\PsiE\ll 1}, Eq.~(\ref{eq:tau_Psi}) is equivalent to the 
flat spacetime evaluation in Eq.~(\ref{eq:tau_beta}). 

\twofigureoutpdf{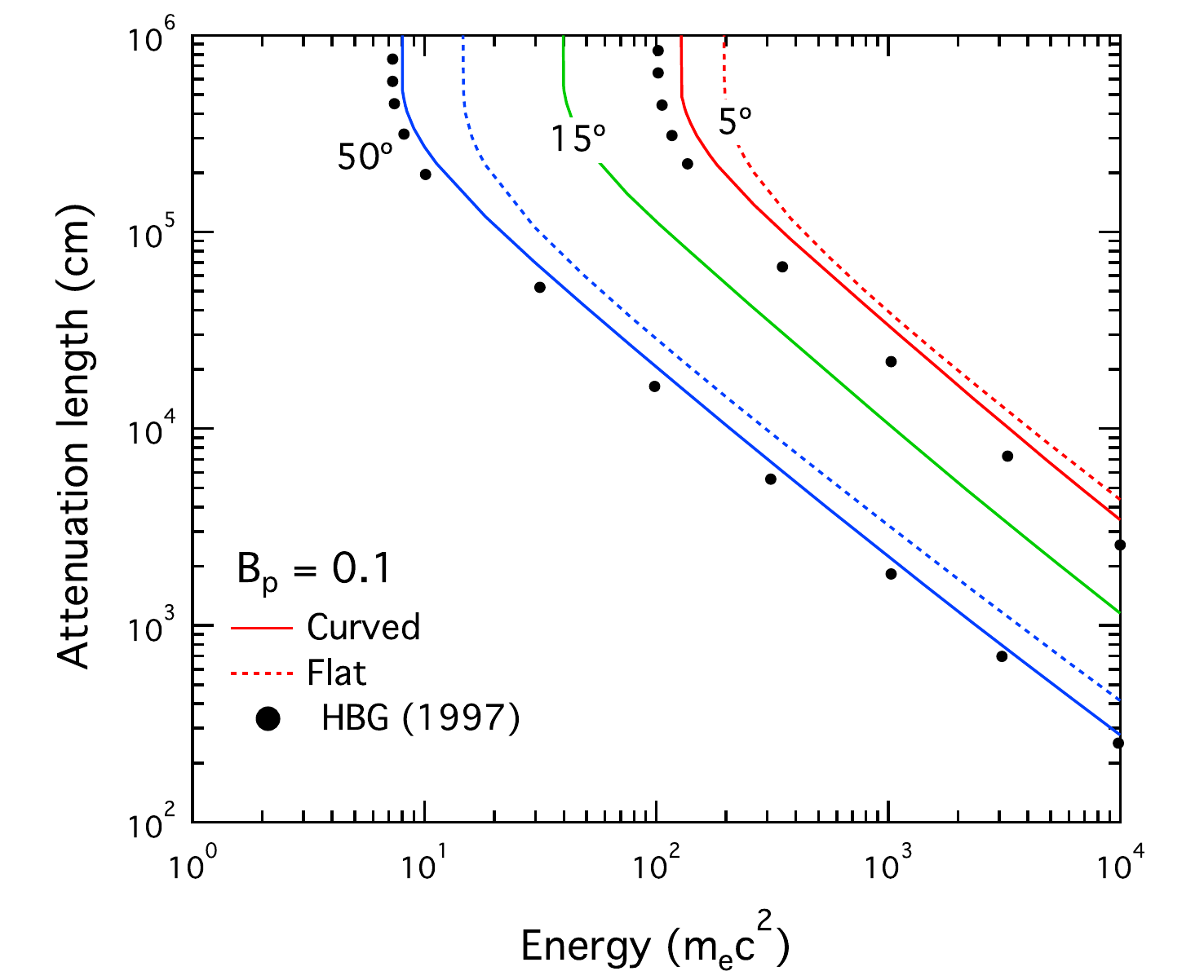}{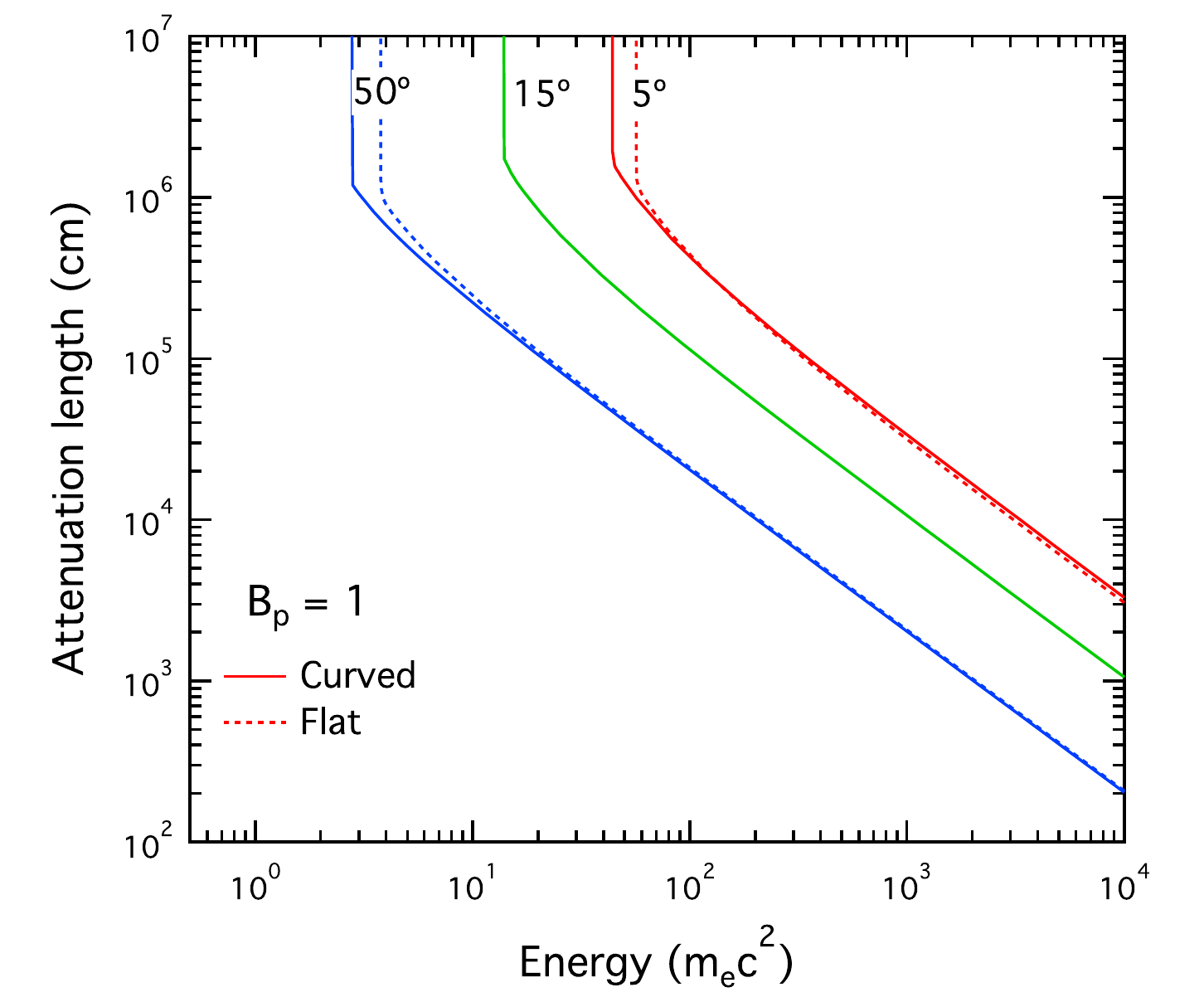}{3.7}{-0.05}{
Attenuation lengths for photons emitted at colatitudes of 5, 15, and 50 degrees, 
as functions of the photon energy \teq{\erg} that is observed at infinity,
for a neutron star with surface polar magnetic field of \teq{B_p = 0.1} (left) 
and \teq{B_p = 1} (right).  These represent the quantity \teq{L}, 
defined in Eq.~(\ref{eq:atten_length}) for curved spacetime (solid curves) and 
Eq.~(\ref{eq:atten_tau_def}) for flat spacetime (dotted lines).  All curves use the threshold-corrected 
attenuation coefficient in Eq. (\ref{eq:B88_asymp}).  The curves turn up and 
asymptotically approach infinity at the escape energy \teq{\eesc}.
The large dots for the \teq{B_p=0.1} case on the left depict pair
attenuation lengths from Fig.~2 of \cite{HBG97}, specifically for emission 
colatitudes \teq{\thetaE} of \teq{5^{\circ}} and \teq{50^{\circ}}.
}{fig:Atten_GR}

Figure \ref{fig:Atten_GR} displays the attenuation lengths computed for 
curved spacetime at two different magnetic fields.  These are 
evaluated specifically for emission from the neutron star surface.  The curves 
are monotonically declining functions of photon energy \teq{\erg} as observed 
at infinity.  At high energies, where \teq{L\ll \rns}, pair attenuation occurs 
very close to the surface and general relativistic effects modify the field structure 
and photon trajectory and redshift in a manner that is essentially independent 
of \teq{\erg}.   Accordingly, for the \teq{B=0.1} example, 
a dependence \teq{L\propto (\erg\thetaE)^{-1}} is approximately realized, 
just like the Minkowski spacetime dependence 
deduced from Eq.~(\ref{eq:tau_small_thetae_Erber_fin}), but with a smaller 
coefficient of proportionality in the GR case.  The influence of curved spacetime 
reduces \teq{L} slightly, primarily because it amplifies both the field strength 
and the photon energy in the LIF.  In the \teq{B=1} example, the GR-corrected and 
flat spacetime attenuation lengths are almost identical because photon conversion
arises very soon after pair threshold ($\omega_\perp = 2$) in the LIF is crossed during 
propagation. The trajectories then sample regimes \teq{\PsiE\ll\Psi_b} before attenuation, 
so that the path length differential in Eq.~(\ref{eq:ds_dPsi}) approximately satisfies 
\teq{ds/dr \approx 1/\sqrt{1-\PsiE}}, using \teq{dr/r = - d\Psi /\Psi}.  Hence the 
post-Newtonian GR correction to the path length \teq{s=L} is precisely that for 
the blueshift of the photon energy in the LIF.  Accordingly, the computation of 
\teq{L} in such threshold-conversion domains is insensitive to general relativistic 
modifications.

At low energies, the curves turn up and asymptotically approach 
infinity at the escape energy \teq{\eesc}.    A small shift in escape energy is 
evident, due largely to the gravitational redshifting of the photon energy.
The monotonic trend of decreasing \teq{L} and \teq{\eesc} with increasing 
colatitude \teq{\thetaE} of emission is a result of increased field line curvature 
away from the magnetic polar regions.  The footpoint  emission colatitude 
\teq{\theta_f \equiv\thetaE} can be coupled to a pulsar rotation period if it is assumed 
to be applicable to the last open field line, \teq{\theta_f\to\theta_p}.
For a dipolar field in flat space-time this ``polar cap'' colatitude is given 
by \teq{\sin^2\theta_p = 2\pi \rns /(Pc) \equiv \rns /\rlc}, where 
\teq{\rlc = Pc/2\pi} is the light cylinder radius.  With general relativistic 
modifications to the field structure, as defined by Eq.~(\ref{eq:dipole_GR}), 
\begin{equation}
   \sin^2\theta_p \; =\; \dover{\psilc\, \xi_r(\psilc )}{\PsiE\, \xi_r(\PsiE)}\quad .
 \label{eq:polar_cp_GR}
\end{equation}
This is Eq.~(27) of \cite{GH94}.  Here \teq{\psilc = r_s/\rlc}, which is 
usually much less than unity for young pulsars, so that 
\teq{\xi_r(\psilc )\approx 1/3}.  Generally, \teq{\PsiE = r_s/\rns} is not 
much less than unity.  Finally, for the \teq{B_p=0.1} case
(left panel), Fig.~\ref{fig:Atten_GR} also displays points corresponding 
to the pair attenuation computations in Fig.~2 of \cite{HBG97}.  Our \teq{L} results 
here range from 10--30\% higher than these older evaluations --- this 
difference is discussed below.

\begin{figure}[h]
\centerline{
 \includegraphics[width=.7\textwidth]{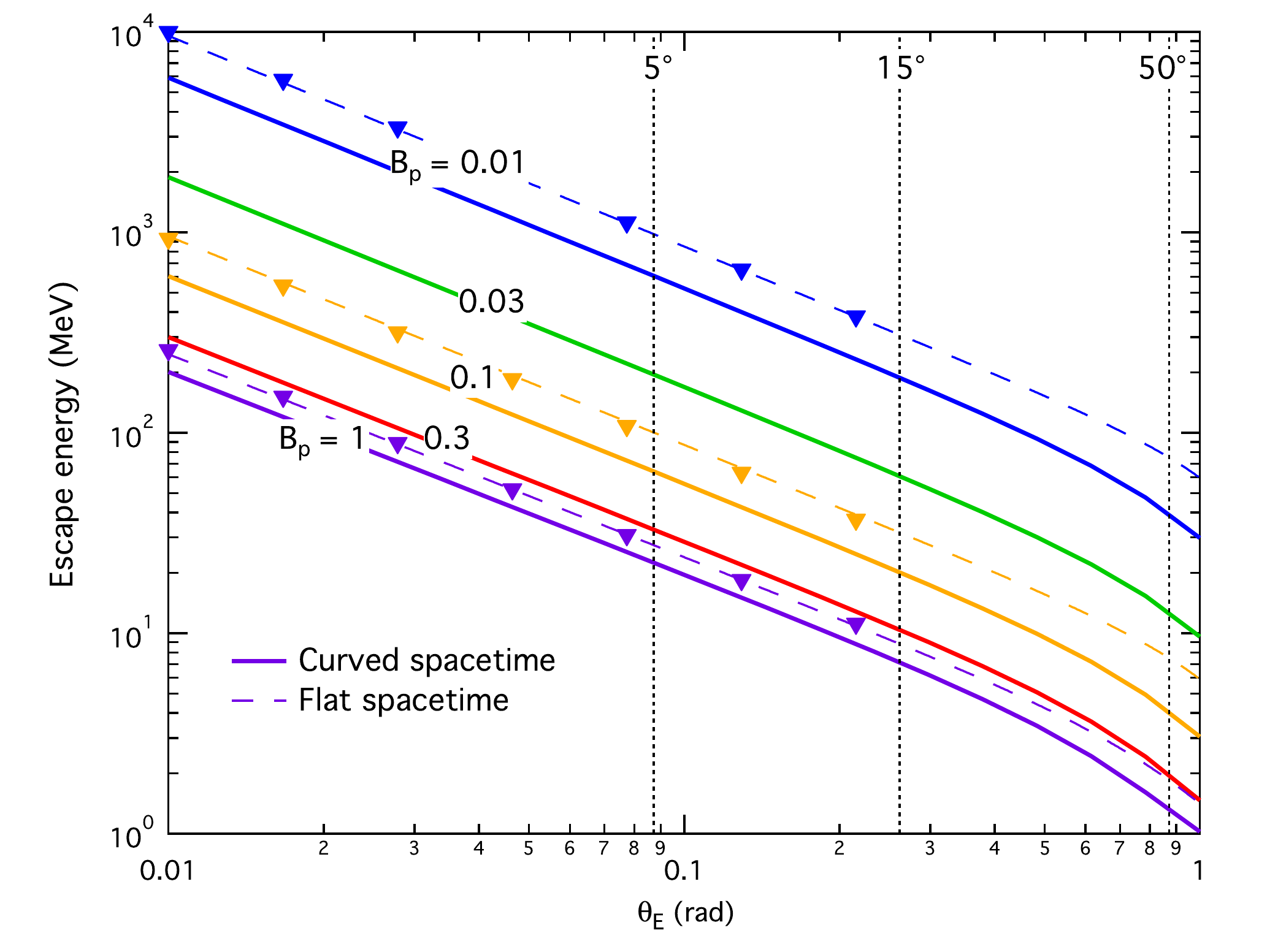}}
\caption{Escape energies from the neutron star surface in curved spacetime 
for the attenuation rate defined in Eq.~(\ref{eq:atten_length}).  The dashed 
curves are taken from the solid lines in Fig. \ref{fig:Eesc_flatspace} and 
represent the flat spacetime escape energies for the same magnetic fields.  
Triangles are the escape energies calculated by \citet{CCH96}, as in 
Fig.~\ref{fig:Eesc_flatspace}.  Colatitudes of 5, 15, and 50 degrees are marked 
for easier comparison with Fig.~\ref{fig:Atten_GR}.  Including GR effects lowers 
the escape energies, but preserves the same slope because the basic 
form of the exponential in the attenuation coefficient is unchanged.}
 \label{fig:Escape_energies_GR}
\end{figure}

The escape energies calculated for the general relativistic analysis are shown in 
Fig.~\ref{fig:Escape_energies_GR}, as functions of the emission colatitude, for different 
polar magnetic field strengths.  Also depicted are the flat space-time equivalents 
for \teq{B_p=0.01, 0.1, 1} (in units of \teq{B_{\rm cr}}), 
clearly demonstrating that GR corrections
have a greater impact for \teq{B_p\ll 1} cases (almost by a factor of two) than for 
\teq{B_p\approx 1} domains.  When \teq{\thetaE\ll 1}, the escape energies 
for curved spacetime simply satisfy \teq{\eesc \propto \thetaE^{-1}}, as expected,
since the form of the argument of the exponential in the pair creation attenuation 
coefficient remains approximately the same as for the flat space-time situation:
photon conversion arises well above pair threshold.  In this low emission 
colatitude regime (\teq{\thetaE\lesssim 0.2}), when the field is highly sub-critical, 
it then also follows that \teq{\eesc \propto B_p^{-1}}, a dependence that is evident in 
the Figure.  Once the polar field approaches and exceeds \teq{B_{\rm cr}}, the 
escape energies become almost independent of the field value, because 
any pair conversion at high altitudes still is fairly near the threshold 
\teq{\omega_{\perp}=2}.

\begin{figure}[h]
\centerline{
 \includegraphics[width=.7\textwidth]{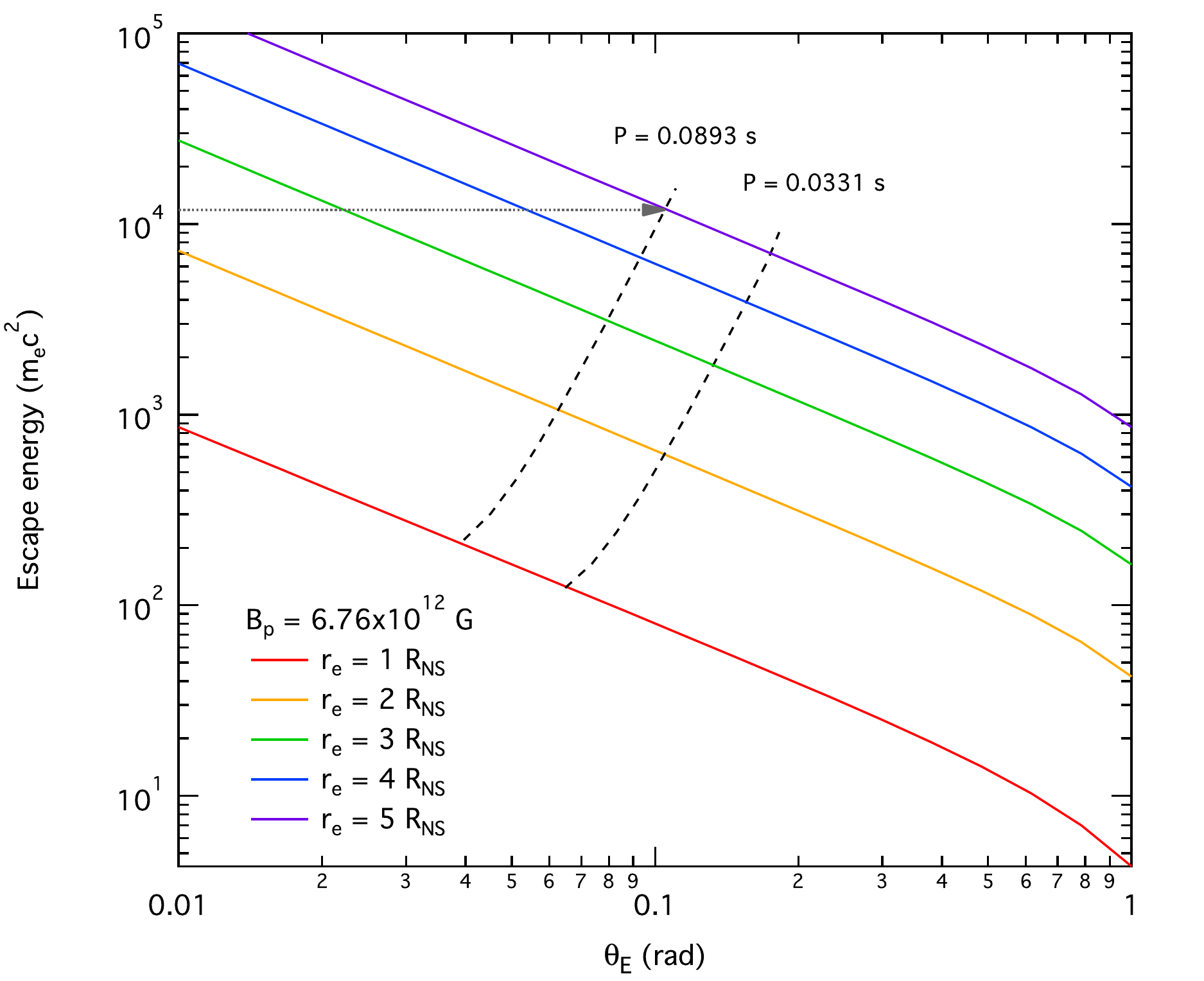}}
\caption{Pair creation escape energies in curved spacetime for five altitudes of emission, 
as a function of emission colatitude.  The unscaled surface polar magnetic field 
\teq{B_p=6.76\times 10^{12}}Gauss is that for the Vela pulsar, and is only
slightly different from the value for the Crab pulsar.  The two dotted curves 
represent the loci of colatitudes appropriate for the last 
open field line for pulsars with periods of 0.033 s (Crab) and 0.089 s (Vela).
Also marked with the horizontal arrow is {\it twice} the cutoff energy \teq{E_c=3.03}GeV 
measured in the {\it Fermi}-LAT phase-averaged spectrum of Vela, identifying the 
minimum altitude of \teq{\rE\approx 5\rns} for super-GeV emission in this pulsar.}
 \label{fig:EescGR}
\end{figure}

In Fig.~\ref{fig:EescGR}, the general relativistic escape energy is again 
plotted as a function of emission colatitude, but now illustrating 
the dependence upon emission altitude \teq{\rE =h\rns}.  This evinces the 
expected increase of \teq{\eesc} as the emission point becomes more 
remote from the stellar surface.  To forge a preliminary connection with 
pulsar observations, contours in this escape energy phase space are 
depicted for the last open field lines pertinent to the Crab and Vela pulsars. 
These employ solutions of Eq.~(\ref{eq:polar_cp_GR}) for the parametric 
locus of this field lines, specifically for the two different pulsar periods, and curve 
somewhat down near the stellar surface since curved space-time reduces the polar 
cap size (e.g. see GH94). 
At low altitudes, the trend of \teq{\eesc \propto h^2} is 
approximately realized along these diagonal contours.  Once the emission altitude 
rises above \teq{h\gtrsim 2}, GR influences are quite small, and the flat 
spacetime trend of \teq{\eesc \propto h^{5/2}} in Eq.~(\ref{eq:e_esc_alt_dep})
is approximately satisfied instead along these contours.  The reduction of the polar cap size is 
primarily responsible for the general relativistic weakening of the altitude dependence near 
the stellar surface.  To find a minimum altitude for emission, one locates the 
point on these contours of constant period where \teq{\eesc = \alpha E_c}, where 
\teq{\alpha =2} and \teq{E_c} is the exponential 
cutoff energy of the observed pulsar spectrum.  The choice for the Vela pulsar,
where \teq{E_c=3.03}GeV for the phase-averaged spectrum \citep{PSRCat2}, is illustrated 
in the upper left, yielding an estimate for the minimum altitude of emission
\teq{r_{\rm min}\approx 5\rns} for Vela.  This bound delineates the range of altitudes for which 
pair transparency is achieved in the magnetosphere of a given pulsar, for emission 
along the last open field line.  This protocol for constraining the emission zones 
of pulsars is discussed at greater length in Section.~\ref{sec:observations}.

The pair production attenuation lengths and escape energies 
computed here differ slightly from those presented in \citet{HBG97} and
\citet{BH01}.  The attenuation lengths in Fig.~\ref{fig:Atten_GR} are
systematically higher by around 10\% than those in the left panel of
Fig.~2 of \cite{HBG97}.  The escape energies in
Fig.~\ref{fig:Escape_energies_GR} are higher than the corresponding
evaluations in \cite{HBG97} by around 20-30\%.  The origin of this
difference is presently unclear.  We observe that there appears to be a
slight disagreement between the values of \teq{\sin\thetakB} computed in
\cite{HBG97} for curved spacetime and those derived in this work and in
\cite{GH94}, with those in \cite{HBG97} being about 15--20\% higher.
This is consistent with the slightly lower values of \teq{L} and
\teq{\eesc} computed in \cite{HBG97} relative to those here.  As noted
above, there is excellent agreement between our geometry and attenuation
coefficient calculations and those presented in \citet{GH94}.  Our
numerical results for the GR case map continuously over to the 
\teq{\PsiE \to 0} flat spacetime cases illustrated in
Sec.~\ref{sec:flatspace}. These latter checks indicate that the curved
spacetime results presented here appear to be robust.

\begin{figure}[h]
\centerline{
 \includegraphics[width=.7\textwidth]{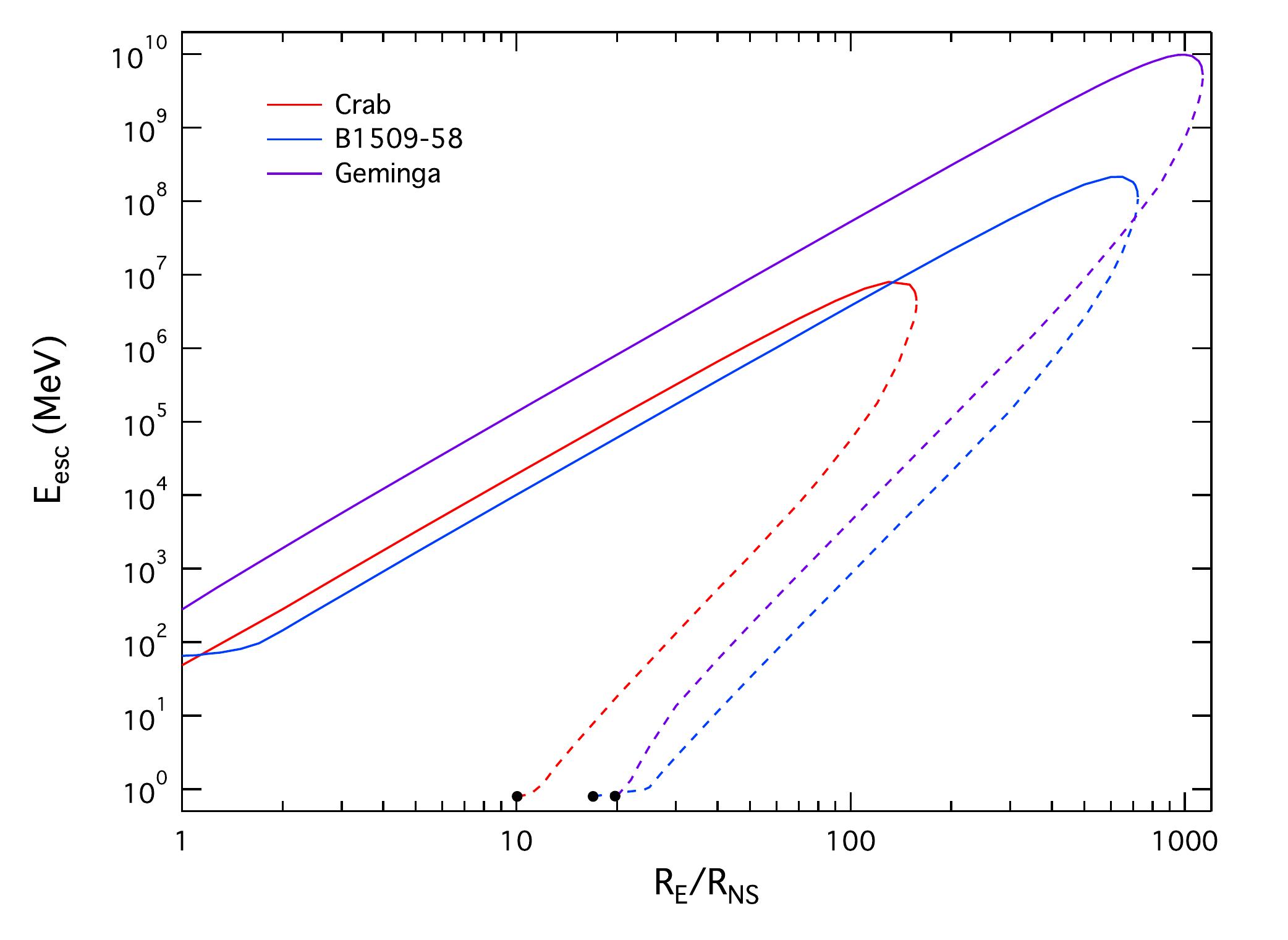}}
\caption{Escape energies in MeV for photons emitted parallel to  
the last open field line for pulsars with periods and surface magnetic  
field strengths equal to those of three observed pulsars.  Solid  
curves indicate a photon traveling in the direction of increasing  
colatitude (\teq{\eta > 0}, the ``upward" direction).  Dashed curves  
indicate a photon traveling in the direction of decreasing colatitude  
(\teq{\eta < 0} with \teq{\eta} defined as in Fig. \ref{fig:prop_geometry},  
the ``downward" direction).  General relativistic effects are included  
in this analysis, but rotational aberration effects are not.  The  
dashed curves come to an end (black dots) where photons emitted from  
the altitude on the x-axis would impact the neutron star.  Note that  
the escape energy at this point may be less than the pair creation  
threshold of \teq{2 m_e c^2} because the energy at infinity will be  
blueshifted above the threshold close to the neutron star.}
 \label{fig:photons_up_down}
\end{figure}

As a concluding focus, the techniques in this Section can be applied to  
downward-traveling photons as well, a consideration that is germane to 
determining polar cap and surface reheating.  A curvature photon emitted from  
an inward-bound electron or positron will experience both stronger  
magnetic fields and larger \teq{\sin\thetakB} along its path than its  
outward-traveling counterpart, so its escape energy will be  
considerably lower.  Fig.~\ref{fig:photons_up_down} shows the escape  
energies for photons emitted along the last open field line in curved  
spacetime.  Solid curves represent upward-traveling photons; dashed  
curves represent downward-traveling photons.  The dashed curves come  
to an end where photons emitted from the altitude on the x-axis would  
impact the neutron star.  The different colored curves represent  
parameters for three different pulsars: Crab (red), B1509-58 (J1513-5908; blue),  
and Geminga (purple).

The downward-traveling curves come to an end when the  
emission location starts to experience the neutron star's shadow.  The  
edge of the shadow can by found by solving for the maximum \teq{\Psi}  
(corresponding to the distance of closest approach to the neutron  
star) for a photon trajectory, and then setting that maximum \teq{\Psi}  
equal to \teq{2M/R} and solving for the emission radius.  Using  
\citet{GH94} Eq. (10) (recast in our preferred variables), we can find  
the maximum \teq{\Psi} by solving the cubic equation in \teq{\Psi} given by
\begin{equation}
   0 \; =\; \left(\frac{d\Psi}{d\theta}\right)^2 \equiv \Psi_b^2 - \Psi^2  
   \left(1-\Psi\right) \quad .
 \label{eq:psi_max_eq}
\end{equation}
\teq{\Psi_b} depends only on the emission location, so using 
Eq.~(\ref{eq:Psi_b_def}) and \citet{GH94} Eq. (27) to get \teq{\Psi_b (h)} for  
emission along the last open field line, we can find \teq{\Psi_{\rm max} (h)}.  
When the trajectory just clips the neutron star, we will have
\begin{equation}
   \Psi_{\rm max} (h) \; =\; \frac{2M}{R} \quad ,
 \label{eq:psi_max_sol}
\end{equation}
and this can be numerically solved for \teq{h} to give the intersection of  
the edge of the shadow and the last open field line.  The escape  
energy for the trajectory that passes closest to the neutron star can  
also be estimated.  This surface-skimming path passes through  
the region where the magnetic field is strongest and \teq{\sin\theta_{kB}}  
is close to 1, so the attenuation rate will be very large.  For all  
the pulsars shown here, the absolute pair creation threshold of  
\teq{\omega = 2 m_e c^2} acts as a wall in this region.  The photons will  
pair-produce as soon as their energy (gravitationally blueshifted in
the local inertial frame) is
above threshold, so the energy at infinity for photons that can escape  
from the edge of the shadow is given approximately by
\begin{equation}
   \erg \; \approx\; 2 \sqrt{1-\Psi_{\rm max}} = 1.513\, 
 \label{eq:omega_psi_max}
\end{equation}
(in units of \teq{m_e c^2}) for \teq{M = 1.44 M_\odot}, \teq{R = 10^6 \rm{cm}}.  This is  
independent of \teq{h}, which is determined using the above protocol.   
Threshold effects are stronger in pulsars with stronger magnetic  
fields, as we showed in Section \ref{sec:eesc_flatspace}, and we can  
see that for the high-field pulsar B1509-58, both the upward-traveling  
and downward-traveling curves begin to flatten out at the lower ends  
where the photons pass closest to the neutron star.  A more extreme  
example of this can be seen in the magnetar case in Fig.  
\ref{fig:magnetar_emax}, where the magnetic field is supercritical and  
the pair creation threshold strongly influences escape energies up to  
high altitudes.

For emission points at small to moderate fractions of the  
light cylinder radius, the escape energies for upward-traveling  
photons show a power law dependence on emission altitude with an index  
of approximately \teq{5/2}.  This is expected since general  
relativistic effects will become negligible very quickly as we move  
away from the surface and Eq.~(\ref{eq:e_esc_alt_dep}) applies.   
Calculating the expected power law index for the downward-traveling  
curves is more difficult, since none of the small-angle approximations  
we used to obtain Eq.~(\ref{eq:e_esc_alt_dep}) are valid at \teq{\Psi =  
\Psi_{\rm max}}, which is where pair attenuation is anticipated to be most
effective.  Then GR contributions cannot be neglected when the photon  
passes close to the neutron star even if it was emitted at a high  
altitude.  However, we can make a naive flat spacetime estimate by  
solving for the value of \teq{\eta} at the point of closest approach to  
the neutron star and performing a series expansion of the argument of  
the exponential in Eq.~(\ref{eq:Erber_asymp}) at that point.  If we  
set the argument of the exponential equal to 1, we can then solve for  
\teq{\omega} in terms of \teq{h}.  This analysis suggests a leading order  
contribution of \teq{\eesc \propto h^4}, with a non-negligible \teq{h^5} term  
as well.  This is not too divergent from the actual approximate power  
law dependence of \teq{\omega \propto h^{9/2}}.

A few global characteristics are easily discerned from this  
analysis.  First, downward-traveling photons are attenuated much more  
strongly than upward-traveling photons, with escape energies orders of  
magnitude lower than their outbound counterparts for the same emission  
location.  Second, even with the stronger attenuation, photons emitted  
in the downward direction from high altitude gaps may still be visible  
in the {\it Fermi}-LAT band if they are not attenuated by other means
such as via \teq{\gamma\gamma\to e^+e^-} pair creation.  For the  
Crab pulsar parameters, for example, downward-traveling 300 MeV  
photons emitted from a gap along the last open field line can escape  
to infinity if they originate above approximately 40 neutron star  
radii.

\section{RELATIVISTIC ABERRATION DUE TO STELLAR ROTATION}
 \label{sec:aberration}

In a pulsar's rapidly rotating magnetosphere, the attenuation rate due
to magnetic pair creation is affected by the deformation of the
magnetic field lines, both at high altitudes where the corotation
velocity is a significant fraction of the speed of light and at low
altitudes near the magnetic pole.  This phenomenon can equivalently be
described as the aberration of the photon momentum and energy from the
rotating stellar field frame.  Since pair opacity considerations are
focused primarily on the inner magnetosphere, it is the polar zone that
is of principal interest in this Section.  The computation here for the
optical depth is first derived for the most general case of arbitrary
emission colatitudes and altitudes in an oblique rotating neutron star,
and then restricted to special cases where one can derive incisive
analytic approximations.  In these calculations, we consider 
only a rotating rigid dipole field and neglect effects near the pole caused 
by sweepback of field lines near the light cylinder \citep[e.g.][]{DH04}.  We will 
also neglect general relativity effects, which have already been explored extensively.

Our overall approach will be to calculate the photon's straight-line
trajectory in the inertial observer frame (denoted throughout by the
subscript ``O"), transform it into the magnetic field rest frame (an
instantaneous non-inertial frame denoted by subscript ``S" for ``star
frame") where it is a curved path, and therein calculate the
instantaneous photon momentum and magnetic field at every point along
this path.  This then yields a straightforward determination of the
angle \teq{\sin\thetakB} from the magnitude of the cross product of the
magnetic field and trajectory vectors.  Since the characteristics of the
rate under Lorentz boosts along the magnetic field are captured in the
form in Eq.~(\ref{eq:pp_general}), with the explicit appearance of the Lorentz
invariant \teq{\omega_{\rm s}\sin\thetakB}, this completely leads to the
specification of the reaction rate in the star frame. To return to the
observer frame, note that Eq.~(23) of \citet{DH83} \citep[see
also][]{DL75b} provides a general transformation law for calculating the
attenuation rate in a non-inertial rotating frame, given the rate in the
inertial observer frame in which both magnetic and rotation-induced
electric fields are present.  One can invert this by interchanging the
roles of the two frames, noting that in the rotating star frame, there
is no electric field and thus the \teq{\mathbf{E} \times \mathbf{B}}
drift velocity is zero.  The attenuation rate in the observer frame 
\teq{R^{\rm pp}_{\hbox{\sixrm O}}} is then just the rate calculated 
in the rotating frame \teq{R^{\rm pp}_{\hbox{\sixrm S}}}, modified by a time
dilation factor of \teq{1/\gamma} for the boost between the two frames, 
i.e. \teq{R^{\rm pp}_{\hbox{\sixrm O}} = R^{\rm pp}_{\hbox{\sixrm S}}/\gamma}.
While the boost depends on the location of the photon along its path to
escape, this protocol is algorithmically simple: it avoids accounting for 
the complex time development of a sequence of Lorentz boosts 
between trajectory points in the non-inertial star frame.

The instantaneous attenuation rate in the rest frame of an inertial
observer employed in this Section is given by the Erber form, and
following Eqs.~(6) and (8) of \citet{DL75b} is
\begin{equation}
   R^{\rm pp}_{\hbox{\sixrm O}}\;\approx\; 
   \frac{1}{\gamma}\frac{3\sqrt{3}}{16\sqrt{2}} 
   \frac{\fsc}{\lambar} B \sin\thetakB \exp\left[-\frac{8}{3\omegaS B \sin\thetakB}\right] \quad .
 \label{eq:pp_rate_Oframe}
\end{equation}
Here \teq{\gamma = 1/\sqrt{1-\beta^2}} is the Lorentz factor corresponding 
to the local corotation velocity \teq{\beta = v/c} and \teq{\omegaS} is the 
photon energy in the star frame.  As before, \teq{\sin\thetakB} is the 
angle between the photon propagation direction and the local magnetic 
field direction in the rotating frame, given by
\begin{equation}
   \sin\thetakB \; =\; \Bigl\vert \mathbf{\hat{k}_S} \times \mathbf{\hat{B}_S} \Bigr\vert\quad .
 \label{eq:bperp}
\end{equation}
\teq{\mathbf{\hat{B}_S}} and \teq{\mathbf{\hat{k}_S}} are, respectively,
the magnetic field direction vector and the direction of photon travel
in the rotating frame.  It is generally straightforward to substitute
the threshold-corrected approximate rate from \citet{BK07} given in
Eq.~(\ref{eq:B88_asymp}), but here we will not do so except for
illustrative purposes in Figure \ref{fig:Eesc_smallthetae}.  The Erber
rate gives an accurate description of the character of the results when
the magnetic field is significantly subcritical, as it is for most very
short-period pulsars.  More importantly, its simpler mathematical form
is more amenable to the analytic approximations developed in this
Section. Corrections imposed by using the BK07 rate will be small in
most cases, and the direction of the changes will be exactly as expected
from Section \ref{sec:eesc_flatspace}: escape energies will increase by
a factor that is an increasing function of the surface polar magnetic
field.

\begin{figure}[h]
 \centerline{\includegraphics[width=.68\textwidth]{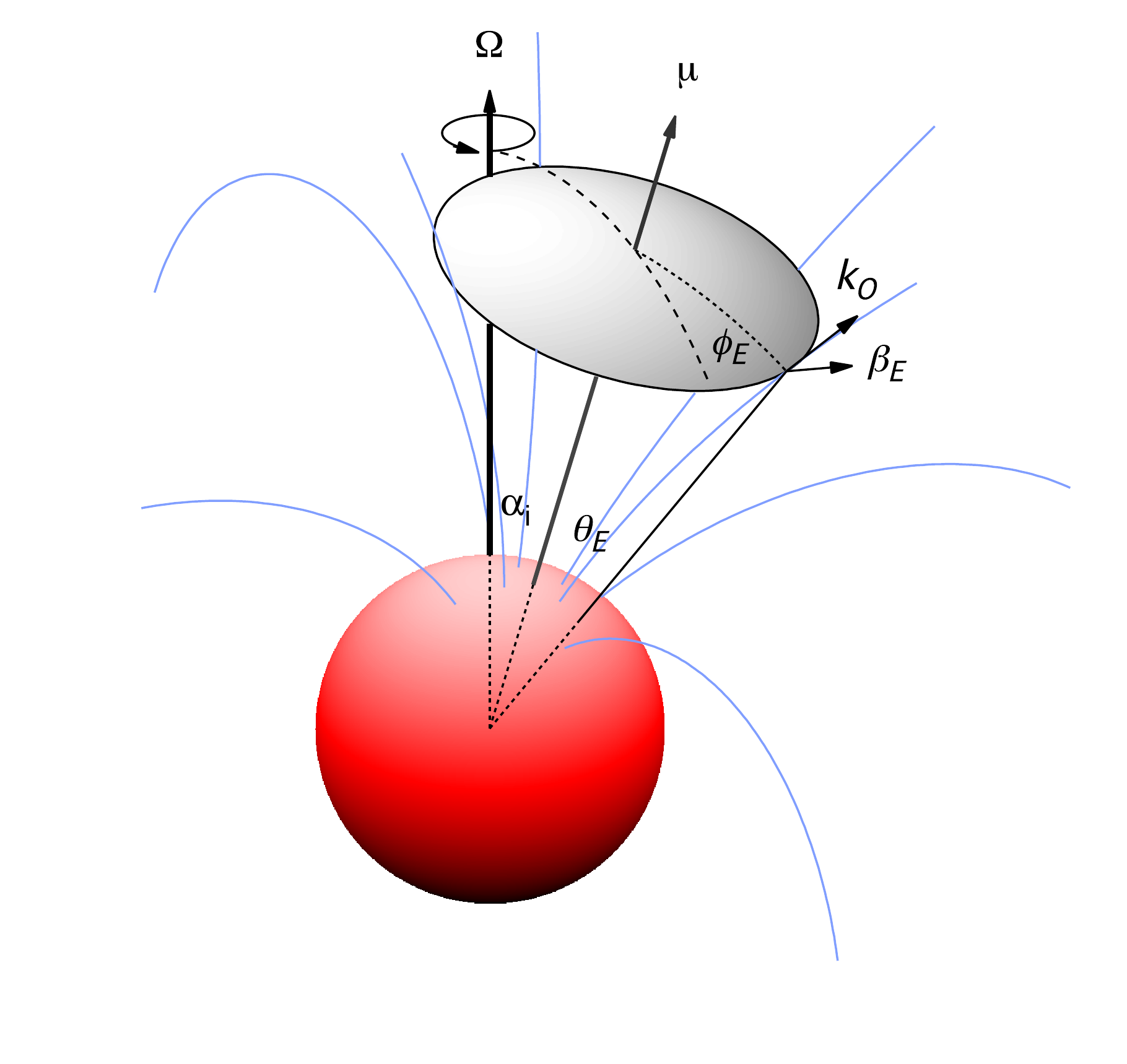}}
\caption{Geometry definitions for photon emission from the magnetosphere
of a rotating neutron star.  The angle \teq{\alpha_i} between the rotation
axis \teq{\mathbf{\Omega}} and the magnetic axis \teq{\boldsymbol{\mu}} is the
pulsar inclination angle.  The coordinates \teq{\thetaE} and \teq{\phiE}, the
colatitude and azimuthal angle of the point of emission, are defined
with respect to the magnetic axis.  Light blue curved lines are magnetic
field lines.  The vectors \teq{\boldsymbol{\betaE}} and \teq{\mathbf{k_O}} are
the scaled velocity and photon trajectory in the observer frame at the
point of emission; \teq{\boldsymbol{\betaE}} is directed into the page and
\teq{\mathbf{k_O}} is directed out of the page.  Though it is not easily
discernable in this diagram, \teq{\mathbf{k_O}} is not exactly tangent to
the magnetic field line that passes through the emission point, but
tilted slightly away from the velocity vector.}
 \label{fig:aberrationgeo}
\end{figure}

The geometry for the photon emission and opacity determination is 
illustrated in Fig.~\ref{fig:aberrationgeo}, where the magnetic dipole axis 
is inclined at an angle \teq{\alpha_i} relative to the spin axis.  In Cartesian 
coordinates with the \teq{z} axis aligned with the rotation axis and the 
\teq{xz} plane defined as that containing the magnetic and rotation axes 
(so that for an orthogonal rotator with \teq{\alpha_i = 90^\circ}, the 
magnetic axis is aligned with the \teq{+x} axis), the magnetic field 
(light blue lines in the Figure) at observer time \teq{t_{\hbox{\sixrm O}}=0} is given by
\begin{equation}
   \mathbf{B_S}(0) \; =\; 
   \begin{bmatrix}
      B_x(0)\\
      B_y(0)\\
      B_z(0)
   \end{bmatrix}
   \; =\; \frac{3 B_p \rns^3}{2 \rS^3}
   \begin{bmatrix} 
      \cos\alpha_i\cos\thetaS\sin\thetaS\cos\phiS+\sin\alpha_i\left(\cos^2\thetaS-\frac{1}{3}\right)\\
      \cos\thetaS\sin\thetaS\sin\phiS\\
      -\sin\alpha_i\cos\thetaS\sin\thetaS\cos\phiS+\cos\alpha_i\left(\cos^2\thetaS-\frac{1}{3}\right)
   \end{bmatrix}.
 \label{eq:Brot}
\end{equation}
By specifying a standard dipole \teq{\muB (3 \cos\thetaS\sin\thetaS ,\,
0,\, 3\cos^2\thetaS- 1)/r_s^3}, where \teq{\muB = B_p \rns^3/2} is the
dipole moment, this form is simply obtained by performing sequential
rotations by \teq{\phiS} about the \teq{z}-axis and then \teq{\alpha_i}
about the \teq{y}-axis, the latter of which can be described by the 
rotation matrix (subscript I denotes the magnetic inclination operator)
\begin{equation}
   T_{\rm I}\; =\; 
   \begin{bmatrix}
      \cos\alpha_i & 0 & \sin\alpha_i\\
      0 & 1 & 0\\
      - \sin\alpha_i & 0 & \cos\alpha_i
   \end{bmatrix} \quad .
 \label{eq:TI_matrixform}
\end{equation}
Throughout this presentation, the subscript S
denotes quantities in the star frame, and the angles in
Eq.~(\ref{eq:Brot}) relate to spherical polar coordinates in that frame.
To derive the time development of this field \teq{\mathbf{B_S}(t_{\hbox{\sixrm O}})}
as the star rotates about the \teq{z} axis while a photon escapes the magnetosphere, 
we multiply this by the appropriate rotation matrix \teq{\mathbf{T}_{\Omega}}:
\begin{equation}
   \mathbf{B_S}(t_{\hbox{\sixrm O}}) \; =\;  \mathbf{T}_{\Omega} \cdot \mathbf{B_S}(0)
   \quad ,\quad
   \mathbf{T}_{\Omega} (t_{\hbox{\sixrm O}}) \; =\;
   \begin{bmatrix}
      \cos\Omega t_{\hbox{\sixrm O}} & - \sin \Omega t_{\hbox{\sixrm O}} & 0\\
      \sin \Omega t_{\hbox{\sixrm O}} & \cos\Omega t_{\hbox{\sixrm O}} & 0\\
      0 & 0 & 1
   \end{bmatrix} \quad .
 \label{eq:Bfield_time_develop}
\end{equation}
Here \teq{t_{\hbox{\sixrm O}}=s/c} is the time an observer determines 
as the photon propagates a distance \teq{s} after emission.  In 
general, \teq{\Omega t_{\hbox{\sixrm O}} \ll 1} for the low altitude 
propagation zones where the greatest contributions to pair opacity are realized.
It is important to note that \teq{\mathbf{B_S}} represents the field vector in the 
{\it instantaneous star frame} at time \teq{t_{\hbox{\sixrm O}}}, and must be 
boosted to the observer frame to define the complete electromagnetic field therein.

For initial conditions, it is assumed that the photon is emitted very
nearly parallel to the magnetic field at the point of emission \emph{in
the star frame}, as was done for the non-rotating case. The spherical
polar coordinates at this point are \teq{\rS = \rE}, \teq{\thetaS =
\thetaE}, and \teq{\phiS = \phiE}, as depicted in Figure
\ref{fig:aberrationgeo}.  Normalizing this vector gives the direction of
the magnetic field, which depends only on the colatitude and azimuthal
angle, and can be obtained directly from Eq.~(\ref{eq:Brot}). In the
inertial observer (O) frame, the magnetic field direction vector and the
photon trajectory vector will no longer be parallel at the point of
emission, due to relativistic aberration precipitated by the rotation.
To specify this, the corotation velocity \teq{\overrightarrow{\betaE}c}
at the point of emission is needed. This depends only on the pulsar
period and the 3D location of the emission point, and is given by
\begin{equation}
   \overrightarrow{\betaE} \; =\; \frac{2\pi \rE}{cP}
   \begin{bmatrix}
      -\sin\thetaE\sin\phiE\\
      \cos\alpha_i\sin\thetaE\cos\phiE+\sin\alpha_i\cos\thetaE\\
      0
\end{bmatrix}\quad .
 \label{eq:betae}
\end{equation}
For \teq{\rE \leq Pc/2\pi} this dimensionless speed is always less than unity, 
except in the special case of emission at the equatorial light cylinder radius 
(\teq{\thetaE=\pi /2}, \teq{\rE = Pc/2\pi}) and for an aligned rotator (\teq{\alpha_i=0}).

We now progress to the construction of the photon's path in the reference frame 
of an inertial observer, which is a straight line in the absence of general relativistic corrections.
Using the photon's starting direction in the star frame, labelled by the unit vector
\teq{\mathbf{\hat{k}_{S,E}} \equiv \mathbf{\hat{B}_{S,E}}} 
derived from Eq.~(\ref{eq:Brot}), and the instantaneous relative velocity between 
the star frame and the observer frame of \teq{\betaEvec c}, we can calculate the 
photon's trajectory vector in the observer frame by performing a Lorentz 
transformation on the photon's 4-momentum in the star frame:
\begin{equation}
   \mathbf{k_O} \; =\; \omegaSE  
      \left\{\mathbf{\hat{B}_{S,E}}+\gammaE\left[\frac{\gammaE}{\gammaE+1}
      \left(\overrightarrow{\betaE}\cdot \mathbf{\hat{B}_{S,E}}\right)-1\right]\overrightarrow{\betaE}\right\} \quad .
 \label{eq:kOdef}
\end{equation}
Following the dimensionless convention adopted throughout, 
the wave vector \teq{\mathbf{k_O}} is scaled in terms of the 
inverse Compton wavelength \teq{m_ec/\hbar}, as is its star frame 
counterpart \teq{\mathbf{k_S}} below.  In the star frame, the 
photon energy at the point of emission, \teq{\omegaSE}, is 
Doppler-shifted relative to the constant photon energy 
\teq{{\ergO}} in the inertial observer frame:
\begin{equation}
   \omegaSE \; =\; \dover{\ergO}{\gammaE\,
        \bigl(1 - \mathbf{\hat{B}_{S,E}} \cdot \betaEvec\bigr) }\quad .
 \label{eq:omegaSE_def}
\end{equation}
This completes the specification of \teq{\mathbf{k_O}}, which is a constant during 
propagation to infinity.  When \teq{\betaE} is small, as it is
for emission relatively close to the neutron star, all quantities that are 
second order in \teq{\betaE} can be neglected, so that  Eq.~(\ref{eq:kOdef}) 
approximately assumes the form
\begin{equation}
   \mathbf{k_O} \;\approx\; \ergO
   \left\{\mathbf{\hat{B}_{S,E}} - \overrightarrow{\betaE}\right\} \quad .
 \label{eq:k_O_lowbeta}
\end{equation}
The small-\teq{\beta} approximation will be considered in more detail later 
in this Section and also in Appendix \ref{sec:abapprox}, for the purposes 
of generating useful analytic developments.

To compute the reaction rates in Eq.~(\ref{eq:pp_rate_Oframe}), one needs
the direction of photon travel in the star frame, where the path is curved, so that 
the value of \teq{\sin\thetakB} can be determined. This direction can be obtained 
at each point by transforming \teq{\mathbf{k_O}} back to the star frame, 
using the instantaneous \teq{\beta} and \teq{\gamma} at each point.  
This transformation preserves the component of momentum \teq{\mathbf{k_{\perp}}
= \mathbf{k_O} - \bigl( \mathbf{k_O} \cdot \hat{\beta} \bigr) \, \hat{\beta} } orthogonal to 
the boost, but 
stretches/contracts the component along \teq{\vec{\beta}} via the relation
\teq{\mathbf{k_S} \cdot \hat{\beta} = \gamma \bigl( \mathbf{k_O} \cdot \hat{\beta}
+ \beta \,\vert \mathbf{k_O} \vert \, \bigr)}, similar to the protocol adopted for Eq.~(\ref{eq:kOdef}).
The momentum vectors of the photon in the two frames at each position 
along the trajectory are then given by the conjugate relations
\begin{eqnarray}
   \mathbf{k_O} & = & \mathbf{k_{S}}+\gamma\left[\frac{\gamma}{\gamma+1}
       \left(\overrightarrow{\beta}\cdot \mathbf{k_{S}}\right)
         - \vert\mathbf{k_S}\vert \right]\overrightarrow{\beta} \quad ,\nonumber\\[-5.5pt]
 \label{eq:kO_kS_Lorentz_pair}\\[-5.5pt]
   \mathbf{k_S} & = & \mathbf{k_{O}}+\gamma\left[\frac{\gamma}{\gamma+1}
       \left(\overrightarrow{\beta}\cdot \mathbf{k_{O}}\right)
         + \vert\mathbf{k_O}\vert \right]\overrightarrow{\beta} \quad .\nonumber
\end{eqnarray}
The first equation in this couplet is just an extension of Eq.~(\ref{eq:kOdef})
to arbitrary altitudes using \teq{\overrightarrow{\betaE} \to \overrightarrow{\beta}}
and \teq{\omegaSE \mathbf{\hat{B}_{S,E}} \to \mathbf{k_{S}}}.
To maintain constancy of \teq{\mathbf{k_O}} along the entire trajectory,
\teq{\vert \mathbf{k_S}\vert = \omegaS} adjusts its value at each point via 
the aberration relation \teq{\gamma \omegaS \bigl(1 - \mathbf{\hat{k}_{S}} 
\cdot \betavec\bigr) = \ergO} that is the analog of 
Eq.~(\ref{eq:omegaSE_def}).  The second relation of this pair is just the 
inversion of the first, under the interchange S\teq{\leftrightarrow}O and 
also \teq{\betavec\to -\betavec}; it is that desired for the computation 
of \teq{\sin\thetakB}.  Observe that forming the dot product
\teq{\omegaS^2 = \vert \mathbf{k_{S}}\vert^2} can be used to derive the 
inverse aberration relation \teq{\omegaS = \gamma \ergO ( 1 + \mathbf{\hat{k}_{O}} \cdot \betavec )}.
Note that in the absence of rotation (\teq{P \rightarrow \infty}, \teq{\beta \rightarrow 0}), 
one recovers \teq{\mathbf{k_S} = \mathbf{k_O} = \ergO \mathbf{\hat{B}_{S,E}}}, 
as expected.

Calculating the corotation velocity \teq{\betavec}
at every point along the photon trajectory is straightforward.  
The straight-line photon path in the observer frame is given simply by
\begin{equation}
   \mathbf{r_O} = \mathbf{\rE} + s\mathbf{\hat{k}_O},
 \label{eq:ro}
\end{equation}
where \teq{s=ct} is the distance traveled, and \teq{\mathbf{\rE}} is the
emission point.  This is conveniently expressed in spherical coordinates 
about the rotation axis, still as a function of \teq{s}:
\begin{equation}
   \begin{bmatrix}
       \rO (s)\\
       \thetaO (s)\\
       \phiO (s)
   \end{bmatrix}
   \; =\;
   \begin{bmatrix}
         \sqrt{\rE^2+s^2+2s \mathbf{\rE} \cdot \mathbf{\hat{k}_O}}\; \\
          \cos^{-1}\dover{\mathbf{r_{E,z}}+s\mathbf{\hat{k}_{O,z}}}{ \rO (s)} \vphantom{\Biggl(} \\
          \tan^{-1} \dover{\mathbf{r_{E,y}}+s \mathbf{\hat{k}_{O,y}}}{ \mathbf{r_{E,x}}+s\mathbf{\hat{k}_{O,x}} }
   \end{bmatrix}
   \quad .
 \label{eq:traj_polar_coord}
\end{equation}
In this form, the coupling between the rotational phase \teq{\phiO} 
and the trajectory is simply displayed.  In general, we do not restrict the emission plane 
to zero phase \teq{\phiO}, i.e. \teq{\mathbf{k_O}}, \teq{\mathbf{\hat{B}_{S,E}}}
and \teq{\overrightarrow{\betaE}} are not coplanar.  The local corotation velocity 
is then given as a function of \teq{s} in Cartesian coordinates by
\begin{equation}
   \overrightarrow{\beta}(s) \; =\;  \frac{\rO (s)\Omega}{c}\, \sin\thetaO
   \begin{bmatrix}
        -\sin\phiO (s)\\
        \cos\phiO (s)\\
        0
     \end{bmatrix}
   \quad .
 \label{eq:vel_corotate}
\end{equation}
Since \teq{\mathbf{k_O}} in Eq.~(\ref{eq:kOdef}) is also expressible in Cartesian coordinates, 
the determination of \teq{\mathbf{k_S}} via Eq.~(\ref{eq:kO_kS_Lorentz_pair}) is routine.

The remaining ingredient needed for the determination of the pair conversion rates is 
the second vector for the calculation of \teq{\sin\thetakB} in Eq.~(\ref{eq:bperp}),
namely the magnetic field vector in the star frame, \teq{\mathbf{B_S}(s)}.  Given 
the simple dipolar form in the star frame, the only complexity is encapsulated 
in the conversion of the straight-line trajectory in the inertial observer frame
into star frame coordinates, where the path is curved.  Given coordinates
\teq{\{x_{\hbox{\sixrm O}}, y_{\hbox{\sixrm O}}, z_{\hbox{\sixrm O}}\} } in the 
observer frame, obtained from Eq.~(\ref{eq:ro}), the photon trajectory 
in the star frame can be expressed via
\begin{equation}
   \begin{bmatrix}
       c t_{\hbox{\sixrm S}}\\
       x_{\hbox{\sixrm S}}\\
       y_{\hbox{\sixrm S}}\\
       z_{\hbox{\sixrm S}}
   \end{bmatrix}
   \; =\;
   \mathcal{L} \cdot 
   \begin{bmatrix}
       c t_{\hbox{\sixrm O}}\\
       x_{\hbox{\sixrm O}}\\
       y_{\hbox{\sixrm O}}\\
       z_{\hbox{\sixrm O}}
   \end{bmatrix} 
   \quad \hbox{for}\quad
    \mathcal{L} \; =\; 
   \begin{bmatrix}
      \gamma & \gamma \beta_x & \gamma \beta_y & 0\\
      \gamma \beta_x & 1+\left(\gamma-1\right) \dover{\beta_x^2}{\beta^2} 
          & \left(\gamma-1\right) \dover{\beta_x\beta_y}{\beta^2} & 0\\
      \gamma \beta_y & \left(\gamma-1\right) \dover{\beta_x\beta_y}{\beta^2} 
          & 1+\left(\gamma-1\right) \dover{\beta_y^2}{\beta^2} & 0\\
      0 & 0 & 0 & 1
   \end{bmatrix}
 \label{eq:boost_spatial}
\end{equation}
being a standard Lorentz transformation matrix, \teq{\mathcal{L} =\mathcal{L}(s)}.   
Since the boost is purely in the \teq{xy} direction and \teq{s = c t_{\hbox{\sixrm O}}}, 
this transformation simplifies to
\begin{equation}
   \begin{bmatrix}
       x_{\hbox{\sixrm S}}\\
       y_{\hbox{\sixrm S}}\\
       z_{\hbox{\sixrm S}}
   \end{bmatrix}
   \; =\;
   \begin{bmatrix}
       \gamma \beta_x s + \biggl(1 + \dover{\gamma^2\beta_x^2}{\gamma +1}\biggr) x_{\hbox{\sixrm O}} 
             + \dover{\gamma^2 \beta_x \beta_y}{\gamma +1} \, y_{\hbox{\sixrm O}} \vphantom{\Biggr)} \\
       \gamma \beta_y s + \dover{\gamma^2 \beta_x \beta_y}{\gamma +1} \, x_{\hbox{\sixrm O}}
             + \biggl(1 + \dover{\gamma^2\beta_x^2}{\gamma +1}\biggr) y_{\hbox{\sixrm O}} \\
       z_{\hbox{\sixrm O}}
\end{bmatrix}
\end{equation}
These Cartesian coordinates are oriented just as in the coordinate configuration 
at the time \teq{t=0} of emission, with the \teq{z}-axis parallel to \teq{\mathbf{\Omega}}.
The star frame magnetic is most easily specified by defining the polar coordinates 
in a specification that possesses the current rotational orientation, and is then 
``de-inclined'' with respect to the rotation axis.
Since the field configuration has evolved slightly due to the rotation of the star, 
we sequentially perform inverse rotation and inverse inclination operations
\begin{equation}
   \rS
   \begin{bmatrix}
       \sin\thetaS\cos\phiS\\
       \sin\thetaS\sin\phiS\\
       \cos\thetaS
   \end{bmatrix}
   \; =\; 
   \begin{bmatrix}
       \overline{x}_{\hbox{\sixrm S}}\\
       \overline{y}_{\hbox{\sixrm S}}\\
       \overline{z}_{\hbox{\sixrm S}}
   \end{bmatrix}
   \; =\; T_{\rm I}^{-1} \cdot T_{\Omega}^{-1}
  \begin{bmatrix}
       x_{\hbox{\sixrm S}}\\
       y_{\hbox{\sixrm S}}\\
       z_{\hbox{\sixrm S}}
   \end{bmatrix}
 \label{eq:cartesian_Sframe_derotated}
\end{equation}
to express the star frame Cartesian coordinates in a form that can be directly 
interpreted for the pertinent spherical polar coordinates --- these are now oriented
such that \teq{\thetaS=0} corresponds to the magnetic axis,
as opposed to the rotation axis.
These (\teq{\rS ,\, \thetaS ,\, \phiS}) coordinates are used to calculate the magnetic field vector 
\teq{\mathbf{B_S}} using Eq.~(\ref{eq:Brot}), which incorporates the inclination 
element.  However, in addition, the time evolution operator \teq{T_{\Omega}} must also then 
be applied, to render the field vector in a form representative of time \teq{t_{\hbox{\sixrm O}}},
i.e. corresponding to Eq.~(\ref{eq:Bfield_time_develop}).  This protocol then 
yields \teq{\mathbf{B_S}} in evolved star frame Cartesian coordinates, which together with 
Eq.~(\ref{eq:kO_kS_Lorentz_pair}) can be employed to
routinely evaluate the cross product \teq{|\mathbf{k_S} \times \mathbf{B_S}|}, 
and therefore \teq{\sin\thetakB}.   Coupled with the aberration relation 
\teq{\omegaS = \gamma \ergO ( 1 + \mathbf{\hat{k}_{O}} \cdot \betavec )},
this completes the ensemble of relations need to compute the 
final rate in Eq.~(\ref{eq:pp_rate_Oframe}).

To obtain the optical depth, we integrate the rate over the path length \teq{s}.  
Just as with the non-rotating flat spacetime calculation, it is expedient 
to change variables from \teq{s} to a path angle \teq{\eta}, defined by
\begin{equation}
   s\; =\; \sqrt{\rE^2+r^2-2r\rE\cos\eta}\quad .
 \label{eq:s_eta_reln}
\end{equation}
This is analogous to that displayed in Fig.~\ref{fig:prop_geometry}, but notably 
with the plane defined by the radius vector and photon momentum at any time being 
noncoindent with the instantaneous plane containing the magnetic field axis and 
radial vector at the time of emission.  Therefore, Eq.~(\ref{eq:theta_fs}) does not apply 
except when \teq{\betaE =0} and the star does not rotate, and the general relationship
between \teq{\eta} and \teq{\thetaE} and \teq{\theta} is quite complicated, and contains 
all the information associated with magnetic field orientation at each rotational phase
along the photon's path to infinity.  One advantage of working 
with this integration variable is that it is always small when \teq{\thetaE} is small, 
which facilitates an asymptotic analysis (detailed below) that serves to check the computations.
Further simplification arises by defining the angle \teq{\deltaE} between the 
emission and radial directions, just as for the non-rotating flat spacetime 
analysis, i.e., 
\begin{equation}
   \deltaE \; =\; \cos^{-1}\biggl( \dover{\mathbf{\rE} \cdot \mathbf{\hat{k}_O}}{|\rE|}\biggr) \quad .
\label{eq:deltaE}
\end{equation}
Then consideration of the appropriate triangles (visualized using 
Fig.~\ref{fig:prop_geometry}) quickly establishes that 
Eq.~(\ref{eq:prop_dist}) applies, so that the optical depth integral becomes
\begin{equation}
   \tau_A \; =\; \rE\sin\deltaE\int_0^{\deltaE} 
      \dover{R_A\, d\eta}{\sin^2 (\deltaE-\eta )} \quad .
 \label{eq:tau_A_final}
\end{equation}
The upper limit of integration is the path angle \teq{\eta} when the 
photon's trajectory crosses the light cylinder; since this 
is essentially at infinity, we have \teq{\eta_{\hbox{\sevenrm max}} 
\approx \deltaE}, as in Section~\ref{sec:flatspace}.

The effects of rotational
aberration on the escape energies of photons emitted from the neutron star surface
are explored in Fig.~\ref{fig:Eesc_smallthetae}, specifically as 
functions of small colatitudes (below around \teq{5^{\circ}}) near the magnetic pole.
The $x$-axis therefore constitutes different locales on the polar cap,
with only a select value corresponding to the footpoint of the last open field line
i.e., \teq{\thetaE=0.08} (\teq{\approx 4.6^{\circ}}) for the Crab pulsar.
The chosen magnetic field is the Crab value, and most of the results (curves) 
are generated using the Erber rate.  Escape energy curves are 
generated for three different pulse periods and two different rotator 
inclinations \teq{\alpha_i}.  Results are also displayed for 
different azimuthal angles \teq{\phiE}, with \teq{\phiE =0} corresponding to 
emission points lying in the plane defined by \teq{\mathbf{k_O}} and \teq{\mathbf{\hat{B}_{S,E}}}.
As \teq{\phiE} is changed to accommodate different rotational phases, 
different local rotational speeds \teq{\betaE} are sampled, 
with  \teq{\phiE >0} defining cases where the non-radial component of 
photon momentum is in the direction of rotation, and \teq{\phiE < 0} 
constituting counter-rotation emission cases.  Note that this 
sign convention is opposite the one used by \citet{Lee10}.
The Figure exhibits the obvious trend that as the period is 
decreased from large values (\teq{P=100}sec, essentially non-rotating), 
the aberration modifications to \teq{\eesc} become larger at small colatitudes, 
and their influence persists to larger colatitude domains.  Mostly, but not 
always, rotation reduces \teq{\eesc} since photons can promptly and
more readily propagate across field lines in the star frame; this 
is best exemplified by the \teq{P=10}ms case.
Note the similarity of the \teq{\phiE = 0} aberration-corrected curves 
in the Figure to the non-rotating \teq{\theta_{kB,0} = 0.01} curves for 
magnetic pair attenuation in Figure~1 of \cite{BH01}. 
 
\begin{figure}[h]
 \centerline{\includegraphics[width=.7\textwidth]{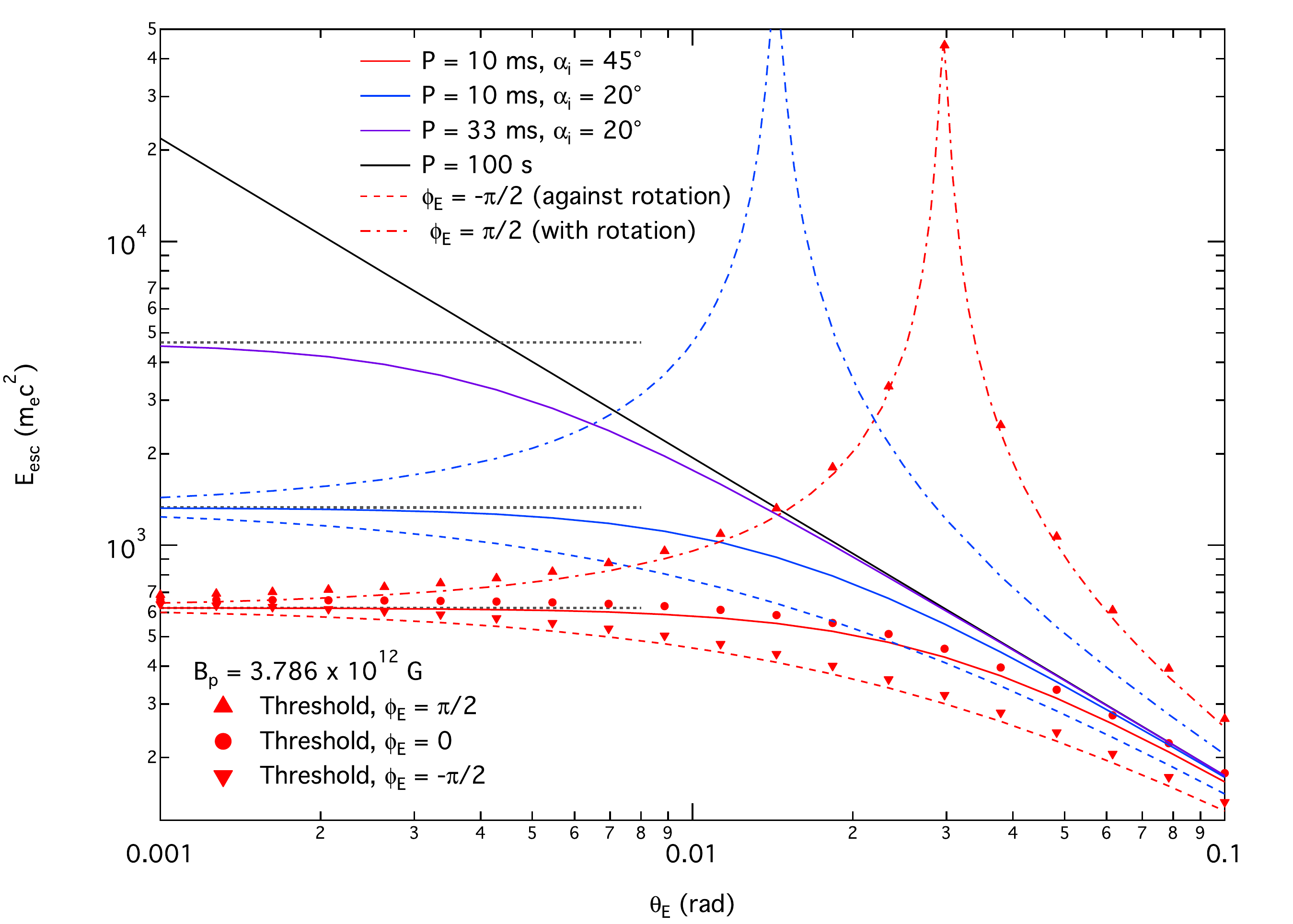}}
\caption{Escape energies from the neutron star surface for colatitudes
near the magnetic pole with three different sets of pulsar parameters,
period \teq{P}, magnetic inclination angle \teq{\alpha_i}, and orbital
phase \teq{\phiE}.  The magnetic field strength is half the surface polar
value for the Crab, and the Erber rate in Eq.~(\ref{eq:Erber_asymp}) is
used for the curve computations.  The black \teq{P=100}sec example
defines a non-aberration case as discussed in
Section~\ref{sec:flatspace}. Solid curves have an azimuthal angle of
\teq{\phiE = 0}, where the emission point lies in the plane defined by
the magnetic axis and the rotation axis. Dot-dash curves represent
emission from an azimuthal angle of \teq{\phiE = \pi /2}, where the
photon is emitted with the direction of neutron star rotation.  Dashed
curves correspond to azimuthal angles of \teq{\phiE = - \pi /2},
against the direction of neutron star rotation. This subdivision 
into three different azimuthal angle values applies only to \teq{P=10}ms
curves.  For the three \teq{P=10}ms, \teq{\alpha_i=45^{\circ}} cases,
the circles and triangles represent escape energies calculated with the
threshold-corrected attenuation rate approximation given in \citet{BK07}.
The gray dotted lines represent the small colatitude analytic
approximation in Eq.~(\ref{eq:tau_zero}) for the 10 ms and 33 ms cases. 
}
 \label{fig:Eesc_smallthetae}
\end{figure}

At very small colatitudes, the escape energies for the same period and 
magnetic inclination all approach the same finite value regardless of the initial
azimuthal angle about the polar cap.  This saturation is a core
characteristic of aberration corrections. When the colatitudes are somewhat
larger but still fairly small, however, the asymmetry between the
leading edge and the trailing edge of the polar cap becomes evident. 
The trailing edge sees a slight decrease in escape energies caused by
the general increase in \teq{\sin\thetakB} as the photons' transit across
field lines is aided by the field's rotation.  The leading edge sees an
increase of well over an order of magnitude in escape energy, a caustic-like
effect that is manifested when the field line sweepback creates a narrow
range in colatitudes where the photon's trajectory is nearly parallel to
the magnetic field lines over a substantial section of the photon path.  
This azimuthal bifurcation of the escape energies is maximized when 
\teq{\thetaE \sim \rns\sin\alpha_i/\rlc},  and declines rapidly at 
larger emission colatitudes.  The large value of \teq{\eesc} implies that this
constrained portion of the surface polar locale is actually visible in the
GeV band that is detected in most {\it Fermi}-LAT pulsars.  These 
``leading edge'' peaks arise at \teq{\thetaE\approx 2 \beta_p}, where 
\teq{\beta_p = \rns \sin \alpha_i/\rlc} is the corotational dimensionless 
speed at the surface magnetic pole, and therefore are proximate to
the polar cap of size \teq{\theta_c\approx \sqrt{\rns/\rlc}}.  
One can therefore infer that a small portion of the 
polar cap corresponding to a particular phase (i.e., \teq{\phiE}) of pulsation 
may actually be transparent to pair attenuation up to a few GeV
in photon energy.

Also depicted in Fig.~\ref{fig:Eesc_smallthetae} are circles, upward-pointing
triangles, and downward-pointing triangles, which represent the escape energies
calculated with the threshold-corrected rate from Eq.~(\ref{eq:B88_asymp}) 
for \teq{\phiE = 0, +\pi/2, -\pi/2} respectively.  For
this subcritical surface magnetic field, the threshold effects raise the
escape energies slightly across the board, but clearly do not alter the general
character of the results.

The saturation limiting case of \teq{\thetaE \to 0} is particularly amenable 
to analytic approximation and should serve as a good indicator of how 
strongly rotational aberration affects maximum energies and minimum 
altitudes of emission (addressed below) for magnetic pair creation 
transparency.  For a non-rotating neutron star, a photon emitted from 
the magnetic pole could never attain a non-zero \teq{\sin\thetakB}, 
and so the attenuation rate approaches zero as \teq{\thetaE \rightarrow 0}.  
For a rotating neutron star, the angle between the photon trajectory and the 
magnetic field at the point of emission is nonzero {\it in the inertial observer frame}, 
even for photons emitted from the magnetic poles.  Thus, the attenuation rate is 
nonzero and minimum emission altitudes and escape energies saturate 
at meaningful values.  Such an asymptotic result can be determined 
by following a procedure much like the one discussed at some length in 
Section~\ref{sec:tauintegration}.  One obtains the following approximation 
for the optical depth at \teq{\thetaE = 0} (with \teq{B_p} scaled by \teq{B_{\rm cr}}):
\begin{equation}
   \tau_A\; \approx\; \dover{3^6\sqrt{3}}{2^{18}} \dover{\fsc \rns}{\lambar \, h^2} 
   \sqrt{\pi \erg}\,  \left(\dover{B_p \rns \sin\alpha_i}{\rlc } \right)^{3/2}
   \; \exp\left\{-\dover{2^{12} h^2 \rlc }{ 3^5
       \erg B_p\, \rns \sin\alpha_i\vphantom{\bigl(} }\right\} \quad ,
 \label{eq:tau_zero}
\end{equation}
where \teq{h=\rE /\rns}, and \teq{\rlc = Pc/2\pi} expresses the pulsar period. 
The details of this development can be found in Appendix~\ref{sec:abapprox}.  
Setting this equal to unity, we can solve for \teq{\erg_{esc}}:
\begin{equation}
   \eesc \; =\; \dover{2^{12}\, h^2}{3^5\, B_p \beta_p} 
        \left[\log_e \dover{\fsc \rns }{\lambar} 
        + \dover{1}{2}\log_e \dover{\eesc}{h^4} + \dover{3}{2} \log_e B_p \beta_p -5.34\right]^{-1} \quad .
 \label{eq:eesc_polar_aberr}
\end{equation}
In this expression, \teq{\beta_p = \rns\sin\alpha_i/\rlc} is again the dimensionless rotation 
speed at the surface polar location in its circular path, and is the key parameter
defining the scale of rotational influences.  Comparison with the corresponding 
non-rotating result in Eq.~(\ref{eq:e_esc_Erber}) indicates that here \teq{\beta_p}
plays a role equivalent to the footpoint colatitude \teq{\theta_f} there, because 
both \teq{\beta_p} and \teq{\theta_f} are proportional to the angle between the 
photon momentum vector and the radial direction at the neutron star surface.
Accordingly, aberration modifications should diminish for these low altitude 
emission considerations when \teq{\thetaE} substantially exceeds \teq{\beta_p}, 
and all cases should coalesce to the static (\teq{P\to\infty}), flat spacetime curve:
this circumstance is clearly evident in Fig.~\ref{fig:Eesc_smallthetae}.
Observe that the choice of azimuthal angle \teq{\phiE} provides a second-order 
correction \teq{O(\thetaE^2)} when at extremely small colatitudes above the magnetic pole, and
so does not appear in these leading-order asymptotic formulae.
 
The transcendental Eq.~(\ref{eq:eesc_polar_aberr}) must be solved numerically, 
in general, though the weak dependence on \teq{\eesc} in the logarithmic term, 
when neglected, permits approximate analytic solution for the escape energy.  
Numerical results are diplayed as the gray dashed line saturation asymptotes in 
Fig.~\ref{fig:Eesc_smallthetae},  for the case of surface emission (\teq{h=1})
and periods 10 ms and 33 ms.  One can immediately see that 
to leading order, as \teq{\thetaE \to 0}, since \teq{\eesc\propto \beta_p^{-1}}, 
one should expect to see \teq{\eesc \propto P} and \teq{\eesc \propto 1/\sin\alpha_i}:
these dependences are clearly exhibited in the Figure.  Moreover, the 
computed numerical \teq{\eesc} curves asymptotically approach the 
solutions of  Eq.~(\ref{eq:eesc_polar_aberr}) as \teq{\thetaE \to 0}.

\begin{figure}[h]
 \centerline{\includegraphics[width=.77\textwidth]{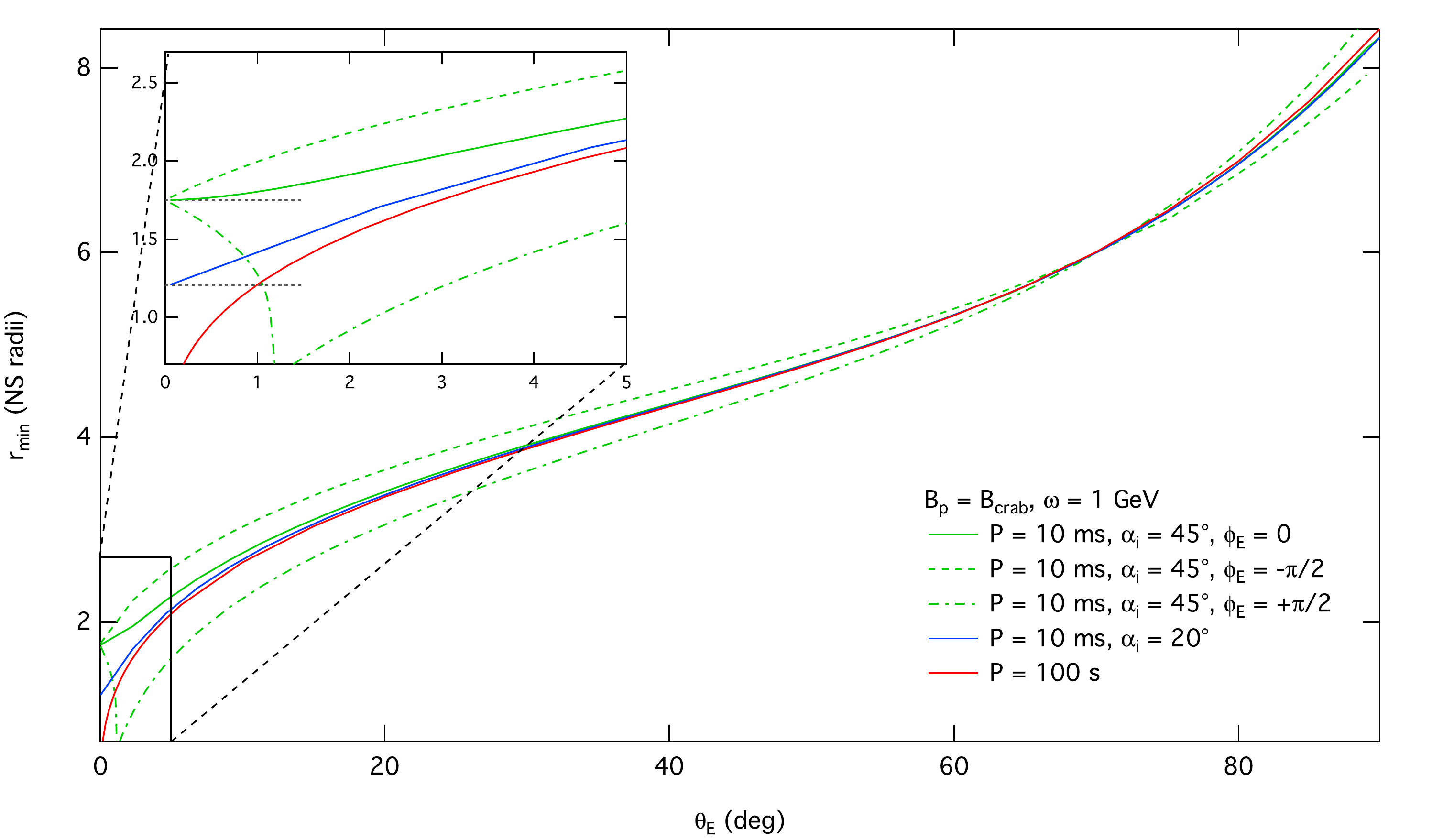}}
\caption{Minimum altitudes of emission from the magnetic pole to the 
magnetic equator, for different sets of pulsar parameters.  The 
Crab field value was adopted.  As in Fig.~\ref{fig:Eesc_smallthetae}, 
the dot-dash curves represent photons emitted on the leading side 
of the magnetic pole (evincing non-monotonic behavior of \teq{r_{\rm min}}
with \teq{\thetaE} -- see inset), and dashed curves represent photons emitted 
on the trailing side of the pole.  The gray dotted lines in the inset represent the 
zero-colatitude analytic approximation in Eq.~(\ref{eq:tau_zero}).  }
 \label{fig:rmin_fullrange}
\end{figure}

Fig. \ref{fig:rmin_fullrange} shows the effect of rotational aberration
on the minimum altitudes of emission of 1 GeV photons; similar behavior
is exhibited for other test photon energies.  The surface polar magnetic
field is again chosen to be the Crab value.  Such altitude bounds are
plotted as functions of emission colatitude, and except for very close
to the pole, are generally well below the altitude of the last open
field line. The red curve is a pulsar with a period of 100 sec, where
aberration effects are minimal.  The blue and green curves represent
photons emitted from an azimuthal angle of zero for pulsars with periods
of 10 seconds and inclination angles of 20 and 45 degrees, respectively.
 Green dotted and dashed lines represent photons emitted from the
leading side of the magnetic pole (with the direction of rotation) and
the trailing side (against the direction of rotation).  Gray dashed
lines again represent the \teq{\thetaE = 0} limit calculated from
Eq.~(\ref{eq:tau_zero}), so that \teq{r_{\rm min}/\rns} would represent
the \teq{h} value that precisely generates an escape energy of 1 GeV in
Eq.~(\ref{eq:eesc_polar_aberr}). It should be noted that like
\citet{Lee10}, these calculations do not take pair threshold effects
into account; by inspection of Fig.~\ref{fig:Eesc_smallthetae}, it is
apparent that corrections produced by accurately treating the threshold
are small overall for the Crab field.  In addition, we have not
considered the full Kerr metric calculation, which would correctly
account for aberration and general relativity together.  Given the small
nature of aberration influences overall on \teq{\gamma\to e^+e^-} pair
opacity considerations at low to moderate altitudes, addressing this
complication is not strongly motivated.

Very near the magnetic pole, i.e. \teq{\thetaE\lesssim 0.3^{\circ}},
aberration modifications increase the minimum emission altitude
substantially for all azimuthal angles. The finite value of \teq{\beta_p} 
causes a photon emitted parallel to the magnetic field in the star frame to
reach the minimum angle for pair production much earlier than for the
non-rotating case, thus sharply decreasing the photon mean free path at
low altitudes. Consequently, the minimum altitude is generally higher
than for the \teq{\beta_p=0} situation. Moving away from the pole, the
asymmetry between the leading and trailing sides of the magnetic pole
becomes apparent.  The trailing side sees a small increase in minimum
altitudes, as field line crossing becomes easier.  The leading side sees
a sharp valley in minimum altitude (see inset), paralleling the peak in escape
energies in Fig.~\ref{fig:Eesc_smallthetae}. Note that \teq{r_{\rm min}}
values below the surface are indeed possible, and indicate transparency
to \teq{\gamma\to e^+e^-} for emission near the pole. For colatitudes 
larger than around \teq{5^{\circ}}, the small absolute changes to \teq{\sin\thetakB} 
are swamped by the rapidly decreasing magnetic field.  Finally, near the
magnetic equator when altitudes are highest and thus aberration effects
are at their strongest (\teq{\betaE\gg \beta_p}), 
the asymmetry between the leading and trailing
sides reappears, while \teq{\phiE=0} curves are barely altered from the
\teq{\beta_p\approx 0} (i.e. 100 ms) case.  In summation, aberration
corrections alter the minimum altitudes relative to the non-rotating 
calculations most significantly at small colatitudes.  

Figure 3 of \citet{Lee10} presents calculations of the minimum altitude
of emission as a function of emission colatitude for different photon
energies and pulsar inclination angles.  Our results differ from theirs
in a number of important respects. For the smallest pulsar inclination
angles \teq{\alpha_i}, the minimum altitudes of emission should approach
the limit for a non-rotating neutron star, regardless of the other
properties of the pulsar; our computations clearly display this
characteristic.  Fig.~\ref{fig:rmin_fullrange} indicates that the
minimum altitude of emission is a monotonically increasing function of
emission colatitude all the way up to equatorial emission at
\teq{\thetaE = 90^\circ}.  It is not clear why the minimum altitudes
obtained by \citet{Lee10} decrease sharply at large colatitudes and
especially why this effect is not noticeably weaker for \teq{\alpha_i =
15^\circ} than for \teq{\alpha_i = 86^\circ}.  In contrast, our
computations realize intuitively sensible behavior: at equatorial
colatitudes, photons more readily cross field lines after emission, and 
so one expects the minimum altitude for pair transparency to increase, 
sampling lower fields at and above the emission locale.

\section{SOURCE CONNECTIONS: GAMMA-RAY PULSARS AND MAGNETARS}
 \label{sec:observations}

In this Section, we explore the implications of the pair transparency
considerations for both gamma-ray pulsars and magnetars. The {\it Fermi}
pulsar catalogues \citep{Catalog,PSRCat2} provide us with a wealth of
data for constraining the source regions for high energy emission from
\teq{\sim 140} pulsars.  The phase lag of the gamma-ray emission with
respect to the radio emission can be used to estimate the altitude of
the gamma-ray emitting region \citesmartpp{Seyffert12}.  Using
relativistic descriptions of polarization position angle evolution, it
can be determined \citesmartpp{BCW91} that the radio emission originates
at typically \teq{10-30\rns} above the magnetic pole.  Since the
gamma-ray emission in outer magnetospheric models is usually
significantly offset from the magnetic polar axis, the phase lag
$\delta$ is a key parameter for narrowing the solution space in combined
radio and gamma-ray light curve modeling.  The gamma-ray peak
separation, available for pulsars with 2 or more gamma-ray peaks, can
also be compared to light curve modeling results to estimate the
altitude of the emitting region \citesmartptwo{Watters}{VJH12}. These
generally conclude that the active zone is at altitudes \teq{0.03 \rlc -
\rlc}, with determinations sensitively depending on the obliquity of the
rotator and the observer's viewing perspective.

Such geometric limits on the emission altitudes generally place the
source of \teq{\sim}GeV gamma rays in the outer magnetosphere 
\citep{PGHG10,WR11,Pierbattista14}.  The complementary emission radii 
lower bounds calculated in this paper from magnetic pair creation are 
mostly at much smaller altitudes than this, yet provide interesting constraints 
in some cases. In particular, in the absence 
of aberration they do not depend on the magnetic inclination of the 
pulsar and the observer viewing angle.  Moreover, the pair transparency 
bounds can be used even if the pulsar is radio quiet (33 LAT pulsars 
as of October 2013) or has only a single gamma-ray peak 
(31 LAT pulsars; \citesmartp{PSRCat2}).

In acceleration gap models of gamma-ray emission from pulsars, the gaps
are generally assumed to be located in a narrow band along the last open
magnetic field line \citep[e.g.,][]{Romani96,MH04,Hirotani07}. This
follows the precedent established in earlier polar cap models
\citesmartpp{DH82} of gamma-ray pulsars.  If we assume that emission
takes place on and parallel to the last open field line, we can then
calculate an approximate minimum radius of emission using the procedure
in Section \ref{sec:eesc_flatspace} with the general relativistic
corrections of Section \ref{sec:GR}.  We therefore use pair transparency
to provide a physical lower bound to \teq{r_{\rm min}}. These minimum
radii depend on assumptions about the neutron star mass and radius to
only a small extent.  For example, doubling the assumed mass of the
neutron star increases the minimum radii by approximately 5\% at most
near the surface, with the change being due to general relativistic
influences. In addition, they are independent of the inclination of the
neutron star when aberration is unimportant (mostly the case). Using the
Erber approximation result in Eq.~(\ref{eq:e_esc_Erber}), it is simply
determined that for fixed \teq{\eesc}, one should expect a correlation
of \teq{r_{\rm min}^{5/2}\propto \eesc B_p\theta_f\propto B_p/\sqrt{P}},
using the approximate polar cap footpoint colatitude dependence
\teq{\theta_f\equiv\theta_p\approx \sqrt{2\pi \rns /(Pc)} \propto
P^{-1/2}}. If one neglects the second logarithmic
(\teq{B_p}-dependent) term inside the parentheses in
Eq.~(\ref{eq:e_esc_Erber}), this flat space-time correlation is approximately
\begin{equation}
   \dover{r_{\rm min}}{\rns} \; \approx\; 0.037 \eesc^{2/5}\, 10^{2\sigma_{\rm IG}/5}
   \quad ,\quad
   \sigma_{\rm IG} \; =\; \log_{10}\left( \dover{B_{12}}{\sqrt{P}} \right)\quad ,
 \label{eq:rmin_sigmaIG_dep}
\end{equation}
where the unscaled polar field is \teq{B_p=B_{12}10^{12}}Gauss, and we set \teq{\rns=10^6}cm.
For the purposes of this discussion, here we adopt the relation 
\teq{B_p \approx 6.4\times 10^{19}\sqrt{P\, {\dot P}} } for the unscaled field, which 
corresponds to the vacuum magnetic dipole moment \teq{\mu = B_p\rns^3/2}
\citep{ST83,UM95}, and is twice the conventional choice of \cite{MT77} 
that is applicable to the equatorial surface field.  Note that non-dipolar
contributions and plasma loading of the magnetosphere modify such 
inferences of pulsar surface fields from \teq{P} and \teq{\dot{P}}. 

\begin{figure}[h]
\centerline{
	\includegraphics[width=.68\textwidth]{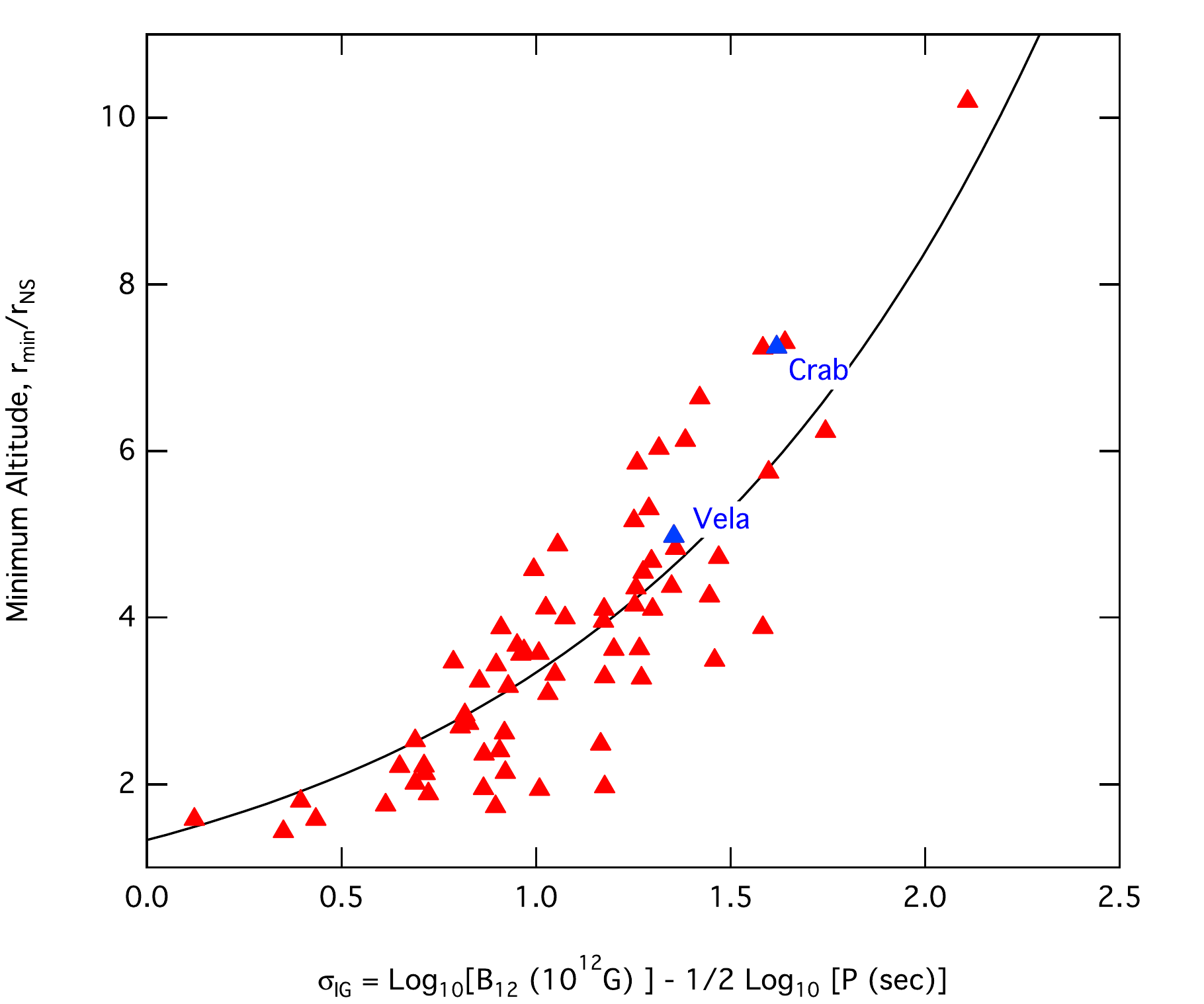}}
\caption{Minimum emission radii for a selection (see text) of young
pulsars among the 117 pulsars in the second {\it Fermi} pulsar catalog
\citep{PSRCat2}.  The minimum radius in neutron star radii is plotted
against a scaling of the logarithm of \teq{B_{12}/\sqrt{P}}. This
couples directly to the dependence suggested in
Eq.~(\ref{eq:rmin_sigmaIG_dep}) for flat space-time, an indication of
which is provided by the black curve, which represents \teq{r_{\rm
min}/\rns = 4/3\times 10^{2\sigma_{\rm IG}/5}}. These minimum radii are
found by solving \teq{\tau(\rE,\erg_c)=1} for \teq{\rE}, using twice the
exponential cutoff energies \teq{E_c} measured by {\it Fermi}-LAT
\citesmartpp{PSRCat2}, and assuming that emission takes place on the
last open field line.  If the same calculation is done for the VERITAS
detection of the Crab pulsar at 120 GeV, we get a minimum radius of
\teq{20\rns}, while the MAGIC claim of pulsed emission up to
\teq{350-400}GeV would raise the bound to \teq{r_{\rm min}\sim 31\rns};
both far exceed the minimum radius of \teq{7.25\rns} obtained from the
{\it Fermi} cutoff energy.}
 \label{fig:fermi_rmins}
\end{figure}

Fig. \ref{fig:fermi_rmins} shows the minimum radii of emission for the
pulsars listed in the second {\it Fermi} catalog, plotted against \teq{\sigma_{\rm IG}},
a proxy for the logarithm of \teq{B_p/\sqrt{P}}.  These minimum radii are
calculated using a test photon energy that is twice the cutoff energy
$E_c$ published in the catalog paper \citesmartpp{PSRCat2}. The choice of
\teq{2E_c} does not exactly represent the highest energy photons detected
from the source, but is slightly above the energy where the \teq{\nu F_\nu}
is at a maximum, modulo a factor that depends on the power-law spectral
index below the cutoff.  The cutoff energies from the fits range from
0.4 to 5.9 GeV for the sources where spectral fitting produces
significant determinations of \teq{E_c}; the Crab pulsar is an exceptional
case that will be isolated below.  The handful of pulsars with poor
spectral fits in Table 9 of \citesmart{PSRCat2} are not included in this
analysis.  

Pulsars with magnetic fields too low for magnetic pair
creation to play a significant role in propagation of GeV-band photons in their
magnetospheres, including nearly all millisecond pulsars, are not constrained
at all by this technique; their pair attenuation ``minimum radii" are
not only inside the neutron star, but inside the Schwarzschild radius,
and therefore not meaningful.
As an example, PSR B1821-24 in the globular cluster M28, with its 
period \teq{P=3.05}ms and period derivative \teq{{\dot P}=1.62 \times 10^{-18}}sec 
sec$^{-1}$, possesses the highest known spin-down luminosity of any MSP \citep{Johnson13}.
Its surface polar field is \teq{B_p\approx 4.5\times 10^9}Gauss, i.e. approximately 
\teq{10^{-4}B_{\rm cr}}.  One can infer from Fig.~\ref{fig:Eesc_flatspace} that
its polar cap angle of \teq{\approx 15^{\circ}} would then yield an escape energy 
of around 30 GeV, a factor of 100 higher than for the \teq{B_p=0.01} case illustrated
therein.  This \teq{\eesc} is above the turnover energy \teq{E_c\sim 6}GeV
determined in the pulsed {\it Fermi}-LAT data in \citet{Johnson13}, so that 
pair transparency for all photons detected by the LAT can 
be presumed at altitudes above the surface.

The \teq{r_{\rm min}} values in Fig.~\ref{fig:fermi_rmins} display a
general increase with \teq{\sigma_{\rm IG}}, but considerable scatter.
The relation in Eq.~(\ref{eq:rmin_sigmaIG_dep}) is schematically
represented in the Figure via the black curve \teq{r_{\rm min}/\rns =
4/3\times 10^{2\sigma_{\rm IG}/5}}, corresponding to a value of
\teq{\eesc \approx 4\times 10^3} (i.e., 2.0 GeV) in
Eq.~(\ref{eq:rmin_sigmaIG_dep}). This defines the general character of
the minimum altitude calculations, so that the scatter expresses the
distribution of \teq{E_c} values in the {\it Fermi} pulsar spectral
database.  The point in the top right corner, with an \teq{r_{\rm min}}
value over \teq{10 \rns}, is for PSR J1119-6127, a high-field pulsar
with a surface polar field well above the quantum critical value.  For
the Crab and Vela pulsars, the minimum radii calculated using the {\it
Fermi}-LAT cutoff energies are indicated on the plot.  If one performs
the same calculation for the VERITAS detection of pulsed emission from
the Crab pulsar at 120 GeV \citesmartpp{Aliu_Crab}, the minimum emission
radius for those 120 GeV photons is approximately \teq{20\rns }.  Even
more interestingly, the MAGIC experiment has displayed evidence of
pulsed emission from the Crab up to 350--400 GeV \citep[see Fig. 4
of][]{Aleksic Crab}, for which the \teq{r_{\rm min}} determination
increases to around \teq{31\rns}, or approximately \teq{0.2\rlc}.  This
is a profound constraint that impacts the discussion of slot gap versus
outer gap models of the Crab, and is completely independent of altitude
inferences from pulse profile geometric analyses.  Observe that if the
lower spin-down field estimate of \cite{MT77} is adopted, this bound
drops only to \teq{r_{\rm min}\approx 23\rns \approx 0.15\rlc}, still
offering an important constraint.  Note also that at these altitudes, one
might expect rotational aberration to be influential, perhaps more 
so in the trailing edge of emission.  By inspection of the altitude 
bound in Fig.~\ref{fig:rmin_fullrange} for lower energy (GeV) photons, 
we anticipate that aberration will provide at most a modest modification 
to this \teq{r_{\rm min}\sim 0.2\rlc} determination.

To provide a striking contrast to this case, the same analysis 
can be performed for the softest {\it Fermi}-LAT gamma-ray pulsar, 
PSR J1513-5908 (B1509-58), whose signal extends only out to around 230 MeV.  
This high-field pulsar has a period of \teq{151}ms, is fairly faint 
in the LAT band \citep{Fermi B1509} and is detected by AGILE \citep{AGILE B1509}, also 
out to around 200-300 MeV.  It was not detected by EGRET on the 
Compton Gamma-Ray Observatory, but was seen by COMPTEL 
at energies below 30 MeV \citep{COMPTEL B1509}.
Using 230 MeV to set \teq{\eesc}, we infer a minimum emission
radius of approximately \teq{2.3\rns}.  While this is not constraining 
for slot gap and outer gap models, it is of significant interest that it is somewhat above the 
surface.  The broad pulse profile \citep{Fermi B1509} implies that a more substantial range of 
emission altitudes can be accommodated than is typical for gamma-ray pulsars.  Moreover,
the fact that one of the gamma-ray peaks very slightly leads the radio pulse peak 
\citep{Fermi B1509} may be an indicator of a polar cap type component to 
the pulsar's signal.  Early interpretation of the EGRET upper limits argued for the
action of photon splitting in the strong fields near the surface of PSR J1513-5908 \citep{HBG97},
which has \teq{B_p = 3.1\times 10^{13}}Gauss.
This suggestion was predicated on the contention that polar cap models could 
account for the emission in this pulsar, and required the gamma-ray 
attenuation by the splitting process to arise proximate to the polar cap.
The pair transparency altitude bound computed here indicates that the magnetic field local to 
the emission of \teq{\sim 230}MeV photons is less than \teq{\sim 2.4\times 10^{12}}Gauss, 
a lower field domain that strongly inhibits photon splitting opacity
\citep[e.g.,][]{BH01} in a neutron star magnetosphere.

It should be noted that if the magnetic field includes higher-order
multipole components, as in \citet{AS79}, the minimum emission radii calculated using this
technique will change.  The addition of a toroidal component to the
magnetic field will increase the magnetic field magnitude and decrease
the field line radius of curvature in most of the inner magnetosphere,
which will make it easier for photons to transit across field lines and
thereby enhance the attenuation of photons by  magnetic pair creation. 
This must then be compensated by moving the emission regions to higher
altitudes where the field is lower.  For this case, therefore, our
\teq{r_{\rm min}} estimates serve as conservative lower limits.  An
off-center dipole magnetic field \citep[e.g.,][]{HM11} will introduce an
asymmetry in field magnitude and radius of curvature, which will bring
\teq{r_{\rm min}} down on the ``stretched" side and raise it up on the
``compressed" side.  This provides a modest range of minimum altitudes
so that our computations serve as approximate guides.  We anticipate
that the modifications introduced by offset dipolar morphology are
tantamount to selecting a field line of slightly different footpoint
colatitude from that for the last open field line in a true dipole
field; interpretation of the numerical results presented here can then be 
adjusted accordingly.

The second observational context considered here concerns magnetars,
namely soft gamma repeaters (SGRs) and anomalous X-ray pulsars (AXPs).
With their supercritical fields, these energetic cousins of pulsars
possess steady pulsed emission that is very different from that of
gamma-ray pulsars. They have prominent thermal X-ray emission below 10
keV, and virtually equally luminous hard X-ray emission
\citep[e.g.,][]{KHM04,Goetz06,denHartog08,dHKH08} that is well-described
by power-law tails. These flat tails cannot extend beyond energies of a
few hundred keV due to constraining upper limits obtained by the COMPTEL
instrument on the Compton Gamma-Ray Observatory. A prominent model for
the tail emission is that it is due to inverse Compton scattering of
surface thermal X-rays by relativistic electrons in the strong fields of
the inner magnetosphere \citep{BH07,NTZ08,BWG11,Beloborodov13}. Such
emission should then be subject to the magnetic pair attenuation that is
the focus of this paper.  Here we determine the energy ranges for which
pair opacity should strongly attenuate the tail spectra. Magnetars also
exhibit flaring episodes, ranging in luminosity up to the giant flares
seen in three soft gamma repeaters.  This emission also does not extend
beyond about 1 MeV.  A comprehensive observational and theoretical
summary of magnetars can be found in \cite{Mereghetti08}.

\begin{figure}[h]
\centerline{\includegraphics[width=.75\textwidth]{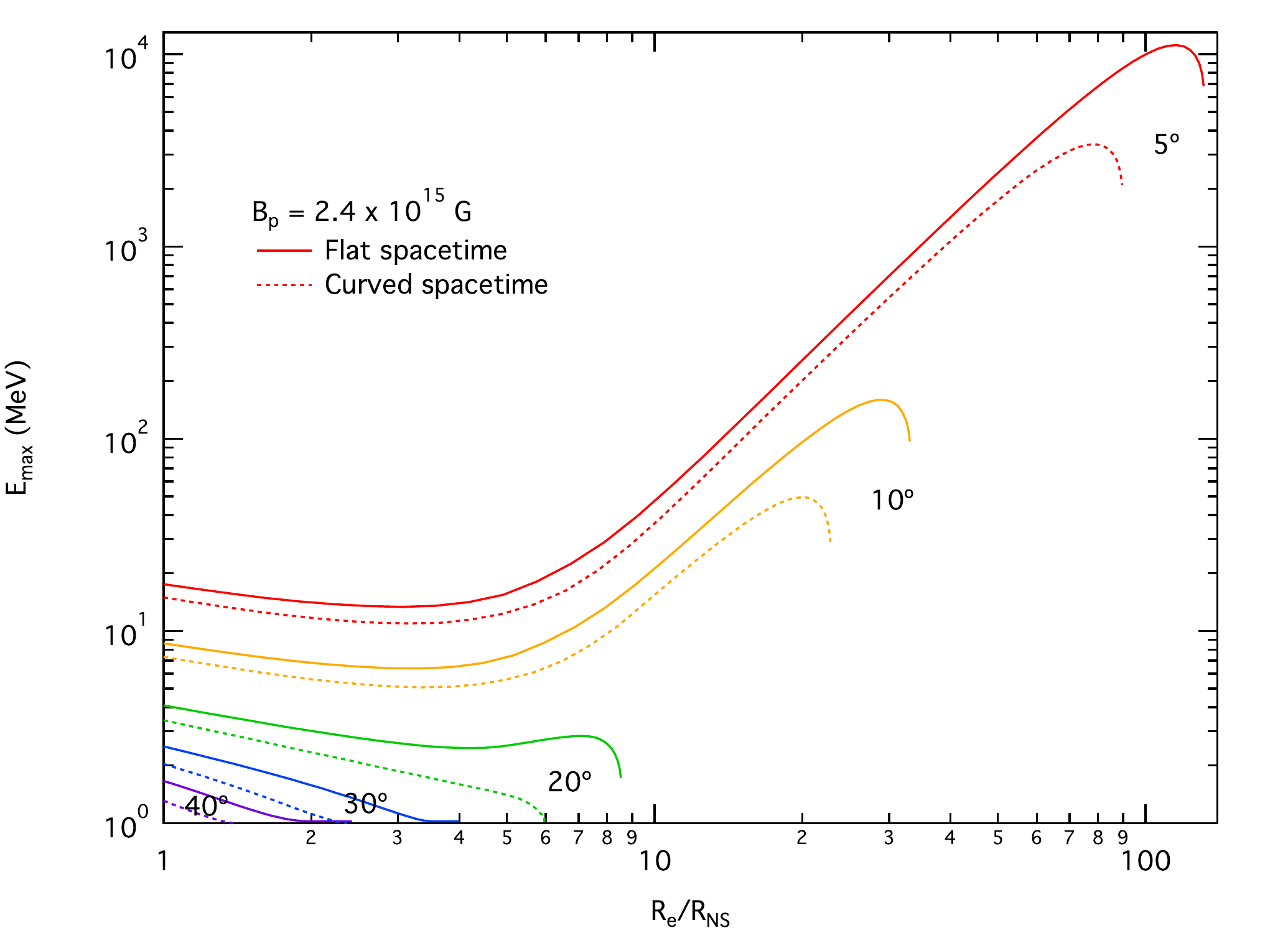}}
\caption{Pair creation escape energies \teq{\eesc} for photons emitted along 
four different field lines  of a magnetar with a surface polar magnetic field of 
\teq{2.4 \times 10^{15}}Gauss.  These are plotted as functions of the radius 
of emission in neutron star radii.  Curves are labeled with the ``footpoint colatitude" 
of the field line, and the selected values 
(\teq{\theta_f=5^{\circ},10^{\circ}, 20^{\circ}, 30^{\circ}, 40^{\circ}}) correspond to 
the closed field line region. Solid curves represent flat spacetime and dashed curves 
fully include general relativistic effects.  For emission below 10 stellar radii, one would not 
expect photons to emerge from the magnetosphere above around 10 MeV.}
\label{fig:magnetar_emax}
\end{figure}

For magnetar applications, the asymptotic approximations to the
attenuation rate given in Eqs.~(\ref{eq:Erber_asymp}) and
(\ref{eq:B88_asymp}) are not applicable at low altitudes, and the full
form of the rate given in \cite{DH83} must be used, i.e. that embodied
in Eqs.~(\ref{eq:tpppar}) and~(\ref{eq:tppperp}).  This is because for
magnetic fields above \teq{B_{\rm cr}}, the pair conversion optical
depth is sensitive to the individual thresholds for each Landau level
transition. We use here a hybrid protocol that calculates the
attenuation rate for the lowest two Landau levels exactly, and at higher
energies uses the BK07 asymptotic form. The magnetic pair creation
escape energies for photons emitted along dipole magnetic field lines
are plotted as a function of altitude of emission in
Fig.~\ref{fig:magnetar_emax}.  Each set of curves is labeled with the
footpoint colatitude of the field line when it intersects the magnetar
surface.  Aberration influences are omitted in these calculations,
although GR modifications are included for the curves indicated. The
differences in \teq{\eesc} values between the general relativistic and
flat spacetime results are manifested for all altitudes because for a
fixed footpoint colatitude, GR distorts the field line at all altitudes,
increasing the curvature so that it crosses the magnetic equator at a
smaller radius. Accordingly, the escape energies are lowered slightly. 
If one were to plot results for the last open field line, the footpoint
colatitudes would differ between flat spacetime and GR cases according
to Eq.~(\ref{eq:polar_cp_GR}), but the \teq{\eesc} curves would merge at
moderate to high altitudes.

The surface polar magnetic field used for this plot is \teq{2.4\times
10^{15}}Gauss, half the estimated surface magnetic field strength for
SGR 1806-20 and somewhat above the corresponding value for SGR 1900+14. 
At typical magnetar periods, the polar cap size (angle \teq{\theta_p})
is less than 1 degree, so all of these curves represent emission taking
place in the closed field line region.  All photons with energies above
these \teq{\tau (\eesc )=1} curves will be absorbed via magnetic pair
creation before they can escape to infinity.  Since these curves are
computed for cases where emission is presumed to be beamed along field
lines, for non-zero initial \teq{\thetakB} angles, the opacity will
increase dramatically and reduce \teq{\eesc} so that these bounds will
become more stringent.

The curves in Fig.~\ref{fig:magnetar_emax} decline at first with
increasing emission altitude, and then exhibit pronounced rises.  The
weak decline is associated with the regime of pair creation near
threshold when the field at \teq{\rE} is near-critical or supercritical,
up to around \teq{\rE/\rns\sim 3-6}.  In this portion of parameter
space, \teq{\eesc} is relatively insensitive to the field strength, and
drops slightly with rising \teq{\rE} since increasing field line
curvature allows photons to more readily propagate across field lines. 
The curves then transition into emission altitudes where the field is
subcritical, and the familiar \teq{\eesc\propto \rE^{5/2}} in
Eq.~(\ref{eq:e_esc_Erber}) is approximately realized --- this is clearly
evident in the \teq{5^{\circ}} footpoint colatitude case.  The turndowns
at the extreme highest altitudes correspond to quasi-equatorial emission
cases where the photons again readily transit across field lines that
exhibit greater flaring. Smaller radii of field line curvature are also
responsible for the trend of declining escape energy with increasing
footpoint colatitude \teq{\theta_f}, with an approximate dependence
\teq{\eesc\propto 1/\theta_f} when \teq{\theta_f\ll 1}.  For somewhat
lower surface polar field cases that are not shown, for near-surface
emission the escape energy rises, but not dramatically, since pair
creation still occurs near the \teq{2m_ec^2} threshold.  This situation
is applicable to the lower field magnetars such as AXP 4U 0142+61 and
AXP 2259+586 which possess \teq{B_p=2.7\times 10^{14}}Gauss and
\teq{B_p=1.2\times 10^{14}}Gauss, respectively.  In particular, the
\teq{\eesc\propto \rE^{5/2}} domain for \teq{\theta_f=5^{\circ},
10^{\circ}} then penetrates to lower emission altitudes in such cases,
and \teq{\eesc} rises at higher \teq{\rE}, as the magnetar environment
starts to transition towards the sub-critical \teq{B_p} regime that is
treated in Fig.~\ref{fig:EescGR}.

This opacity phase space plot clearly indicates that magnetars should 
not be visible to the {\it Fermi}-LAT telescope if their activation takes 
place in the inner magnetosphere, which is the preferred paradigm 
for twisted magnetosphere models \citep[e.g.,][]{Beloborodov09,PBH13} 
of magnetars.  We can see, for example, that 100 MeV
photons cannot escape at all from the region bounded by the closed field
line that crosses the neutron star surface at a colatitude of
\teq{\theta_f=10^\circ}, and can only escape from the region bounded by the field
line with a footpoint colatitude of \teq{5^\circ} if they originate more
than 10 neutron star radii from the center of the star.  This may
explain the non-detection of 13 magnetars at {\it Fermi}-LAT energies
(\citealt{Fermi magnetar}; see also \citealt{SG10})
--- any emission above 100 MeV must emanate 
from high altitude zones in the outer magnetosphere.  This would have to be 
the case if the claim of a \teq{5\sigma} detection of pulsed emission above 200 MeV
in a 4 year collection of {\it Fermi}-LAT public archive data for AXP 1E 2259+586 
\citep{Wu13} is confirmed.  Such evidence of pulsation in this source is not 
found in the analysis of \cite{Fermi magnetar}, and no public confirmation of the \cite{Wu13} 
claim has been offered by the {\it Fermi}-LAT Collaboration to date.
If emission from select magnetars at energies above 100 MeV is eventually 
verified, the pair attenuation studies presented here have important implications 
for the paradigms surrounding magnetar quiescent emission.  If it is not confirmed, 
then pair opacity provides a natural explanation why magnetars are not seen in 
hard gamma-rays.

\section{CONCLUSIONS}
In this paper, single-photon (magnetic) pair creation transparency
conditions for neutron star magnetospheres have been calculated,
beginning with the simplest case of a non-rotating star in the absence
of general relativistic effects, and progressing to sequentially
consider curved spacetime and then the influences of rotational
aberration.  Optical depths are calculated for arbitrary photon emission
points in neutron star magnetospheres in the special case where photons
are initially parallel to the magnetic field.  Such initial propagation
conditions are expected in pulsars because electrons that emit curvature
radiation gamma rays have such high Lorentz factors that all the photons
will be subject to very strong relativistic beaming along {\bf B}.  A similar
situation exists for magnetars, if their hard X-ray signals are
generated by inverse Compton scattering using extremely relativistic
electrons.  The optical depth determinations enable the presentation of
attenuation lengths, minimum altitudes of emission for a given energy,
and escape energies for a given emission altitude, at various stages of
the paper.  In developing these results, we have presented a set of
analytic and semi-analytic forms that greatly simplify the computation
of magnetic pair creation opacities. These approximations can be applied
to any emission altitude and most colatitudes, and were discussed in the
contexts of gamma-ray pulsars and magnetars.  

The principal findings of our analysis are as follows.  Flat spacetime
computations reproduce extant results in the literature nicely, and
indicate that pair attenuation escape energies anti-correlate with the
colatitude of emission and the surface polar field strength, as expected
due to the mathematical character of the pair attenuation coefficient. 
The introduction of general relativistic modifications modestly reduces both
attenuation lengths and escape energies for pair creation at
low altitudes near the stellar surface, but provides almost negligible
alteration from flat spacetime results above \teq{\sim 5} stellar radii.
 For the inner magnetospheric considerations germane to \teq{\gamma\to
e^+ e^-} opacity, rotational corrections to the static pair attenuation
analyses are generally found to be quite limited, except for a small
domain where photons are emitted approximately directly above the magnetic
pole.  In such cases, rotational aberration forces photons to propagate
across field lines almost immediately after emission, as viewed by the
distant, static observer.  For emission colatitudes below around 3
degrees, depending
on the directional phase of photon emission, aberration can either 
increase or decrease the instantaneous pair conversion rates 
relative to those for the non-rotating case, and accordingly the escape energy
can either rise or decline.

For young pulsars, the paper generates estimates of the minimum altitude
\teq{r_{\rm min}} that permits pair transparency out to the maximum
gamma-ray energies detected by {\it Fermi}-LAT.  These \teq{r_{\rm min}}
values are one of the few constraints available on the emission location
in gamma-ray pulsars with a single peak, and they have the advantage of
being a physics-based constraint that is not solely dependent on the
geometry of the emitting region.  The minimum emission altitudes that we
calculate from magnetic pair creation are normally far below those
obtained from pulse profile fitting with slot gap or outer gap models,
which are typically \teq{r_{\rm min} > 0.05\rlc} for two-peaked
pulsars.  They become much more useful constraints on
curvature-radiation-based models for the significant number of
single-peaked young pulsars, for which pulse profile modeling is not
effective in the absence of a pulsed radio counterpart.  Interestingly,
for the Crab pulsar, the altitude bound rises to around \teq{r_{\rm
min}\sim 20\rns} due to its energetic emission confirmed by both VERITAS and MAGIC 
out to 120 GeV; this bound is raised further to \teq{r_{\rm
min}\sim 31\rns} when using the report of pulsed emission in MAGIC data out to 350--400 GeV.  
This most extreme bound is around 20\% of the Crab's light cylinder radius;
it was obtained when omitting rotational aberration effects, 
which will modify the limit somewhat, but not drastically.
Our results are clearly not applicable to millisecond pulsars, where the
surface magnetic fields are too low for magnetic pair creation opacity
to be significant.  The pair creation calculations presented are
germane to magnetars, and they indicate that soft gamma repeaters and
anomalous X-ray pulsars should not be detectable above 100 MeV 
by {\it Fermi}-LAT unless their emission regions are generally at altitudes of 
around 10 stellar radii or higher.

\vskip 10pt
\acknowledgments 
We thank Alice Harding and Peter Gonthier for helpful discussions, and 
for comments following a careful reading of the manuscript.   We also thank the 
referee for some suggestions helpful to the polishing of the paper.
We are grateful for the generous support of the National Science Foundation
through grant AST-1009725, and the NASA Astrophysics Theory and {\it Fermi} Guest Investigator Programs
through grants NNX09AT79G, NNX10AC59A and NNX11AO12G.

\newpage

\appendix

\section{Appendix A: Approximating the Photon Trajectory Curvature Integral}

The photon trajectory in curved spacetime is defined by colatitude \teq{\theta}
expressed as an integral over the propagation altitude parameter \teq{\Psi = r_s/r}, 
where \teq{r_s} is the neutron star's Schwarzschild radius.  
The angle in Eq.~(11) of GH94 is the difference  
between the angle (in the local inertial frame) of the photon momentum vector 
to the radial vector at the point of emission, and the angle of the photon 
trajectory to the local radial vector at a point defined by 
\teq{\Psi}; it relates to \teq{\theta} as follows:
\begin{equation}
   \theta(\Psi) \; \equiv\; \thetaE + \Delta\theta \; =\; 
   \thetaE + \int_{\Psi}^{\PsiE} \frac{d\Psi_r}{\sqrt{\Psi_b^2-\Psi_r^2(1-\Psi_r)}}\quad .
 \label{eq:curved_traj_app}   
\end{equation}
Since \teq{\Psi \leq\PsiE} in this construction, as the photon
propagates out from the star, then the change in colatitude
\teq{\Delta\theta} is necessarily positive as the altitude \teq{r}
increases. Also, \teq{\Psi_b=r_s/b} expresses the general relativistic
impact parameter \teq{b} for the unbound photon path.

Computation of the trajectory using numerical integration is expensive
in terms of time, particular for repeated applications in Monte Carlo
simulations of magnetospheric cascades, so it is expedient to derive an
analytic approximation to the integral in Eq.~(\ref{eq:curved_traj}).
Using manipulations outlined in Chapter 17 of Abramowitz \& Stegun
(1965), this integral can be expressed in terms of elliptic functions. 
Such a step does not facilitate its evaluation, since the parameter
\teq{\PsiE/\Psi_b} is not necessarily small, a condition that would
render series expansion of elliptic functions more amenable.  In our
neutron star cases, \teq{\PsiE\lesssim 0.4} is generally realized, and
this suggests a series expansion in this parameter.  To effect such, we
have designed an expansion algorithm (not unique) that is motivated by
the flat spacetime limit \teq{\PsiE\to 0} of the integral. Define
\begin{equation}
   \rho_f\; =\; \sqrt{\Psi_b^2- \Psi_r^2}
   \quad ,\quad
   \rho_c\; =\; \sqrt{\Psi_b^2- \Psi_r^2(1- \Psi_r )}
 \label{eq:rhof_rhoc_def}
\end{equation}
as flat and curved spacetime forms, respectively, of the denominator of the integrand of the 
trajectory integral.  A Taylor series expansion for \teq{\rho_c/\rho_f = \sqrt{1+ \Psi_r^3/\rho_f^2}}
can be developed in the generally small parameter \teq{\Psi_r^3/\rho_f^2}.  Note that 
this parameter is not much less than unity for near-surface, equatorial cases.  This 
protocol results in a series expansion in \teq{\Psi_b} for the integral:
\begin{equation}
   \Delta\theta \; =\; \int^{\PsiE}_{\Psi} \dover{d\Psi_r}{\rho_f}
   \left\{ 1 - \dover{\Psi_r^3}{2\rho_f^2} + \dover{3\Psi_r^6}{8\rho_f^4}
   - \dover{5\Psi_r^9}{16\rho_f^6} + \dover{35\Psi_r^{12}}{128\rho_f^8} \right\} + O \left( \PsiE^5 \right) \quad .
 \label{eq:traj_scaled_approx}
\end{equation}
Now define the scaled parameters
\begin{equation}
   \Upsilon\; =\; \dover{\Psi}{\Psi_b}
   \quad ,\quad 
   \Upsilon_{\hbox{\sixrm E}}\; =\; \dover{\PsiE}{\Psi_b}\quad .
 \label{eq:Psi_scaled}
\end{equation}
The integrals in Eq.~(\ref{eq:traj_scaled_approx}) are all analytically tractable, and yield 
a useful analytic approximation for the photon trajectory:
\begin{equation}
   \Delta\theta \; \approx\; \delthetaapp\;\equiv\; 
   \biggl[ \arcsin \upsilon  - \Psi_b\, f_1(\upsilon ) + \Psi_b^2\, f_2(\upsilon ) 
   - \Psi_b^3 \, f_3(\upsilon )\biggr]^{\Upsilon_{\rm E}}_{\Upsilon}  \quad ,
 \label{eq:traj_approx_final}
\end{equation}
when retaining only the first four terms in the integrand of Eq.~(\ref{eq:traj_scaled_approx}).
Here
\begin{eqnarray}
   f_1(\upsilon) & = & \dover{2-\upsilon^2}{2\sqrt{1-\upsilon^2}} - 1\nonumber\\[-0.5pt]
   f_2(\upsilon) & = & \dover{15}{16}\, \arcsin \upsilon 
               - \dover{\upsilon\, (15 - 20\upsilon^2 + 3 \upsilon^4)}{16\, (1-\upsilon^2)^{3/2}}
 \label{eq:}\\[-0.5pt]
   f_3(\upsilon) & = & \dover{128 - 320 \upsilon^2 + 240 \upsilon^4 - 40 \upsilon^6 - 5 \upsilon^8}{
               48 (1 - \upsilon^2)^{5/2}} - \dover{8}{3}\quad . \nonumber
\end{eqnarray}
This provides an alternative to the Beloborodov (2002) approximation. As
constructed, for small arguments \teq{\upsilon}, the functions employed
in the approximation scale as \teq{f_n(\upsilon )\propto
\upsilon^{1+3n}}. This regime is sampled for \teq{\Psi_b\gg 1}, the low
impact parameter cases appropriate for circumpolar colatitudes. 
Accordingly, the series implied by extension of
Eq.~(\ref{eq:traj_approx_final}) to higher order terms is nicely
convergent even when \teq{\Psi_b} is large.

\begin{figure}[h]
 \centerline{
  \includegraphics[width=.65\textwidth]{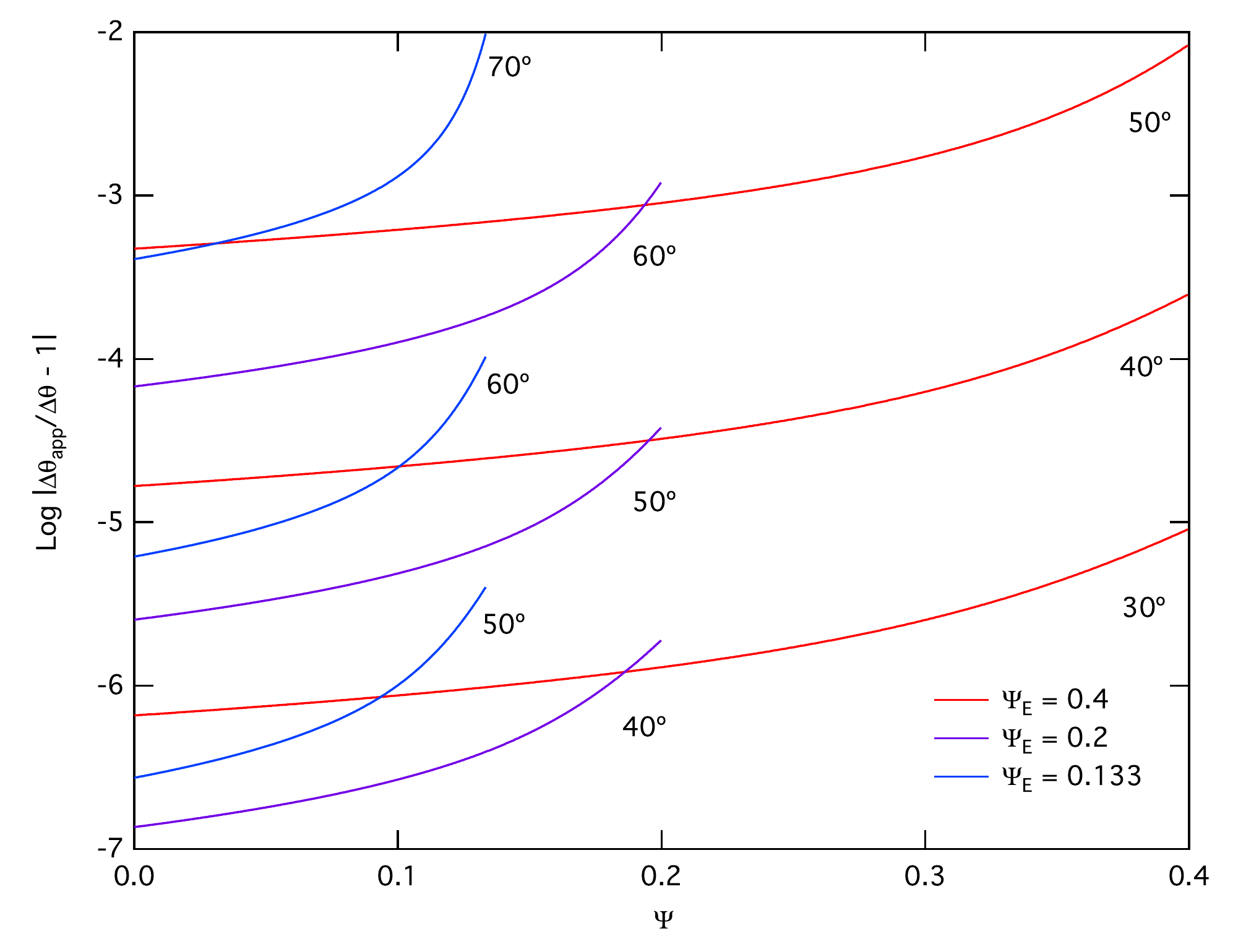}
  }
  \caption{The fractional precision \teq{\vert \delthetaapp /\Delta\theta -1\vert} 
of the approximation in Eq.~(\ref{eq:traj_approx_final}) to the full 
trajectory integral in Eq.~(\ref{eq:curved_traj}) for photon propagation 
in curved spacetime.  Three groups of curves, color-coded, are illustrated
for altitude parameters \teq{\PsiE=0.4, 0.2, 0.133}, as marked, corresponding 
to emission at the neutron star surface and at two and three stellar radii.
The range of altitudes \teq{0 \leq \Psi \leq \PsiE} spans from the 
emission locale all the way out to infinity.  Within each group are three 
curves for emission colatitudes \teq{\thetaE}, as labelled, illustrating how 
the precision of the approximation improves nearer the magnetic axis.
}
 \label{fig:traj_integ_approx}
\end{figure}

For photon emission from the neutron star surface, with
\teq{\PsiE\approx 0.4}, this approximation for the transit in colatitude
is accurate to better than 0.1\% at all subsequent altitudes for
emission colatitudes \teq{\thetaE\lesssim \pi/4}.  Raising the altitude
of emission, i.e. reducing \teq{\PsiE} below \teq{0.1} substantially
improves this.  This level of precision is entirely suitable for the
pertinent pulsar parameter space, where footpoint colatitudes are
usually inferior to \teq{\theta_f\lesssim 30^{\circ}}. To illustrate
this, in Fig.~\ref{fig:traj_integ_approx} we plot the fractional
precision \teq{\vert \delthetaapp /\Delta\theta -1\vert} of the
approximation in Eq.~(\ref{eq:traj_approx_final}) relative to the exact
integral in Eq.~(\ref{eq:curved_traj}) or
Eq.~(\ref{eq:curved_traj_app}), as a function of \teq{\Psi} for
different colatitudes \teq{\thetaE} and altitudes \teq{\PsiE}
representative of the locales sampled in the pair attenuation
calculations. Given an emission altitude parameter \teq{\PsiE}, the
emission colatitude \teq{\thetaE} can then be used to define the impact
parameter \teq{\Psi_b} in Eq.~(\ref{eq:Psi_b_def}).  The precision
clearly is degraded at high colatitudes, though is always better than
1\% for the colatitudes illustrated.  Yet, because the general
relativistic curvature is diminished at higher emission altitudes, the
range of colatitudes \teq{\thetaE} yielding a given level of precision
increases as \teq{\PsiE} declines.  To retain 0.1\% precision, it is
best to restrict use of the approximation to field line footpoint
colatitudes \teq{\theta_f\lesssim 45^{\circ}}.

\section{Appendix B: Approximating the Optical Depth for Zero-Colatitude Emission in a Rotating Magnetosphere}
 \label{sec:abapprox}

To facilitate semi-analytic checks on escape energies and minimum  
altitude numerics, we consider here the special case of  
aberration-corrected emission from directly above the magnetic pole  
(\teq{\thetaE = 0}), for which the initial velocity and magnetic field  
vectors assume particularly simple forms. Using a combination of  
Eq.~(\ref{eq:kOdef}) and Eqs.~(\ref{eq:Brot}) and~(\ref{eq:betae}) the  
direction of photon travel in the inertial observer frame simplifies to
\begin{equation}
\mathbf{\hat{k}_O} \; \approx\;
    \begin{bmatrix}
       \sin\alpha_i\\
       -\betaE\\
       \cos\alpha_i
    \end{bmatrix},
  \label{eq:kO_polarcase}
\end{equation}
where \teq{\betaE = 2\pi \rE \sin\alpha_i /cP \equiv h\beta_p} is the  
magnitude of the corotation velocity divided by \teq{c}.  In this  
result, we have used the fact that \teq{\betaE\ll 1} at the low to  
moderate altitudes of interest for this5B development.  Remember that  
\teq{\beta_p = \rns\sin\alpha_i/\rlc} is the corotation speed of the  
magnetic pole at the stellar surface. Then the angle between the  
photon trajectory and the radial direction in the observer frame,  
given by Eq.~(\ref{eq:deltaE}), reduces to the simple form
\begin{equation}
    \delta_{E,0} \;\approx\; \arcsin \left( \dover{2\pi \rE  
\sin\alpha_i}{Pc}\right) \quad .
  \label{eq:deltaE0_approx}
\end{equation}
Since \teq{\betaE \ll 1}, this is approximately
\begin{equation}
    \delta_{E,0} \;\approx\;  \dover{2\pi \rE \sin\alpha_i}{Pc}  
\;\equiv\; h\beta_p\quad .
  \label{eq:deltaE0_approx_v2}
\end{equation}
As in Eq.~(\ref{eq:chi_def}) we define \teq{\chi} to be the radial  
distance of the photon from the center of the neutron star divided 
by the radius of emission.  Applying the law of sines, we have
\begin{equation}
    \chi \; =\; \dover{\sin\delta_{E,0}}{\sin(\delta_{E,0}-\eta)}
    \;\approx\; \dover{1}{1-\eta/\delta_{E,0}} \quad ,
 \label{eq:chi_def_appB}
\end{equation}
since both \teq{\delta_{E,0}} and \teq{\eta} are small.  Accordingly,  
escape to infinity corresponds to \teq{\eta \to \delta_{E,0}}.  
Similarly, for \teq{s}, the law of sines and small-angle  
approximations give
\begin{equation}
s \approx \rE \left(\chi-1\right).
\end{equation}
These can be carried through the coordinate transformations to express the
approximate coordinates for the photon path in the star frame in terms  
of \teq{\eta}:
\begin{equation}
    \rS \;\approx\; \frac{\rE}{1-\eta/\delta_{E,0}}
    \quad ,\quad
    \thetaS \;\approx\; \eta 
    \quad ,\quad
    \phiS \;\approx\; -\dover{\pi}{2}   \quad .
  \label{eq:thetae0path}
\end{equation}
Observe that \teq{\phiS \approx -\pi/2} is an angular restriction that follows from  
the rotation velocity \teq{\vec{\betaE} = \mathbf{\Omega} \times  
\mathbf{\rE}} at the point of emission being approximately orthogonal  
to the plane defined by \teq{\vec{\mu}} and \teq{\mathbf{\Omega}},  
since \teq{\mathbf{\rE}} is nearly parallel to \teq{\vec{\mu}}.  With these approximations 
one can obtain a relatively compact form for the magnetic field and  
photon trajectory vector in the star frame.  For the field we have
\begin{equation}
    \mathbf{B_S} \;\approx\; \dover{3B_p \rns^3}{2\rE^3}  
\left(1-\dover{\eta}{\delta_{E,0}}\right)^3
    T_{\Omega}(t_{\hbox{\sixrm O}})  \cdot
    \begin{bmatrix}
       \dover{2}{3} \,\sin\alpha_i\\
       - \eta \\
       \dover{2}{3}\, \cos\alpha_i
    \end{bmatrix} \quad ,
\end{equation}
which, using \teq{\Omega t_{\hbox{\sixrm O}} \approx s/\rlc \approx 
\delta_{E,0}\eta /[(\delta_{E,0}-\eta )\,\sin\alpha_i ]\ll 1} and 
Eqs.~(\ref{eq:deltaE0_approx}), (\ref{eq:deltaE0_approx_v2}) 
and~(\ref{eq:chi_def_appB}), is easily shown to be 
approximately equivalent to
\begin{equation}
    \mathbf{B_S} \;\approx\; \dover{B_p}{h^3}  
\left(1-\dover{\eta}{\delta_{E,0}}\right)^3
    \begin{bmatrix}
       \sin\alpha_i\\
       -\dover{3}{2}\, \eta + \dover{\delta_{E,0} \eta}{\delta_{E,0}-\eta}\\
       \cos\alpha_i
    \end{bmatrix} \quad ,
\end{equation}
The photon momentum vector is
\begin{equation}
    \mathbf{k_S} \; \approx\; \omega
    \begin{bmatrix}
       \sin\alpha_i\\
       \dover{\delta_{E,0}\eta}{\delta_{E,0}-\eta}\\
       \cos\alpha_i
    \end{bmatrix} \quad ,
 \label{eq:ks_approx}
\end{equation}
where the \teq{y} component is far inferior to the other two.
The magnitude of the cross product $|\mathbf{k_S} \times  
\mathbf{B_S}|$ then has a leading order term given by
\begin{equation}
    \Bigl\vert \mathbf{k_S} \times \mathbf{B_S}\Bigr\vert
    \; \approx\; \dover{3 \omega B_p}{2h^3} \,
       \dover{\eta \left(\delta_{E,0}-\eta\right)^3}{\delta_{E,0}^3} \quad .
\end{equation}
Note that in this identity, and also in Eq.~(\ref{eq:ks_approx}),
we can replace \teq{\omega} by the observer frame photon energy 
\teq{\erg} since the Doppler shift in Eq.~(\ref{eq:omegaSE_def})
simplifies for this \teq{\betaE\ll 1} case.

With all the quantities in Eq.~(\ref{eq:pp_rate_Oframe}) defined in terms
of the small variable \teq{\eta}, the integral can then be approximated as
in Section~\ref{sec:tauintegration} using the method of steepest descents.
The peak of the integrand occurs at
\begin{equation}
    \eta_{pk} \; =\; \dover{\delta_{E,0}}{4} \quad ,
  \label{eq:eta_peak}
\end{equation}
as we found for the non-rotating, flat spacetime case.  In practice, the attenuation  
rate drops to almost zero well below \teq{\eta_{max} \approx \delta_{E,0}}, 
due to the rapid decline in the field strength at high altitudes, and therefore
we set \teq{\eta_{\max}\to\infty} in the steepest descents protocol.  
This then yields
\begin{equation}
    \tau\; =\; \int_0^{\eta_{max}} \overline{R} \frac{ds}{d\eta} d\eta
    \; \approx \; \dover{3^6\sqrt{3}}{2^{18}} \,\dover{\fsc}{\lambar}\,
        \dover{B_p \rns^3}{\rE^2} \sqrt{\frac{8 \pi^4 B_p \rns^3  
        \erg \sin^3\alpha_i}{c^3 P^3}} \;
    \exp\left\{-\frac{2^{11} c P \rE^2}{3^5 \erg B_p \pi R^3 \sin\alpha_i}\right\} \quad ,
 \label{eq:tau_zero_app}
\end{equation}
which can be recast slightly as Eq.~(\ref{eq:tau_zero}). This satisfyingly 
simple result emerging from the complexity of aberration considerations
is only applicable in the limit of non-relativistic boosts at low altitudes 
between the star and observer frames: it can be reliably applied when 
\teq{\rlc/\rns \gtrsim 30}, which includes all {\it Fermi} young 
gamma-ray pulsars.  To determine  
the threshold for pair transparency, we set this integral equal to  
unity, and solve for one of the variables in terms of the others. From  
this analysis, it is clear that when \teq{\thetaE \rightarrow 0}, the  
minimum altitude for pair transparency for a given photon energy will  
attain a non-zero value and the escape energy for a photon emitted  
from a given altitude will remain finite.

\end{document}